\documentclass[11pt]{article}
\usepackage{amssymb}
\usepackage{amsmath}
\usepackage{graphics}

\catcode`@11
\def\seceqaa{\@addtoreset{equation}{section}
\def\theequation{A\arabic{equation}}}
\def\seceqbb{\@addtoreset{equation}{section}
\def\theequation{B\arabic{equation}}}
\def\seceqcc{\@addtoreset{equation}{section}
\def\theequation{C\arabic{equation}}}
\def\seceqdd{\@addtoreset{equation}{section}
\def\theequation{D\arabic{equation}}}
\def\seceqee{\@addtoreset{equation}{section}
\def\theequation{E\arabic{equation}}}
\def\seceqff{\@addtoreset{equation}{section}
\def\theequation{F\arabic{equation}}}
\def\seceqgg{\@addtoreset{equation}{section}
\def\theequation{G\arabic{equation}}}
\catcode`@11
\begin{document}
\begin{titlepage}
\begin{center}
{\large \bf On Aspects of Holographic  Thermal QCD at Finite Coupling}\\
Karunava Sil\footnote{e-mail: krusldph@iitr.ac.in}  and
 Aalok Misra\footnote{e-mail: aalokfph@iitr.ac.in}
 \\
Department of Physics, Indian Institute of Technology,
Roorkee - 247 667, Uttaranchal, India

 \date{\today}
\end{center}
\thispagestyle{empty}
\begin{abstract}
In the context of \cite{metrics}'s string theoretic dual of thermal QCD-like theories at finite gauge/string coupling (as part of the `MQGP' limit of \cite{MQGP}), we obtain the QCD deconfinement temperature compatible with lattice results for the right number of light flavors $N_f=3$,  and the correct mass scale of the light (first generation) quarks. The type IIB background of \cite{metrics} is also shown to be thermodynamically stable. Further, we show that the temperature dependence of DC electrical conductivity  mimics a one-dimensional Luttinger liquid, and the requirement of the Einstein relation (ratio of electrical conductivity and charge susceptibility equal to the diffusion constant) to be satisfied requires a specific dependence of the Ouyang embedding parameter on the horizon radius. These results arise due to the non-K\"{a}hlerity and non-conformality of the type IIB background. On the geometrical side we quantify the former (non-K\"{a}hlerity)  by evaluating the $SU(3)/G_2$-structure torsion classes of the local type IIA mirror/M-theory uplift. Analogous to what was shown for the type IIB background  in \cite{transport-coefficients}, we first show that the type IIA delocalized SYZ mirror (after fine tuning) can also be approximately supersymmetric. We then work out the $G_2$-structure torsion classes of the local M-theory uplift of the mirror type $\rm IIA$ metric  - in the large-$N$ limit at finite coupling, $G_2$ structure approaches $G_2$ holonomy.
\end{abstract}
\end{titlepage}.

\section{Introduction and Motivation}

In recent years it has been realized that the problem of strongly coupled gauge theories are best tackled by the gauge/string duality. One of the remarkable examples of this duality is the AdS/CFT correspondence \cite{maldacena} conjectured by Maldacena in 1997. According to this correspondence type IIB superstring theory in $AdS_5\times S^5$ is dynamically equivalent to the four dimensional $SU(N)$ Yang-Mills theory with large $N$ and $\mathcal{N}=4$ supersymmetry. This correspondence is actually based on the so called holographic principle: information of the bulk of dimension $d$ is mapped to a $d-1$ dimensional theory living on the boundary. A generalization of the AdS/CFT correspondence was required to gain a deeper insight into QCD. In particular, efforts have been made to relax some of the constraints such as  conformal symmetry of the gauge theory which was necessary for the validity of the correspondence.  In fact it is believed that strongly coupled thermal QCD `laboratories' like strongly coupled Quark Gluon Plasma (sQGP), apart from having a large t'Hooft coupling,  {\it  must also be characterized by finite gauge coupling \cite{Natsuume-sQGP}}. It is hence important to have a framework in the spirit of gauge-gravity duality, to be able to address this regime in string theory. {\it Finite gauge coupling would under this duality translate to finite string coupling hence necessitating addressing the same from an $M$ theory perspective.} This was initiated in \cite{MQGP} and \cite{transport-coefficients}.

In this work, using the top-down holographic thermal QCD model of \cite{metrics}, we have discussed some QCD-related properties at finite temperature, and most importantly, {\it at finite gauge coupling}\footnote{Note however, this is not a paper on QGP.}. It is largely in this respect that through this paper we will attempt to fill in an important gap by studying {\it at finite gauge coupling} (as part of the `MQGP limit' of \cite{MQGP}) for the first time:
 \begin{itemize}
 \item
 Physics-related issues such as:
 \begin{itemize}
 \item
  evaluation of lattice-compatible $T_c$ for the right number and masses of light quarks,
  \item
  demonstrating the thermodynamical stability of \cite{metrics},
  \item
  obtaining the temperature dependence of electrical conductivity $\sigma$, charge susceptibility $\chi$ and hence seeing the constraints which the Einstein's law (relating $\frac{\sigma}{\chi}$ to the diffusion constant) imposes on the holomorphic Ouyang embedding of $D7$-branes into the resolved warped deformed conifold geometry of \cite{metrics};
  \end{itemize}
  \item
  Math-related issues such as:
  \begin{itemize}
  \item
  quantifying the non-K\"{a}hlerity (which is what influences the Physics issues alluded to above) of the delocalized Strominger Yau Zaslow (SYZ) type IIA mirror of \cite{metrics} constructed in \cite{MQGP} by evaluating the $SU(3)$ structure torsion classes (the same for the type IIB background of \cite{metrics} were evaluated in \cite{transport-coefficients}),
  \item
  evaluating the $G_2$-structure torsion classes, and hence obtain for the first time, an explicit $G_2$-structure of the $M$-theory uplift of the type IIB holographic model of \cite{metrics}.
  \end{itemize}
  \end{itemize}
The Math issues, as explained a bit later in this section and elaborated upon towards the end of Sections {\bf 3} and {\bf 5.1} as well as {\bf 5.2}, are not only a precise way of helping one understand the inherent non-K\"{a}hlerity of the holographic model of \cite{metrics} and its mirror constructed in \cite{MQGP} which is what largely influences the Physics issues, but also explicitly shows the existence of approximate supersymmetry in the MQGP limit justifying the construction of the delocalized SYZ type IIA mirror in \cite{MQGP}. {\it This two-pronged approach in understanding large-$N$ thermal QCD with fundamental quarks at finite gauge coupling, entirely absent in the literature thus far, we feel is unique to our work.}

   We now provide a section-wise description of the {\bf motivation} and {\bf summary of the main results} of this paper.

 \begin{itemize}
 \item
 {\bf [Section 3] Lattice-compatible $T_c$ with the right light quark flavors from and thermodynamical stability of the top-down holographic thermal QCD dual of \cite{metrics} }

 A black hole with temperature T can radiate energy due to quantum fluctuations and become unstable. A black hole is unstable in an asymptotically flat space time due to its negative specific heat. However stability can be achieved at high temperature in asymptotically AdS black-hole background, while at low temperature the (thermal) AdS solution is preferred. There exists a first order phase transition between these two regimes at a temperature $T_c$, known as the Hawking-Page phase transition \cite{Hawking-Page_1983}. In the dual gauge theory this corresponds to the confinement/deconfinement phase transition.  {\it Using the Mia-Dasgupta et al's setup \cite{metrics}, one of the things we do in this paper is to calculate the QCD deconfinement temperature as explained in Section {\bf 3}.}
This  is {\bf motivated} by the following query. From a holographic dual of thermal QCD, at a finite baryon chemical potential, is it possible to simultaneously (within the same holographic dual):
\begin{itemize}
\item
obtain a $T_c$ compatible with lattice QCD results for the right number of light quark flavors,

\item
obtain the mass scale of the light quarks,

\item
incorporate the right mass of the lightest vector meson,

\item
obtain a $T_c$ which increases with decrease of $N_f$ (as required by lattice computations \cite{dTcoverdNfnegative}),

\item
ensure thermodynamical stability?

\end{itemize}
Needless to say, if a proposed holographic dual of thermal QCD  is able to satisfy all the above requirements (in addition to the requirements of UV conformality, IR confinement, etc.), it could be treated as a viable dual. {\it It is our aim to demonstrate, to the best of our knowledge for the first time, that the UV complete holographic dual of thermal QCD as proposed in \cite{metrics} answers all the above in the affirmative, and this is the reason why the results of this section comprise one of the major sets of results in this paper.}

A particularly interesting issue in this context is the incorporation of $N_f$ $D7$ branes in the resolved warped deformed conifold background geometry. The inclusion of quark matter, as was shown in \cite{Ouyang}, is achieved by these $D7$-brane probes. The details vis-a-vis the holographic dual of \cite{metrics}, are summarized in Appendix A.  The gauge theory has a global
$U(N_f)\simeq SU(N_f)\times U(1)$ symmetry in presence of the $N_f$ flavors. This global symmetry in the gauge theory
corresponds to the $U(N_f)$ local symmetry on the world volume of the $D7$ brane. The conserved current of the $U(N_f)$
symmetry acts as a source of the gauge field on the $D$-brane. As the $U(1)$ charge corresponds to the number of
baryons, the chemical potential $\mu_C$ or finite baryon density $n_q$ in the gauge theory can be introduced from the
$U(1)\subset U(N_f)$ gauge field on the $D$-brane. Now {\it at finite baryon density, we  show in this paper that the
confinement/deconfinement phase transition occurs at a temperature around $175 MeV$, which is consistent with the
lattice QCD result}. In deriving the deconfinement temperature we use the mass $m_{\rho}$ of the lightest vector boson as
an input which is around $760 MeV$ from lattice QCD results. Also the consistency of the result demands the number
of light flavors $N_f$  to be equal to 2 or 3 with their masses around $5.6 MeV$, not far from the actual value of the first generation quark masses.

\item
{\bf [Section 4] Temperature dependence of Electrical Conductivity and Charge Susceptibility, 1-D Luttinger Liquid, Einstein's relation and the {\it consequent} dependence of the Ouyang parameter on the horizon radius}

This section is {\bf motivated} by the following queries.

 \begin{itemize}
 \item
  What is the temperature dependence of (transport coefficients such as) the electrical conductivity, charge susceptibility and hence the Einstein's relation (relating their ratio to the diffusion constant)  in the top-down holographic thermal QCD dual of \cite{metrics}?

   \item
   In particular, does the temperature dependence referred to above, mimic some known (e.g. condensed matter) systems?
   \end{itemize}
Needless to say, answers to the above queries would serve as an important guide in understanding and classifying large-$N$ thermal QCD at finite coupling.

 Considering  non-abelian gauge field fluctuations using the gauge-gravity duality prescription, we obtain the $SU(2)$ EOM for $N_f=2$. In Section {\bf 4} we  investigate the temperature dependence of the electrical conductivity as well as charge susceptibility along with the Einstein relation relating their ratio to the diffusion constant, and show that the Ouyang embedding parameter is required to have a non-trivial dependence on the horizon radius.

 Further, we will see that the temperature dependence of electrical conductivity resembles a one-dimensional Luttinger liquid for appropriately tuned Luttinger interaction parameter. This resemblance in a future publication \cite{AM+KSil to appear}, will be seen to be further reinforced by looking at the temperature dependence of thermal conductivity and hence the Wiedemann-Franz law.

\item
{\bf [Section 5]Torsion class chasing or quantifying the non-K\"{a}hlerity of the type IIA SYZ mirror and its M-theory uplift, and seeing existence of approximate SUSY}

 We choose to discuss both, the aforementioned Physics-related issues  and Math-related issues of classification of the delocalised type IIA mirror and its M-theory uplift \cite{MQGP},\cite{transport-coefficients} by working out, respectively,
their $SU(3)$-structure and $G_2$-structure torsion classes, in the same paper. The reason and {\bf motivation} are two-fold.
 \begin{itemize}
 \item
   We are able to, e.g., reproduce a $T_c$ compatible with lattice calculations because of the inherent non-K\"{a}hlerity (apart from non-conformality) of the type IIB background. This is elaborated upon in Section {\bf 3}.  It is hence desirable to see the reflection of this under delocalized SYZ mirror symmetry by explicitly working out the $G$-structure (before and) after the application of delocalized SYZ mirror symmetry and the M theory uplift of the same. Quantifying the notion of non-K\"{a}hlerity and approximate supersymmetry via $G$-structure (torsion classes), is a very natural language for doing precisely that.
   \item
   The construction of the delocalized Strominger Yau Zaslow type IIA mirror of the type IIB holographic model of \cite{metrics} relies on both backgrounds being supersymmetric. The same is shown to be approximately true by evaluation of the $SU(3)$ structure torsion classes.
   \end{itemize}

Consequently,  the latter portion of this paper involves a discussion on Math-related issues regarding the delocalized SYZ type IIA mirror of the  type $\rm IIB$ background of \cite{metrics}, and its M-theory uplift. For the type IIA mirror, by working out the $SU(3)$ structure torsion classes, in the spirit of \cite{Butti et al [2004]}, we show signature of approximate supersymmetry. Given that the M-theory uplift of this type IIA mirror, is expected to involve a seven-fold with $G_2$ structure and four-form fluxes, we then work out, for the first time, {\it a local $G_2$ structure via the $G_2$ structure torsion classes of the M-theory uplift of the holographic large-$N$ thermal QCD type IIB dual model of \cite{metrics}}.

\end{itemize}

The paper is organized as follows. In Sec. {\bf 2} (via five sub-sections), after a brief review of construction of (non-)supersymmetric gauge theories involving (de-)singular(ized) conifolds, we briefly motivate and discuss the type IIB dual of the large-$N$ thermal QCD. In the same section, we discuss the `MQGP limit' and its utility and hence reason for being considered, as well as summarize the results of \cite{MQGP,transport-coefficients} to set the background for the current work and to make this paper self-contained.  In Sec. {\bf 3}, in the MQGP limit of \cite{MQGP}, performing the angular integral in the DBI action pertaining to considering the $U(1)$-subgroup of $U(N_f)$ corresponding to embedding of $N_f\ D7$ branes, and then taking the UV limit of the resultant (incomplete) elliptic integrals, with the mass of the lightest vector meson as an input, we show it is possible to obtain the QCD deconfinement temperature consistent with the lattice results,
as well as the mass scale of the light (first generation) quarks, ensuring the thermodynamical stability of the type $\rm IIB$ background.
Sec. {\bf 4} has a discussion on the equations of motion and their solutions near the asymptotic boundary for the baryon chemical potential [and the isospin gauge field ($N_f=2$)] and obtaining the expressions for the transport coefficients: electrical conductivity and charge susceptibility as functions of temperature. In Sec. {\bf 5}, by appropriate small-$\theta_{1,2}$ limits of the local Type IIA mirror metric, we improve upon our arguments of \cite{MQGP} and show that one can ensure that
$G^{IIA}_{\theta_1\theta_2}=0$ in the MQGP limit for any $r$ in the UV thereby indicating the possibility that the local mirror of a warped deformed conifold could locally be a warped resolved conifold. We also work out the $SU(3)$-structure torsion classes of the local type IIA mirror demonstrating approximate supersymmetry and the $G_2$-structure torsion classes of the local $M$-theory uplift of \cite{MQGP}.  Sec. {\bf 6} has a summary and significance of the results obtained. All technical details are relegated to seven appendices.

\section{Background - A  Review}

In this section, via five sub-sections we will:

\begin{itemize}
\item
provide a short review of the type IIB background of \cite{metrics} (reviewing/discussing a host of related facts scattered in the literature) which is supposed to provide a UV complete holographic dual of large-$N$ thermal QCD, as well as their precursors in subsection {\bf 2.1},

\item
discuss the 'MQGP' limit of \cite{MQGP} and the motivation for considering the same in subsection {\bf 2.2},

\item
discuss some aspects of type IIB and M-theory thermodynamics in subsection {\bf 2.4},

\item
provide a summary in subsection {\bf 2.5}, of a host of transport coefficients from two-point energy momentum/current correlation functions  pertaining to metric/gauge fluctuations as discussed in \cite{transport-coefficients}.
\end{itemize}

\subsection{\cite{metrics}'s Type IIB Dual of Large-$N$ Thermal QCD}

Let us first motivate the necessity of the construction of \cite{metrics} and hence its use in this paper. A bit of a history is hence in order.

\noindent $\bullet$ {\bf Zero\ Temperature\ Klebanov-Witten} \cite{kw} $\longrightarrow$ {\bf Klebanov-Tseytlin} \cite{KT} $\longrightarrow$ ({\bf Klebanov-Strassler} \cite{ks}, {\bf Pando Zayas-Tseytlin} \cite{PT})
$\longrightarrow$ {\bf Non-Zero\ Temperature\ Buchel} \cite{Buchel}  $\longrightarrow$ {\bf Klebanov et al} \cite{Gubser-et-al-finitetemp}:

 The Klebanov-Witten model \cite{kw} involving only $N\ D3$-branes at the tip of a singular conifold yielded an ${\cal N}=1$ $SU(N)\times SU(N)$ gauge theory which though was UV conformal but was not IR confining. The non-conformal Klebanov-Tseytlin model \cite{KT} in addition to the $N\ D3$-branes also included $M\ D5$-branes (fractional $D3$-branes) wrapping the vanishing $S^2$ in the $T^{1,1}$ of the singular conifold yielding an ${\cal N}=1$ $SU(M+N)\times SU(N)$ gauge theory. However, on flowing towards the IR, the ten-dimensional warp factor becomes negative  signalling that the gravity and gauge theories required a new IR completion.  The non-conformal Klebanov-Strassler \cite{ks} resolved the singularity via IR dynamics (gaugino-condensation after extremization of the Afflect-Dine-Seiberg superpotential)  and gave a geometric realization of confinement; at the end of the duality cascade, the branes dissolve into the geometry deforming the conifold and in the process reducing the rank of the gauge group and  one ends up with an ${\cal N}=1$ $SU(M)$ gauge theory which is IR confining. By the way, Pando-Zayas and Tseytlin \cite{PT} proposed an alternative to the deformed conifold resolution of the conifold geometry, the resolved conifold in which the $M\ D5$-branes wrap the blown-up $S^2$. However, as the three-form fluxes $G_3$ are neither primitive nor only of the (2,1)-type (it also possesses a (1,2) component), their solution breaks supersymmetry. As the (1,2)-component vanishes if the resolution parameter $a$ is set to zero, one sees that $S^2$-resolution of the conifold geometry can break supersymmetry; for a small $a$ and in the UV, this will be helpful in arguing the existence of approximate supersymmetry in \cite{metrics} later in this section. The holomorphic embedding of flavor $D7$-branes in a singular conifold geometry was considered in \cite{Ouyang} and in a resolved conifold was considered in \cite{Franche_thesis} (using the complex structure of the resolved conifold as given iv \cite{Knauf+Gwyn[2007]}):
 \begin{equation}
 \label{eq:Ouyang_embedding_RC}
 \left(\rho^6 + 9 a^2\rho^4\right)^{\frac{1}{4}}e^{\frac{i}{2}(\psi-\phi_1-\phi_2)}\sin\frac{\theta_1}{2}\sin\frac{\theta_2}{2} = \mu,
 \end{equation}
 where the redefined radial coordinate $\rho$ is defined via: $r = \left(\frac{2}{3}\right)^{\frac{3}{4}}\left(\rho^6 + 9 a^2 \rho^4\right)^{\frac{1}{4}}$ and $\mu$ is a complex Ouyang embedding parameter. Conventionally, in the $\mu\rightarrow0$-limit, the flavor $D7$-branes are embedded along either of the two branches: $\theta_1=\phi_1=0$, i.e., wrapping a non-compact four-cycle coordinatized by $(\theta_2,\phi_2,\psi,\rho)$ and $\theta_2=\phi_2=0$, i.e., wrapping a non-compact four-cycle coordinatized by $(\theta_1,\phi_1,\psi,\rho)$.   All the aforementioned constructs were at zero temperature. In \cite{Buchel}, finite-temperature/non-extremal version of the abovementioned KT solution was considered with the proposition that the aforementioned KT singularity is cloaked behind $r=r_h$(horizon radius) making therefore Seiberg duality cascade, unnecessary. Unfortunately, the solution was not regular as the non-extremality/black hole function and the ten-dimensional warp factor vanished simultaneously at the horizon radius $r_h$. The authors of \cite{Gubser-et-al-finitetemp} were able to construct a supergravity dual of $SU(M+N)\times SU(N)$ gauge theory which approached the abovementioned KT solution asymptotically and possessed a well-defined horizon. The same was characterized by: modification of $T^{1,1}$ via a `squashing factor' of the $U(1)_\psi$ fiber, non-constancy of the dilaton and non self-duality of the fluxes. But it was valid only for large temperatures with no fundamental quark flavors.

\noindent$\bullet$ {\bf A UV complete holographic dual of large-$N$ thermal QCD - Dasgupta-Mia et al} \cite{metrics}:\\
 \noindent (a) \underline{Brane construction}

 In order to include fundamental quarks at non-zero temperature in the context of type IIB string theory, to the best of our knowledge, the following model proposed in \cite{metrics} is the closest to a UV complete holographic dual of large-$N$ thermal QCD. The KS (duality cascade) and QCD have similar IR behavior: $SU(M)$ gauge group and IR confinement. However, they differ drastically in the UV as the former yields a logarithmically divergent gauge coupling (in the UV) - Landau pole. This necessitates modification of the UV sector of KS apart from inclusion of non-extremality factors. With this in mind and building up on all of the above, the type IIB holographic dual of
 \cite{metrics} was constructed. The setup of \cite{metrics} is summarized below.

 \begin{itemize}
 \item
  From a gauge-theory perspective, the authors of \cite{metrics} considered  $N$ black $D3$-branes placed at the tip of six-dimensional conifold, $M\ D5$-branes wrapping the vanishing two-cycle and $M\ \overline{D5}$-branes  distributed along the resolved two-cycle and placed at the outer boundary  of the IR-UV interpolating region/inner boundary of the UV region \footnote{Let us make some remarks about the stability of $M$ $D5$ and $M\ \overline{D5}$-branes. Conceptually, the gravtitational and RR-attraction between the $D5$ and $\overline{D5}$-branes balance the RR-repulsion between the
 resultant bound state of $D3$-branes.
Consider $N_1$ $D$-branes corresponding to a vector bundle $E_1$ and $N_2$ $\overline{D}$-branes corresponding to a vector bundle
$E_2$, and both wrapping a manifold $X^{(d)}, {\rm dim}_{\mathbb{C}}(X^{(d)})=d$.
    Even with same $N_1$ and $N_2$, due to different twistings, one can be left with a residual charge, which are the lower dimensional $BPS$ $D$-branes that survive after tachyon condensation. This can be understood in the language of stability of vector bundles and the triple:
$(E_1,E_2,T)$ where the tachyon $T$ can be thought of as the map $T:E_1\rightarrow E_2$ \cite{Pantev_et_al}. Imposing holomorphy of $T$ and gauge fields, the solutions to the low energy EOMs on $X^{(d)}$ were shown in \cite{Pantev_et_al} to be equivalent to the condition of stability of the triple. So, taking $N_1=N_2=1$ (for simplicity) wrapping the small $S^2$ of a warped resolved deformed conifold and $E_{1,2}$ being $U(1)$ bundles over $S^2$, it was shown in \cite{brane-stability-Tatar} that one generates
 the WZ term for a $D3$-brane: $\int_{\mathbb{R}^{1,3}}C_4$ if
$c_1(E_1) - c_1(E_2) = 1.$
 In other words one could turn on a unit flux on the world-volume of the $D5$-brane and none on the $\overline{D5}$ and generate a $D3$-brane after tachyon condensation. This can be shown to be compatible with the stability-of-triples argument. Alternatively, one can absorb the $M\ \overline{D5}$-branes as world-volume two-form fluxes on the $D7$-branes' world volume, i.e., one can turn on two-form fluxes on the world volume of the $D7$-branes in such a way so as to generate a negative $D5$-brane charge via $\int F_3$ where the two constants of integration that appear in the solutions to the EOM for the gauge field (corresponding to the aforementioned two-form fluxes) are chosen such that there is no net $D5$-brane charge, i.e. $\lim_{r\rightarrow\infty}\int F_3\sim M_{\rm eff}(r\rightarrow\infty)=0$ \cite{Mia_no_F3_UV}. {\bf The main point of this footnote is that the configuration of $N\ D3$-branes and $M\ D5,\ M\ \overline{D5}$ branes, is equivalent to $M+N D3$-branes in the UV.}}.

  \item
  More specifically, the $M\ \overline{D5}$ are distributed around the antipodal point relative to the location of $M\ D5$ branes on the blown-up $S^2$. If  the $D5/\overline{D5}$ separation is given by ${\cal R}_{D5/\overline{D5}}$, then this provides the boundary common to the outer UV-IR interpolating region and the inner UV region. The region $r>{\cal R}_{D5/\overline{D5}}$ is the UV.  In other words, the radial space, in \cite{metrics} is divided into the IR, the IR-UV interpolating region and the UV. To summarize the above:
  \begin{itemize}
  \item
  $r<r_0$: IR with $r\sim\Lambda$: deep IR where the $SU(M)$ gauge theory confines

  \item
  $r_0<r<{\cal R}_{D5/\overline{D5}}$: the IR-UV interpolating region

  \item
  $r>{\cal R}_{D5/\overline{D5}}$: the UV region.

  \end{itemize}

\item
$N_f\ D7$-branes, via Ouyang embedding,  are holomorphically embedded in the UV (asymptotically $AdS_5\times T^{1,1}$), the IR-UV interpolating region and dipping into the (confining) IR (up to a certain minimum value of $r$ corresponding to the lightest quark)  and $N_f\ \overline{D7}$-branes present in the UV and the UV-IR interpolating (not the confining IR). This is to ensure turning off of three-form fluxes, constancy of the axion-dilaton modulus and hence conformality and absence of Landau poles in the UV.

\item
The resultant ten-dimensional geometry hence involves a resolved warped deformed conifold. Back-reactions are included, e.g., in the ten-dimensional warp factor. Of course, the gravity dual, as in the Klebanov-Strassler construct, at the end of the Seiberg-duality cascade will have no $D3$-branes and the $D5$-branes are smeared/dissolved over the blown-up $S^3$ and thus replaced by fluxes.
\end{itemize}

The delocalized S(trominger) Y(au) Z(aslow) type IIA mirror of the aforementioned type IIB background of \cite{metrics} and its M-theory uplift had been obtained in \cite{MQGP,transport-coefficients}, and  newer aspects of the same will be looked into in this paper.



\noindent (b) \underline{Seiberg duality cascade, IR confining $SU(M)$ gauge theory at finite temperature and}\\ \underline{$N_c = N_{\rm eff}(r) + M_{\rm eff}(r)$}

\begin{enumerate}
\item
{\bf IR Confinement after Seiberg Duality Cascade}: Footnote numbered 3 shows that one effectively adds on to the number of $D3$-branes in the UV and hence, one has $SU(N+M)\times SU(N+M)$ color gauge group (implying an asymptotic $AdS_5$) and $SU(N_f)\times SU(N_f)$ flavor gauge group, in the UV: $r\geq {\cal R}_{D5/\overline{D5}}$. It is expected that there will be a partial Higgsing of $SU(N+M)\times SU(N+M)$ to $SU(N+M)\times SU(N)$ at $r={\cal R}_{D5/\overline{D5}}$  \cite{K. Dasgupta et al [2012]}. The two gauge couplings, $g_{SU(N+M)}$ and $g_{SU(N)}$ flow  logarithmically  and oppositely in the IR:
\begin{equation}
\label{RG}
4\pi^2\left(\frac{1}{g_{SU(N+M)}^2} + \frac{1}{g_{SU(N)}^2}\right)e^\phi \sim \pi;\
 4\pi^2\left(\frac{1}{g_{SU(N+M)}^2} - \frac{1}{g_{SU(N)}^2}\right)e^\phi \sim \frac{1}{2\pi\alpha^\prime}\int_{S^2}B_2.
\end{equation}
  Had it not been for $\int_{S^2}B_2$, in the UV, one could have set $g_{SU(M+N)}^2=g_{SU(N)}^2=g_{YM}^2\sim g_s\equiv$ constant (implying conformality) which is the reason for inclusion of $M$ $\overline{D5}$-branes at the common boundary of the UV-IR interpolating and the UV regions, to annul this contribution. In fact, the running also receives a contribution from the $N_f$ flavor $D7$-branes which needs to be annulled via $N_f\ \overline{D7}$-branes. The gauge coupling $g_{SU(N+M)}$ flows towards strong coupling and the $SU(N)$ gauge coupling flows towards weak coupling. Upon application of Seiberg duality, $SU(N+M)_{\rm strong}\stackrel{\rm Seiberg\ Dual}{\longrightarrow}SU(N-(M - N_f))_{\rm weak}$ in the IR;  assuming after repeated Seiberg dualities or duality cascade, $N$ decreases to 0 and there is a finite $M$, {\bf one will be left with $SU(M)$ gauge theory with $N_f$ flavors that confines in the IR - the finite temperature version of the same is what was looked at by \cite{metrics}}.

 \item
{\bf Obtaining $N_c=3$, and Color-Flavor Enhancement of Length Scale in the IR}:  So, in the IR, at the end of the duality cascade, what gets identified with the number of colors $N_c$ is $M$, which in the `MQGP limit' to be discussed below, can be tuned to equal 3. One can identify $N_c$ with $N_{\rm eff}(r) + M_{\rm eff}(r)$, where $N_{\rm eff}(r) = \int_{\rm Base\ of\ Resolved\ Warped\ Deformed\ Conifold}F_5$ and $M_{\rm eff} = \int_{S^3}\tilde{F}_3$ (the $S^3$ being dual to $\ e_\psi\wedge\left(\sin\theta_1 d\theta_1\wedge d\phi_1 - B_1\sin\theta_2\wedge d\phi_2\right)$, wherein $B_1$ is an asymmetry factor defined in \cite{metrics}, and $e_\psi\equiv d\psi + {\rm cos}~\theta_1~d\phi_1 + {\rm cos}~\theta_2~d\phi_2$) where $\tilde{F}_3 (\equiv F_3 - \tau H_3)\propto M(r)\equiv 1 - \frac{e^{\alpha(r-{\cal R}_{D5/\overline{D5}})}}{1 + e^{\alpha(r-{\cal R}_{D5/\overline{D5}})}}, \alpha\gg1$  \cite{IR-UV-desc_Dasgupta_etal}. The effective number $N_{\rm eff}$ of $D3$-branes varies between $N\gg1$ in the UV and 0 in the deep IR, and the effective number $M_{\rm eff}$ of $D5$-branes varies between 0 in the UV and $M$ in the deep IR (i.e., at the end of the duality cacade in the IR). Hence, the number of colors $N_c$ varies between $M$ in the deep IR and a large value [even in the MQGP limit of (\ref{limits_Dasguptaetal-ii}) (for a large value of $N$)] in the UV.  {\bf Hence, at very low energies, the number of colors $N_c$ can be approximated by $M$, which in the MQGP limit is taken to be finite and can hence be taken to be equal to three. }

Let us now explain how in the IR, in the MQGP limit, with the inclusion of terms higher order in $g_s N_f$  in (\ref{three-form-fluxes}) and the NLO terms in (\ref{h_i}), there occurs an IR color-flavor enhancement of the length scale as compared to a Planckian length scale in KS for ${\cal O}(1)$ $M$, thereby showing that quantum corrections will be suppressed.

Unlike large-$N_c$ gauge theories, we are dealing with large-$N$ thermal QCD-like theories and their gravity duals (which by the way, are not of the $AdS_5\times S^5$-type but involve a warped product of a non-extremal resolved warped deformed conifold and $\mathbb{R}^{1,3}$ with a black hole). Note, the ten-dimensional warp factor $h$ of (\ref{eq:h}), disregarding the angular part,  can be written in terms of the five-form flux $N_{\rm eff}$ as \cite{Ouyang}:
\begin{equation}
h = \frac{4\pi g_s}{r^4}\Biggl[N_{\rm eff}(r) + \frac{9 g_s M^2_{\rm eff} g_s N_f^{\rm eff}}{2\left(2\pi\right)^2}\log r \Biggr],
\end{equation}
where \cite{metrics}
\begin{eqnarray}
\label{NeffMeffNfeff}
& & N_{\rm eff}(r) = N\left[ 1 + \frac{3 g_s M_{\rm eff}^2}{2\pi N}\left(\log r + \frac{3 g_s N_f^{\rm eff}}{2\pi}\left(\log r\right)^2\right)\right],\nonumber\\
& & M_{\rm eff}(r) = M + \frac{3g_s N_f M}{2\pi}\log r + \sum_{m\geq1}\sum_{n\geq1} N_f^m M^n f_{mn}(r)\equiv M + M^\prime + \tilde{M} \equiv M + \tilde{\tilde{M}},\nonumber\\
& & N^{\rm eff}_f(r) = N_f + \sum_{m\geq1}\sum_{n\geq0} N_f^m M^n g_{mn}(r).
\end{eqnarray}
The terms in the double summation in $M_{\rm eff}$ in (\ref{NeffMeffNfeff}) arise, e.g., from the terms higher order in $g_s N_f$ in (\ref{three-form-fluxes}) and the NLO terms in (\ref{h_i}), both of which though in principle calculable from the solutions to the IIB supergravity equations of motion, are very cumbersome to work out.
 Seiberg duality is then effected via $r\rightarrow r e^{-\frac{2\pi}{3g_s (M + M^\prime)}}$ \cite{Ouyang}, under which $N_{\rm eff}\rightarrow N_{\rm eff} - M + \frac{M^2}{(M + M^\prime)^2}N_f$. For $r=\Lambda: \log \Lambda \ll {\frac{2\pi}{3g_sN_f}}$, $\frac{M^2}{(M + M^\prime)^2}N_f = N_f\Biggl\{1 - \frac{3g_sN_f}{\pi}\log\Lambda + {\cal O}\left[\left(\frac{3 g_sN_f}{2\pi}\log\Lambda\right)^2\right]\Biggr\}$. Hence, up to ${\cal O}\left(g_sN_f^2\log\Lambda\right)$, $N_{\rm eff}\rightarrow N_{\rm eff} - (M - N_f)$. Continuing this process until, as written earlier, one cascades almost (as one has to consider higher order terms in $\frac{3g_sN_f}{\pi}\log\Lambda$ in the MQGP limit that involves $g_s\stackrel{\sim}{<}1$ and $N_f\sim{\cal O}(1)$ and $\Lambda: \log \Lambda{<}{\frac{2\pi}{3g_sN_f}}$) the entire $N_{\rm eff}$ away, i.e., $N_{\rm eff}(\Lambda)\approx0$, one ends up with:
\begin{eqnarray}
& & h(\Lambda) \sim \frac{4\pi g_s }{r^4}\Biggl\{\frac{3g_s}{2\pi}\left[(2 M \tilde{\tilde{M}} + \tilde{\tilde{M}}^2)\left(\log \Lambda + \frac{3 g_s}{2\pi}(N_f + \tilde{N}_f)(\log \Lambda)^2\right) + \frac{3g_s M^2\tilde{N}_f}{2\pi}(\log \Lambda)^2\right]\nonumber\\
& & + \left(\frac{3g_s}{2\pi}\right)^2\left((2 M \tilde{\tilde{M}} + \tilde{\tilde{M}}^2)(N_f + \tilde{N}_f)+ M^2\tilde{N}_f\right)\frac{\log \Lambda}{2} \Biggr\}\nonumber\\
& & \ni\frac{4\pi g_s}{r^4}\Biggl[\tilde{\tilde{M}}^2\tilde{N}_f\log \Lambda\Biggr]=\frac{4\pi g_s}{r^4}M^2N_f^3\left(\frac{3 g_s}{2\pi}\sum_{m\geq0}\sum_{n\geq0}N_f^mM^nf_{mn}(\Lambda)\right)^2\sum_{l\geq0}\sum_{p\geq0}N_f^lM^p g_{lp}(\Lambda).\nonumber\\
& &
\end{eqnarray}
Hence, the length scale of the OKS-BH metric in the IR will be given by:
\begin{eqnarray}
\label{length-IR}
& & L_{\rm OKS-BH}\sim\sqrt{M}N_f^{\frac{3}{4}}\sqrt{\left(\sum_{m\geq0}\sum_{n\geq0}N_f^mM^nf_{mn}(\Lambda)\right)}\left(\sum_{l\geq0}\sum_{p\geq0}N_f^lM^p g_{lp}(\Lambda)\right)^{\frac{1}{4}}g_s^{\frac{1}{4}}\sqrt{\alpha^\prime}\nonumber\\
& & \equiv N_f^{\frac{3}{4}}\left.\sqrt{\left(\sum_{m\geq0}\sum_{n\geq0}N_f^mM^nf_{mn}(\Lambda)\right)}\left(\sum_{l\geq0}\sum_{p\geq0}N_f^lM^p g_{lp}(\Lambda)\right)^{\frac{1}{4}} L_{\rm KS}\right|_{\Lambda:\log \Lambda{<}{\frac{2\pi}{3g_sN_f}}},
\end{eqnarray}
which implies that  {\bf in the IR, relative to KS, there is a color-flavor enhancement of the length scale in the OKS-BH metric}. Hence,  in the IR, even for $N_c^{\rm IR}=M=3$ and $N_f=6$ upon inclusion of of $n,m>1$  terms in
$M_{\rm eff}$ and $N_f^{\rm eff}$ in (\ref{NeffMeffNfeff}), $L_{\rm OKS-BH}\gg L_{\rm KS}(\sim L_{\rm Planck})$ in the MQGP limit involving $g_s\stackrel{\sim}{<}1$. As a reminder one will generate higher powers of $M$ and $N_f$ in the double summation in $M_{\rm eff}$ in (\ref{NeffMeffNfeff}), e.g., from the terms higher order in $g_s N_f$ in (\ref{three-form-fluxes}) that become relevant for the aforementioned values of $g_s, N_f$.

 \item
  Further, the global  flavor group in the UV-IR interpolating and UV regions, due to presence of $N_f$ $D7$ and $N_f\ \overline{D7}$-branes, is $SU(N_f)\times SU(N_f)$, which is broken in the IR to $SU(N_f)$ as the IR has only $N_f$ $D7$-branes.

\end{enumerate}

Hence, the following features of the type IIB model of \cite{metrics} make it an ideal holographic dual of thermal QCD:

\begin{itemize}
\item
the theory having quarks transforming in the fundamental representation, is UV conformal and IR confining with the required chiral symmetry breaking in the IR and restoration at high temperatures

\item
the theory is UV complete with the gauge coupling remaining finite in the UV (absence of Landau poles)

\item
the theory is not just defined for large temperatures but for low and high temperatures

\item
(as will become evident in Sec. {\bf 3}) with the inclusion of a finite baryon chemical potential, the theory provides a lattice-compatible QCD confinement-deconfinement temperature $T_c$ for the right number of light quark flavors and masses, and is also thermodynamically stable; given the IR proximity of the value of the lattice-compatible $T_c$,  after the end of the Seiberg duality cascade, the number of quark flavors approximately equals $M$ which in the `MQGP' limit of (\ref{limits_Dasguptaetal-ii}) can be tuned to equal 3

\item
in the MQGP limit (\ref{limits_Dasguptaetal-ii}) which requires considering a finite gauge coupling and hence string coupling, the theory was shown in \cite{MQGP} to be holographically renormalizable from an M-theory perspective with the M-theory uplift also being thermodynamically stable.

\end{itemize}


\noindent (d) \underline{Supergravity solution on resolved warped deformed conifold}

The working metric is given by :
\begin{equation}
\label{metric}
ds^2 = \frac{1}{\sqrt{h}}
\left(-g_1 dt^2+dx_1^2+dx_2^2+dx_3^2\right)+\sqrt{h}\biggl[g_2^{-1}dr^2+r^2 d{\cal M}_5^2\biggr].
\end{equation}
 $g_i$'s are black hole functions in modified OKS(Ouyang-Klebanov-Strassler)-BH (Black Hole) background and are assumed to be:
$ g_{1,2}(r,\theta_1,\theta_2)= 1-\frac{r_h^4}{r^4} + {\cal O}\left(\frac{g_sM^2}{N}\right)$
where $r_h$ is the horizon, and the ($\theta_1, \theta_2$) dependence come from the
${\cal O}\left(\frac{g_sM^2}{N}\right)$ corrections. The  $h_i$'s are expected to receive corrections of
${\cal O}\left(\frac{g_sM^2}{N}\right)$ \cite{K. Dasgupta  et al [2012]}. We assume the same to also be true of the `black hole functions' $g_{1,2}$.  The compact five dimensional metric in (\ref{metric}), is given as:
\begin{eqnarray}
\label{RWDC}
& & d{\cal M}_5^2 =  h_1 (d\psi + {\rm cos}~\theta_1~d\phi_1 + {\rm cos}~\theta_2~d\phi_2)^2 +
h_2 (d\theta_1^2 + {\rm sin}^2 \theta_1 ~d\phi_1^2) +   \nonumber\\
&&  + h_4 (h_3 d\theta_2^2 + {\rm sin}^2 \theta_2 ~d\phi_2^2) + h_5~{\rm cos}~\psi \left(d\theta_1 d\theta_2 -
{\rm sin}~\theta_1 {\rm sin}~\theta_2 d\phi_1 d\phi_2\right) + \nonumber\\
&&  + h_5 ~{\rm sin}~\psi \left({\rm sin}~\theta_1~d\theta_2 d\phi_1 +
{\rm sin}~\theta_2~d\theta_1 d\phi_2\right),
\end{eqnarray}
$r\gg a, h_5\sim\frac{({\rm deformation\ parameter})^2}{r^3}\ll  1\forall r \gg({\rm deformation\ parameter})^{\frac{2}{3}}$ in the UV.  The $h_i$'s appearing in internal metric as well as $M, N_f$ are not constant and up to linear order depend on $g_s, M, N_f$ are given as below:
\begin{eqnarray}
\label{h_i}
& & \hskip -0.45in h_1 = \frac{1}{9} + {\cal O}\left(\frac{g_sM^2}{N}\right),\  h_2 = \frac{1}{6} + {\cal O}\left(\frac{g_sM^2}{N}\right),\ h_4 = h_2 + \frac{a^2}{r^2},\nonumber\\
& & h_3 = 1 + {\cal O}\left(\frac{g_sM^2}{N}\right),\ h_5\neq0,\
 L=\left(4\pi g_s N\right)^{\frac{1}{4}}.
\end{eqnarray}
One sees from (\ref{RWDC}) and (\ref{h_i}) that one has a non-extremal resolved warped deformed conifold involving
an $S^2$-blowup (as $h_4 - h_2 = \frac{a^2}{r^2}$), an $S^3$-blowup (as $h_5\neq0$) and squashing of an $S^2$ (as $h_3$ is not strictly unity). The horizon (being at a finite $r=r_h$) is warped squashed $S^2\times S^3$. In the deep IR, in principle one ends up with a warped squashed $S^2(a)\times S^3(\epsilon),\ \epsilon$ being the deformation parameter. Assuming $\epsilon^{\frac{2}{3}}>a$ and given that $a={\cal O}\left(\frac{g_s M^2}{N}\right)r_h$ \cite{K. Dasgupta  et al [2012]}, in the IR and in the MQGP limit, $N_{\rm eff}(r\in{\rm IR})=\int_{{\rm warped\ squashed}\ S^2(a)\times S^3(\epsilon)}F_5(r\in{\rm IR})\ll   M = \int_{S^3(\epsilon)}F_3(r\in{\rm IR})$; we have a confining $SU(M)$ gauge theory in the IR.

 The warp factor that includes the back-reaction, in the IR is given as:
\begin{eqnarray}
\label{eq:h}
&& \hskip -0.45in h =\frac{L^4}{r^4}\Bigg[1+\frac{3g_sM_{\rm eff}^2}{2\pi N}{\rm log}r\left\{1+\frac{3g_sN^{\rm eff}_f}{2\pi}\left({\rm
log}r+\frac{1}{2}\right)+\frac{g_sN^{\rm eff}_f}{4\pi}{\rm log}\left({\rm sin}\frac{\theta_1}{2}
{\rm sin}\frac{\theta_2}{2}\right)\right\}\Biggr],
\end{eqnarray}
where, in principle, $M_{\rm eff}/N_f^{\rm eff}$ are not necessarily the same as $M/N_f$; we however will assume that up to ${\cal O}\left(\frac{g_sM^2}{N}\right)$, they are. Proper UV behavior requires \cite{K. Dasgupta et al [2012]}:
\begin{eqnarray}
\label{h-large-small-r}
& & h = \frac{L^4}{r^4}\left[1 + \sum_{i=1}\frac{h_i\left(\phi_{1,2},\theta_{1,2},\psi\right)}{r^i}\right],\ {\rm large}\ r;
\nonumber\\
& & h = \frac{L^4}{r^4}\left[1 + \sum_{i,j; (i,j)\neq(0,0)}\frac{h_{ij}\left(\phi_{1,2},\theta_{1,2},\psi\right)\log^ir}{r^j}\right],\ {\rm small}\ r.
\end{eqnarray}


  In the IR, up to ${\cal O}(g_s N_f)$ and setting $h_5=0$, the three-forms are as given in \cite{metrics}:
\begin{eqnarray}
\label{three-form-fluxes}
& & \hskip -0.4in (a) {\widetilde F}_3  =  2M { A_1} \left(1 + \frac{3g_sN_f}{2\pi}~{\rm log}~r\right) ~e_\psi \wedge
\frac{1}{2}\left({\rm sin}~\theta_1~ d\theta_1 \wedge d\phi_1-{ B_1}~{\rm sin}~\theta_2~ d\theta_2 \wedge
d\phi_2\right)\nonumber\\
&& \hskip -0.3in -\frac{3g_s MN_f}{4\pi} { A_2}~\frac{dr}{r}\wedge e_\psi \wedge \left({\rm cot}~\frac{\theta_2}{2}~{\rm sin}~\theta_2 ~d\phi_2
- { B_2}~ {\rm cot}~\frac{\theta_1}{2}~{\rm sin}~\theta_1 ~d\phi_1\right)\nonumber \\
&& \hskip -0.3in -\frac{3g_s MN_f}{8\pi}{ A_3} ~{\rm sin}~\theta_1 ~{\rm sin}~\theta_2 \left(
{\rm cot}~\frac{\theta_2}{2}~d\theta_1 +
{ B_3}~ {\rm cot}~\frac{\theta_1}{2}~d\theta_2\right)\wedge d\phi_1 \wedge d\phi_2, \nonumber\\
& & \hskip -0.4in (b) H_3 =  {6g_s { A_4} M}\Biggl(1+\frac{9g_s N_f}{4\pi}~{\rm log}~r+\frac{g_s N_f}{2\pi}
~{\rm log}~{\rm sin}\frac{\theta_1}{2}~
{\rm sin}\frac{\theta_2}{2}\Biggr)\frac{dr}{r}\nonumber \\
&& \hskip -0.3in \wedge \frac{1}{2}\Biggl({\rm sin}~\theta_1~ d\theta_1 \wedge d\phi_1
- { B_4}~{\rm sin}~\theta_2~ d\theta_2 \wedge d\phi_2\Biggr)
+ \frac{3g^2_s M N_f}{8\pi} { A_5} \Biggl(\frac{dr}{r}\wedge e_\psi -\frac{1}{2}de_\psi \Biggr)\nonumber  \\
&&  \wedge \Biggl({\rm cot}~\frac{\theta_2}{2}~d\theta_2
-{ B_5}~{\rm cot}~\frac{\theta_1}{2} ~d\theta_1\Biggr). \nonumber\\
\end{eqnarray}
The asymmetry factors in (\ref{three-form-fluxes}) are given by: $ A_i=1 +{\cal O}\left(\frac{a^2}{r^2}\ {\rm or}\ \frac{a^2\log r}{r}\ {\rm or}\ \frac{a^2\log r}{r^2}\right) + {\cal O}\left(\frac{{\rm deformation\ parameter }^2}{r^3}\right),$ $  B_i = 1 + {\cal O}\left(\frac{a^2\log r}{r}\ {\rm or}\ \frac{a^2\log r}{r^2}\ {\rm or}\ \frac{a^2\log r}{r^3}\right)+{\cal O}\left(\frac{({\rm deformation\ parameter})^2}{r^3}\right)$.    As in the UV, $\frac{({\rm deformation\ parameter})^2}{r^3}\ll  \frac{({\rm resolution\ parameter})^2}{r^2}$, we will assume the same three-form fluxes for $h_5\neq0$.

Further, to ensure UV conformality, it is important to ensure that the axion-dilaton modulus approaches a constant implying a vanishing beta function in the UV. This is discussed in Appendix B.

\subsection{The `MQGP Limit' of \cite{MQGP}}

In \cite{MQGP}, we had considered the following two limits:
\begin{eqnarray}
\label{limits_Dasguptaetal-i}
&   & \hskip -0.17in (i) {\rm weak}(g_s){\rm coupling-large\ t'Hooft\ coupling\ limit}:\nonumber\\
& & \hskip -0.17in g_s\ll  1, g_sN_f\ll  1, \frac{g_sM^2}{N}\ll  1, g_sM\gg1, g_sN\gg1\nonumber\\
& & \hskip -0.17in {\rm effected\ by}: g_s\sim\epsilon^{d}, M\sim\left({\cal O}(1)\epsilon\right)^{-\frac{3d}{2}}, N\sim\left({\cal O}(1)\epsilon\right)^{-19d}, \epsilon\ll  1, d>0
 \end{eqnarray}
(the limit in the first line  though not its realization in the second line, considered in \cite{metrics});
\begin{eqnarray}
\label{limits_Dasguptaetal-ii}
& & \hskip -0.17in (ii) {\rm MQGP\ limit}: \frac{g_sM^2}{N}\ll  1, g_sN\gg1, {\rm finite}\
 g_s, M\ \nonumber\\
& & \hskip -0.17in {\rm effected\ by}:  g_s\sim\epsilon^d, M\sim\left({\cal O}(1)\epsilon\right)^{-\frac{3d}{2}}, N\sim\left({\cal O}(1)\epsilon\right)^{-39d}, \epsilon\lesssim 1, d>0.
\end{eqnarray}

Let us now elaborate upon the motivation for considering the MQGP limit. There are principally two.
\begin{enumerate}
\item
Unlike the AdS/CFT limit wherein $g_{\rm YM}\rightarrow0, N\rightarrow\infty$ such that $g_{\rm YM}^2N$ is large, for strongly coupled thermal systems like sQGP, what is relevant is $g_{\rm YM}\sim{\cal O}(1)$ and $N_c=3$. From the discussion in the previous paragraphs specially the one in point (c) of sub-section {\bf 2.1}, one sees that in the IR after the Seiberg duality cascade, effectively $N_c=M$ which in the MQGP limit of (\ref{limits_Dasguptaetal-ii})  can be tuned to 3. Further, in the same limit, the string coupling $g_s\stackrel{<}{\sim}1$. The finiteness of the string coupling necessitates addressing the same from an M theory perspective. This is the reason for coining the name: `MQGP limit'. In fact this is the reason why one is required to first construct a type IIA mirror, which was done in \cite{MQGP} a la delocalized Strominger-Yau-Zaslow mirror symmetry, and then take its M-theory uplift.

\item
From the perspective of calculational simplification in supergravity, the following are examples of the same and constitute therefore the second set of reasons for looking at the MQGP limit of (\ref{limits_Dasguptaetal-ii}):
\begin{itemize}
\item
\underline{$(M_{\rm eff}, N_{\rm eff}, N_f^{\rm eff})\stackrel{\rm MQGP}{\longrightarrow}(M, N, N_f)$}: The effective number of $D3$-branes, is given by \ref{NeffMeffNfeff} at $r=r_c$ where the ten-dimensional warp factor changes from the first expression (large $r$) to the second (small $r$) in (\ref{h-large-small-r}). Hence, in the UV, in the MQGP limit of (\ref{limits_Dasguptaetal-ii}), $N_{\rm eff}\sim N$; similarly $M_{\rm eff}\sim M, N^{\rm eff}_f\sim N_f$.

\item
\underline{Asymmetry Factors $A_i, B_j$(in three-form fluxes)$\stackrel{MQGP}{\rightarrow}1$ }: Referring to the asymmetry factors $A_i, B_j$ that figure in the three-form fluxes (\ref{three-form-fluxes}), given that $a^2 = {\cal O} \left(\frac{g_sM^2}{N}\right)r_h^2 + {\cal O}\left(\frac{g_sM^2}{N}(g_sN_f)\right)r_h^4$ \cite{K. Dasgupta et al [2012]}, taking the MQGP limit, $A_i=B_i\approx1$ in the IR/UV.

\item
\underline{Simplification of ten-dimensional warp factor and non-extremality function in MQGP limit}: The ten-dimensional warp factor, in the IR as given in (\ref{eq:h}) or for arbitrary $r$ as given in (\ref{h-large-small-r}), are simplified in the MQGP limit. For large $r$, the following approximation for $h$ is considered in \cite{metrics}:
\begin{eqnarray}
\label{h_large_r}
h={L^4}\Biggl[\frac{1}{r^{4-\epsilon_1}} + \frac{1}{r^{4-2\epsilon_2}}
- \frac{2}{r^{4-\epsilon_2}} + \frac{1}{r^{4-r^{\frac{\epsilon_2^2}{2}}}}\Biggr]\equiv\sum_{\alpha=1}^4\frac{L_{(\alpha)}^4}{r^4_{(\alpha)}},
\end{eqnarray}
where $\epsilon_1\equiv \frac{3g_sM^2}{2\pi N} + \frac{g_s^2M^2N_f}{8\pi^2N} + \frac{3g_s^2M^2N_f}{8\pi N}ln\left(sin\frac{\theta_1}{2} sin\frac{\theta_2}{2}\right)$,
$\epsilon_2\equiv\frac{g_sM}{\pi}\sqrt{\frac{2N_f}{N}}$, $r_{(\alpha)}\equiv r^{1-\epsilon_{(\alpha)}}, \epsilon_{(1)}=\frac{\epsilon_1}{2}, \epsilon_{(2)}\equiv\epsilon_{(3)}=\frac{\epsilon_2}{2}; L_{(1)}=L_{(2)}=L_{(4)}
= L^4, L_{(3)}=-2L^4$. It is conjectured in \cite{metrics} that as $r\rightarrow\infty, \alpha\in[1,\infty)$. It is evident that in the MQGP limit, (\ref{h_large_r}) is greatly simplified.

In fact for $N_f=0$, working with the ansatz:
\begin{eqnarray}
\label{h+g}
& & h = h_{(\ref{eq:h})} + \frac{L^4}{r^4}\left(A_0(r) + A_1(r)\log \left(\frac{r}{r_0}\right) + A_2(r) \log^2 \left(\frac{r}{r_0}\right) \right);\nonumber\\
& &  g = 1 - \frac{r_h^4}{r^4} + G_0(r) + G_1(r) \log \left(\frac{r}{r_0}\right) + G_2(r) \log^2 \left(\frac{r}{r_0}\right),
\end{eqnarray}
 it was shown in \cite{K. Dasgupta  et al [2012]} that $A_1=A_2=G_1=G_2=0$ and
 \begin{equation}
 \left(\begin{array}{c}A_0(r)\\G_0(r)\end{array}\right) = {\cal O}\left(\frac{g_s M^2}{N}, \frac{M}{N}\right)\sum_{k=1}\left(\begin{array}{c}a_0^k \\ g_0^k\end{array}\right)\left(\frac{r_h}{r}\right)^k\ll  1\ {\rm in\ MQGP\ Limit}.
 \end{equation}
  Hence, yet again in the MQGP limit, the expressions are greatly simplified. We will assume that: $h_i,h_{ij}\sim{\cal O}\left(\frac{g_sM^2}{N}\right)\ll  1$ in (\ref{h-large-small-r}) in the MQGP limit.

\end{itemize}
\end{enumerate}

 With ${\cal R}_0$ denoting the boundary common to the UV-IR interpolating region and the UV region, $\tilde{F}_{lmn}, H_{lmn}=0$ for $r\geq {\cal R}_0$ is required to ensure conformality in the UV \footnote{In fact, as we will explain in Section {\bf 3}, ${\cal R}_0$ gets identified with the $D5/\overline{D5}$ separation ${\cal R}_{D5/\overline{D5}}$ in \cite{metrics}.}.  Near the $\theta_1=\theta_2=0$-branch, assuming: $\theta_{1,2}\rightarrow0$ as $\epsilon^{\gamma_\theta>0}$ and $r\rightarrow r_\Lambda\rightarrow\infty$ as $\epsilon^{-\gamma_r <0}, \lim_{r\rightarrow\infty}\tilde{F}_{lmn}=0$ and  $\lim_{r\rightarrow\infty}H_{lmn}=0$ for all components except $H_{\theta_1\theta_2\phi_{1,2}}$; in the MQGP limit and near $\theta_{1,2}=\pi/0$-branch, $H_{\theta_1\theta_2\phi_{1,2}}=0/\left.\frac{3 g_s^2MN_f}{8\pi}\right|_{N_f=2,g_s=0.6, M=\left({\cal O}(1)g_s\right)^{-\frac{3}{2}}}\ll  1.$ So, the UV nature too is captured near $\theta_{1,2}=0$-branch in the MQGP limit. This mimics addition of $\overline{D5}$-branes in \cite{metrics} to ensure cancellation of $\tilde{F}_3$.

\subsection{Construction of  the Delocalized SYZ IIA Mirror and Its M-Theory Uplift in the MQGP Limit}

A central issue to \cite{MQGP,transport-coefficients} has been implementation of delocalized mirror symmetry via the Strominger Yau Zaslow prescription according to which the mirror of a Calabi-Yau can be constructed via three T dualities along a special Lagrangian $T^3$ fibered over a large base in the Calabi-Yau. This sub-section is a quick review of precisely this.

{ To implement the quantum mirror symmetry a la S(trominger)Y(au)Z(aslow) \cite{syz}, one needs a special Lagrangian (sLag) $T^3$ fibered over a large base (to nullify contributions from open-string disc instantons with boundaries as non-contractible one-cycles in the sLag).    Defining delocalized T-duality coordinates, $(\phi_1,\phi_2,\psi)\rightarrow(x,y,z)$ valued in $T^3(x,y,z)$ \cite{MQGP}:
\begin{equation}
\label{xyz defs}
x = \sqrt{h_2}h^{\frac{1}{4}}sin\langle\theta_1\rangle\langle r\rangle \phi_1,\ y = \sqrt{h_4}h^{\frac{1}{4}}sin\langle\theta_2\rangle\langle r\rangle \phi_2,\ z=\sqrt{h_1}\langle r\rangle h^{\frac{1}{4}}\psi,
\end{equation}
using the results of \cite{M.Ionel and M.Min-OO (2008)} it can be shown \cite{transport-coefficients},\cite{AM+KSil to appear} that the following conditions are satisfied:
\begin{eqnarray}
\label{sLag-conditions}
& & i^* J \approx 0,\nonumber\\
& & \Im m\left( i^*\Omega\right) \approx 0,\nonumber\\
& & \Re e\left(i^*\Omega\right)\sim{\rm volume \ form}\left(T^3(x,y,z)\right),
\end{eqnarray}
separately for the $T^2$-invariant sLags of \cite{M.Ionel and M.Min-OO (2008)} for a resolved/deformed conifold implying thus: $\left.i^* J\right|_{RC/DC}\approx0, \Im m\left.\left( i^*\Omega\right)\right|_{RC/DC} \approx 0, \Re e\left.\left(i^*\Omega\right)\right|_{RC/DC}\sim{\rm volume \ form}\left(T^3(x,y,z)\right)$. Hence, if the resolved warped deformed conifold is predominantly either resolved or deformed, the local $T^3$ of (\ref{xyz defs}) is the required sLag to effect SYZ mirror construction.}

{Interestingly, in the `delocalized limit' \cite{M. Becker et al [2004]}  $\psi=\langle\psi\rangle$, under the coordinate transformation :
\begin{equation}
\label{transformation_psi}
\left(\begin{array}{c} sin\theta_2 d\phi_2 \\ d\theta_2\end{array} \right)\rightarrow \left(\begin{array}{cc} cos\langle\psi\rangle & sin\langle\psi\rangle \\
- sin\langle\psi\rangle & cos\langle\psi\rangle
\end{array}\right)\left(\begin{array}{c}
sin\theta_2 d\phi_2\\
d\theta_2
\end{array}
\right),
\end{equation}
and $\psi\rightarrow\psi - \cos\langle{\bar\theta}_2\rangle\phi_2 + \cos\langle\theta_2\rangle\phi_2 - \tan\langle\psi\rangle ln\sin{\bar\theta}_2$
the $h_5$ term becomes $h_5\left[d\theta_1 d\theta_2 - sin\theta_1 sin\theta_2 d\phi_1d\phi_2\right]$, $e_\psi\rightarrow e_\psi$, i.e.,  one introduces an isometry along $\psi$ in addition to the isometries along $\phi_{1,2}$. This clearly is not valid globally - the deformed conifold does not possess a third global isometry}.

{ To enable use of SYZ-mirror duality via three T dualities, one also needs to ensure a large base (implying large complex structures of the aforementioned two two-tori) of the $T^3(x,y,z)$ fibration. This is effected via \cite{F. Chen et al [2010]}:
\begin{eqnarray}
\label{SYZ-large base}
& & d\psi\rightarrow d\psi + f_1(\theta_1)\cos\theta_1 d\theta_1 + f_2(\theta_2)\cos\theta_2d\theta_2,\nonumber\\
& & d\phi_{1,2}\rightarrow d\phi_{1,2} - f_{1,2}(\theta_{1,2})d\theta_{1,2},
\end{eqnarray}
for appropriately chosen large values of $f_{1,2}(\theta_{1,2})$. The three-form fluxes
 remain invariant. The fact that one can choose such large values of $f_{1,2}(\theta_{1,2})$, was justified in \cite{MQGP}.  The guiding principle is that one requires that the metric obtained after SYZ-mirror transformation applied to the non-K\"{a}hler  resolved warped deformed conifold is like a non-K\"{a}hler warped resolved conifold at least locally. Then $G^{IIA}_{\theta_1\theta_2}$ needs to vanish \cite{MQGP}.  This is shown to be true anywhere in the UV in Appendix {\bf C}.}


{
The mirror type IIA metric after performing three T-dualities, first along $x$, then along $y$ and finally along $z$, utilizing the results of \cite{M. Becker et al [2004]} was worked out in \cite{MQGP}. We can get a one-form type IIA potential from the triple T-dual (along $x, y, z$) of the type IIB $F_{1,3,5}$ in \cite{MQGP} and using which the following $D=11$ metric was obtained in \cite{MQGP}:
\begin{eqnarray}
\label{M3}
& &\hskip -0.6in   ds^2_{11} =  e^{-\frac{2{\phi}^{IIA}}{3}} \Biggl[
\frac{1}{\sqrt{h\left(r,\theta_1,\theta_2\right)}}\biggl(-g_1 dt^2+dx_1^2+dx_2^2+dx_3^2\biggr)+ \sqrt{h\left(r,\theta_1,\theta_2\right)}\frac{dr^2}{g_2} +  ds^2_{IIA}({\theta_{1,2},\phi_{1,2},\psi})\Biggr]\nonumber\\
 & & \hskip -0.2in + e^{\frac{4{\phi}^{IIA}}{3}}\biggl(dx_{11} + A^{F_1}+A^{F_3}+A^{F_5}\biggr)^2.
\end{eqnarray}
As in Klebanov-Strassler construction, a single T-duality along a direction orthogonal to the $D3$-brane world volume, e.g., $z$ of (\ref{xyz defs}), yields $D4$ branes straddling a pair of $NS5$-branes consisting of world-volume coordinates $(\theta_1,x)$ and $(\theta_2,y)$. Further, T-dualizing along $x$ and then $y$ would yield a Taub-NUT space  from each of the two $NS5$-branes \cite{T-dual-NS5-Taub-NUT-Tong}. The $D7$-branes yield $D6$-branes which get uplifted to Kaluza-Klein monopoles in M-theory \cite{KK-monopoles-A-Sen} which too involve Taub-NUT spaces. Globally, probably the eleven-dimensional uplift would involve a seven-fold of $G_2$-structure, analogous to the uplift of $D5$-branes wrapping a two-cycle in a resolved warped conifold \cite{Dasguptaetal_G2_structure}. This $G_2$-structure, locally, will be explicitly worked out in section {\bf 5} of this paper.

Now, analogous to the $F_3^{IIB}(\theta_{1,2})$ (with non-zero components being $F_{\psi\phi_1\theta_1}, F_{\psi\phi_2\theta_2}, F_{\phi_1\phi_2\theta_1}$ and $F_{\phi_1\phi_2\theta_2}$) in Klebanov-Strassler background corresponding to $D5$-branes wrapped around a two-cycle which homologously is given by $S^2(\theta_1,\phi_1) - S^2(\theta_2,\phi_2)$, in the delocalized limit of \cite{M. Becker et al [2004]}, in \cite{transport-coefficients}, e.g., \\ $\left.\int_{C_4(\theta_{1,2},\phi_{1/2},x_{10})}G_4\right|_{\phi_{2/1}=\langle\phi_{2/1}\rangle,\psi=\langle\psi\rangle,\langle r\rangle}$ was estimated to be very large. There is a two-fold reason for the same. First, using the local $T^3$-coordinates of (\ref{xyz defs}), this large flux is estimated in the MQGP limit to be $\left(g_s N\right)^{\frac{1}{4}}$ (as, using (\ref{xyz defs}), $G_{\phi_1\ {\rm or}\ \phi_2\ {\rm or}\ \psi\bullet\bullet\bullet}\sim \left(g_s N\right)^{\frac{1}{4}}G_{x\ {\rm or}\ y\ {\rm or}\ z\bullet\bullet\bullet}$ where the bullets denote directions other than $\phi_1,\phi_2,\psi$). This in the MQGP limit, is large. The second is the following. Now, $G_4 = H\wedge (A^{F_1+F-3+F_5} - dx_{10})$ \cite{MQGP} where $A^{F_1+F_3+F_5}$ is the type IIA one-form gauge field obtained after SYZ mirror construction via triple T dualities on the type IIB $F_{1,3,5}$.  As the $S^2(\theta_1,\phi_1)$ is a vanishing two-sphere, to obtain a finite $\int_{S^2(\theta_1,\phi_1)}B_2$  - that appears in the RG equation (\ref{RG}) - one requires a large $B_2$ (From \cite{metrics} one sees that such a large contribution to $B_2$ is obtained near the $\theta_1=\theta_2=0$ branch.) Therefore, this too contributes to a large $G_4$ via a large $H$.

Locally, the uplift (\ref{M3}) can hence be thought of as black $M3$-brane metric, which in the UV, can be thought of as black $M5$-branes wrapping a two cycle homologous to:
$n_1 S^2(\theta_1,x_{10}) + n_2 S^2(\theta_2,\phi_{1/2}) + m_1 S^2(\theta_1,\phi_{1/2}) + m_2 S^2(\theta_2,x_{10})$ for some large $n_{1,2},m_{1,2}\in\mathbb{Z}$ \cite{transport-coefficients}.  In the large-$r$ limit, the $D=11$ space-time is a warped product of $AdS_5(\mathbb{R}^{1,3}\times\mathbb{R}_{>0})$ and ${\cal M}_6(\theta_{1,2},\phi_{1,2},\psi,x_{10})$
\begin{equation}
\hskip -0.4in
\label{M_6}
\begin{array}{cc}
&{\cal M}_6(\theta_{1,2},\phi_{1,2},\psi,x_{10})   \longleftarrow   S^1(x_{10}) \\
&\downarrow  \\
{\cal M}_3(\phi_1,\phi_2,\psi) \hskip -0.4in & \longrightarrow  {\cal M}_5(\theta_{1,2},\phi_{1,2},\psi)   \\
&\downarrow  \\
 & \hskip 0.9in {\cal B}_2(\theta_1,\theta_2)  \longleftarrow  [0,1]_{\theta_1}  \\
 & \downarrow  \\
& [0,1]_{\theta_2}
\end{array}.
\end{equation}

{The $D=11$ SUGRA EOMs/Bianchi identity \cite{M.S. Bremer [1999]} were shown in \cite{transport-coefficients} to be  satisfied near the $\theta_{1,2}=0,\pi$-branches in the MQGP limit:
\begin{eqnarray*}
\label{G4EOM}
& & \hskip -0.4in { R^{{\cal M}}_{MN}=\frac{1}{12}\left (G_{MPQR}G_N^{PQR}-\frac{1}{12} G^{{\cal M}}_{MN} G_{PQRS}G^{PQRS} \right)}\nonumber\\
& & { +\kappa_{11}^2 \left (T_{MN}-\frac{1}{9}G^{{\cal M}}_{MN} T^{Q}_{Q}\right)}\nonumber\\
& & \hskip -0.4in { d*_{11}G_4 + G_4\wedge G_4 = -2 \kappa_{11}^2T_5 (H_3 - A_3)\wedge *_{11}J_6},\nonumber\\
& & \hskip -0.4in { dG_4 = 2\kappa_{11}^2T_5 *_{11}J_6,}
\end{eqnarray*}
where $M5$-brane current $J_6\sim\frac{dx^0\wedge dx^1\wedge dx^2\wedge dx^3\wedge d\theta_1\wedge d\phi_1}{\sqrt{-G^{\cal M}}}$, the space-time energy momentum tensor $T_{MN}$ for a single  M5-brane  wrapped around $S^2(\theta_1,\phi_1)$ is given by:\\ {\small $ T^{MN}(x)  = \int_{{\cal M}_{6} }d^{6}\xi \sqrt{{\rm -det}~{G^{M5}_{\mu\nu}}} G^{(M5)\mu\nu}\partial_{\mu}X^{M} \partial_{\nu}X^N \frac{\delta^{11}(x-X(\xi))}{\sqrt{-{\rm det} \ G^{{\cal M}}_{MN}}}$}
where $X=0,1,...,11$ and $\mu,\nu=0,1,2,3,\theta_1,\phi_1$.}

%

%
%
%
%
%

\subsection{Type IIB and M-Theory Thermodynamics}

Building up on the material reviewed  before and in {\bf 2.1}, we will now briefly review the relevant type IIB and M-theory thermodynamics as worked out in \cite{MQGP}, relevant to the type IIB background of \cite{metrics} and its local M-theory uplift in \cite{MQGP} oriented towards demonstrating the thermodynamical stability of both.


{Let us start with the black $M3$-brane temperature. Now, in the MQGP limit, $G_{00}^{\cal M}, G_{rr}^{\cal M}$ have no angular dependence and hence
the black $M3$-brane temperature is given by  $T=\frac{\partial_r G_{00}}{4\pi \sqrt{G_{00}G_{rr}}}$ \cite{Kovtun-Son-Starinets [2003]}, and works out to:
\begin{eqnarray}
\label{T}
& &\hskip-0.4in T =\frac{\sqrt{2}}{{r_h} \sqrt{\pi} \sqrt{\frac{{g_s} \left(18 {g_s}^2 {N_f} ln ^2({r_h}) {M_{\rm eff}}^2+3 {g_s} (4 \pi -{g_s} {N_f}
   (-3+ln (2))) ln ({r_h}) {M_{\rm eff}}^2+8 N \pi ^2\right)}{{r_h}^4}}}\stackrel{\rm Both\ limits}{\longrightarrow}\frac{r_h}{\pi L^2}. \nonumber\\
   \end{eqnarray}
}


{Despite working at a finite temperature, the type IIB background of \cite{metrics}, possesses approximate supersymmetry. (This, after all, is very important for implementing SYZ mirror symmetry transformation via three T-dualities.) This is for the following reason. The deviation from  $G_3$ being imaginary self dual is estimated to be: $\left|iG_3 - *_6G_3\right|^2\propto \frac{a^4}{r^4}$ \cite{K. Dasgupta  et al [2012]}. Assuming a negligible bare resolution parameter, $a$ in turn is related to the horizon radius $r_h$ via: $a^2 = {\cal O} \left(\frac{g_sM^2}{N}\right)r_h^2 + {\cal O}\left(\frac{g_sM^2}{N}(g_sN_f)\right)r_h^4$  : very small in MQGP limit. We return to this issue in section {\bf 5} from the point of view of explicitly showing that the three-form fluxes $G_3$ are of the (2,1)-type in the UV for a small resolution parameter near the $\theta_1=\theta_2=0$-branch. The amount of near-horizon supersymmetry was determined in \cite{MQGP} by solving for the killing spinor $\epsilon$ by the vanishing superysmmetric
variation of the gravitino in $D=11$ . The near-horizon black $M3$-brane solution,  near $\theta_{1,2}=0,\pi$,
 possesses 1/8 supersymmetry \cite{MQGP}  reminiscent of \cite{M. Cvetic et al [2001]}. We will elaborate more on this in Sec {\bf 5}.}


Let us now summarize the results of \cite{MQGP} as regard the thermodynamical stability of \cite{metrics} and the M-theory uplift of its delocalized SYZ type IIA mirror worked out in \cite{MQGP}. The baryon chemical potential  $\mu_C$ corresponding to a $U(1)$(of $U(N_f)=U(1)\times SU(N_f)$) gauge field living on the world volume of $N_f$ $D7$-branes embedded supersymmetrically inside a conifold via the Ouyang embedding involving a non-zero real embedding parameter, was worked out in \cite{MQGP}. {\it However, the $r>|\mu|^{\frac{2}{3}} >1$ - limit inside the DBI action (after the MQGP limit) was taken in \cite{MQGP} before performing the angular integration.} We will return to this issue in section {\bf 3} and see that taking the UV limit after performing the angular integration has highly non-trivial consequences. The thermodynamical stability of the type IIB background was then demonstrated in \cite{MQGP} by explicitly verifying $ \left.\frac{\partial \mu_C}{\partial T}\right|_{N_f}<0$ and $\left.\frac{\partial \mu_C}{\partial N_f}\right|_{T}>0$.


As the MQGP limit requires taking a finite (close to but smaller than one) gauge coupling, it necessitates addressing the same from an M theory perspective. We now summarize thermodynamical stability from D=11 supergravity point of view. The D=11 supergravity action we considered in \cite{MQGP} included the bulk Einstein-Hilbert (EH), $G_4$-flux  and ${\cal O}(R^4)$ higher order curvature terms, and the boundary Gibbons-Hawking-York surface term. The action, apart from being divergent in the UV, also possesses pole-singularities near $\theta_{1,2}=0,\pi$. We regulate the second divergence in \cite{MQGP} by taking a
small $\theta_{1,2}$-cutoff $\epsilon_\theta$, $\theta_{1,2}\in[\epsilon_\theta,\pi-\epsilon_\theta]$, and demanding $\epsilon_\theta
\sim\epsilon^\gamma$, for an appropriate $\gamma$ such that   the UV-finite part of the action turns out to be independent of this cut-off $\epsilon/\epsilon_\theta$. The holographic renormalization required that the counter-term ${\cal S}^{\rm ct}$ required to be added such that the  action ${\cal S}_{E}$ is finite \cite{R. Monteiro et al [2009]+R. B. Mann R. Mcnees [2009]} are boundary EH, cosmological and flux counter terms and were constructed in \cite{MQGP} to cancel the UV divergence in the D=11 supergravity action. It was then argued that the entropy density $s\sim  r_h^3$ and the specific heat is positive - implying a stable uplift!

Having reviewed the construction in \cite{MQGP,transport-coefficients} of the delocalized type IIA mirror of the type IIB background of \cite{metrics} as well its M-theory uplift, in {\bf 2.5}, we will briefly review the general gauge-gravity duality techniques of \cite{[10]} of obtaining two-point functions involving  energy momentum tensor/currents and summarize a host of results of \cite{transport-coefficients} obtained as a result of its application. But, before doing so, we first review the results of \cite{MQGP} pertaining to evaluation of the shear viscosity $\eta$ and diffusion constant $D$ using the techniques of \cite{Kovtun-Son-Starinets [2003]} in supergravity.

\subsection{Transport Coefficients}

In this subsection we summarize  our results from \cite{MQGP} and \cite{transport-coefficients} - the latter in the form of a table -  pertaining to obtaining values of the shear-viscosity-entropy-density ratio and diffusion constant  and  \cite{transport-coefficients} as regard evaluation of a variety of two-point correlation functions relevant to evaluation of DC electrical conductivity, charge susceptibility, Einstein's relation relating the two, R-charge diffusion constant and shear viscosity.

\noindent (a) From Supergravity \cite{MQGP}


{ Freezing the angular dependence
on $\theta_{1,2}$ (there being no dependence on $\phi_{1,2},\psi,x_{10}$ in the MQGP limit), noting that $G_{00,rr,{\bf R}^3}^{IIA/{\cal M}}$ are independent of the angular coordinates (additionally possible to tune the chemical potential $\mu_C$ to a small value \cite{MQGP}), using
 the result of  \cite{Kovtun-Son-Starinets [2003]}:
 \begin{equation}
 \frac{\eta}{s}=T\frac{\sqrt{|G^{IIA/{\cal M}}|}}{\sqrt{|G^{IIA/{\cal M}}_{tt}G^{IIA/{\cal M}}_{rr}|}}\Biggr|_{r=r_h}
 \int_{r_h}^\infty dr\frac{|G^{IIA/{\cal M}}_{00}G^{IIA}_{rr}|}{G^{IIA/{\cal M}}_{{\bf R}^3}\sqrt{|G^{IIA/{\cal M}}|}}=\frac{1}{4\pi}.
 \end{equation}}

{ In the notations of \cite{Kovtun-Son-Starinets [2003]} one can pull out a common $Z(r)$ in the angular-part of the metrics as:
$Z(r)K_{mn}(y)dy^i dy^j$, (which for the type IIB/IIA backgrounds, is $\sqrt{h}r^2$) in terms of which:
\begin{eqnarray*}
& & \hskip -0.4in D = \frac{\sqrt{|G^{IIB/IIA}|}Z^{IIB/IIA}(r)}{G^{IIb/IIA}\sqrt{|G_{00}^{IIB/IIA}G_{rr}^{IIB/IIA}|}}\Biggr|_{r=r_h}\int_{r_h}^\infty dr
\frac{|G_{00}^{IIB/IIA}G_{rr}^{IIB/IIA}|}{\sqrt{|G^{IIB/IIA}|}Z^{IIB/IIA}(r)}\nonumber\\
& & \hskip -0.4in = \frac{1}{2\pi T}
\end{eqnarray*}}

\noindent (b) Using gauge-gravity duality \cite{transport-coefficients}

We first summarize the prescription to calculate the Minkowskian correlators in AdS/CFT correspondence. We follow \cite{[10]} for this. A solution of the linearized field equation for any field $\phi(u,x)$ choosing $q^\mu=(w,q,0,0)$ is given as,
 \begin{equation}
 \phi(u,x)=\int\frac{d^4q}{(2\pi)^4}e^{-i wt + i qx}f_{q}(u)\phi_{0}(q)
 \end{equation}
where $ f_{q}(u)$ is normalized to 1 at the boundary and satisfies the incoming wave boundary condition at $u=1$, and $\phi_{0}(q)$ is determined by,
 \begin{equation}
 \phi(u=0,x)=\int\frac{d^4q}{(2\pi)^4}e^{-i wt + i qx}\phi_{0}(q).
\end{equation}
If the kinetic term for $\phi(u,x)$ is given by: $\frac{1}{2}\int d^4x du A(u)\left(\partial_u\phi(x,u)\right)^2$, then using the equation of motion for $\phi$ it is possible to reduce an on-shell action to the surface terms as,
\begin{equation}
S=\int\frac{d^4q}{(2\pi)^4}\phi_{0}(-q)\mathcal{F}(q,u)\phi_{0}(q)|^{u=1}_{u=0}
\end{equation}
where the function
\begin{equation}
\label{F}
\mathcal{F}(q,u) = A(u) f_{\pm q}(u)\partial_{u}f_{\pm q}(u).
\end{equation}
 Finally, the retarded Green's function is given by the formula proposed in \cite{[10]}:
\begin{equation}
G^{R}(q)=-2\mathcal{F}(q,u)|_{u=0}.
\end{equation}
We consider the  metric and  gauge field fluctuations  of the background. The retarded Green's functions are defined as
 \begin{equation}
 G^{R\ T}_{\mu\nu,\rho\sigma}(q)=-i\int d^4x e^{-i wt + i qx}\theta(t) \langle[T_{\mu\nu}(x),T_{\rho\sigma}(0)]\rangle,
\end{equation}
with $\langle[T_{\mu\nu},T_{\rho\sigma}]\rangle\sim\frac{\delta^2S}{\delta h_{\mu\nu}\delta h_{\rho\sigma}}$ and
 \begin{equation}
G^{R\ J}_{\mu\nu}(q)=-i\int d^4x e^{-i wt + i qx}\theta(t) \langle[J_{\mu}(x),J_{\nu}(0)]\rangle
\end{equation}
with $\langle[J_{\mu}(x),J_{\nu}(0)]\rangle\sim\frac{\delta^2S}{\delta A_\mu \delta A_\nu}$,
as the energy-momentum tensor $T_{\mu\nu}(x)$ and the current $J_{\mu}(x)$ couple respectively to the metric and gauge field, respectively. So, as examples: the shear viscosity would be given by the Kubo formula $\eta=-\lim_{w\rightarrow0}\frac{1}{w}\left(\lim_{q\rightarrow0}\Im m G^{R\ T}_{xy,xy}(w,q)\right)$ corresponding to vector-mode of metric fluctuation $h_{xy}$ or $\eta=-\lim_{w\rightarrow0}\frac{1}{w}\left(\lim_{q\rightarrow0}\Im m G_{yz,yz}(w,q)\right)$ corresponding to tensor-mode metric fluctuation. Similarly, the DC electrical conductivity is given by $\sigma=\lim_{w\rightarrow0}\frac{\Im m G^{R\ J}_{xx}(w,0)}{w}$.

%

\section{Baryon Chemical Potential and $T_c$ Consistent with Lattice Results and First Generation Quark Masses}

In this section we discuss the evaluation of the QCD confinement-deconfinement transition temperature $T_c$ in the presence of a finite baryon chemical potential/charge density and a constant axion-dilaton modulus. The motivation for this section was spelt out in Section {\bf 1}.

Here is first, an outline of how the calculations in this section will proceed.
\begin{enumerate}
\item

For starters, we revisit our calculation of \cite{MQGP} of the baryon chemical potential  generated via $D7$-brane gauge fields in the background of \cite{metrics}. The temporal component of bulk $U(1)$ field on the $D7$-brane world-volume is related to chemical potential which is defined in a gauge-invariant manner as follows:  $\mu_C=\int_{r_h}^\infty dr F_{rt}$. The field strength's only non-zero component, $F_{rt}$, can be evaluated by solving the Euler-Lagrange equation of motion for DBI Action. Instead of taking the UV-limit of the DBI action for $D7$-branes wrapping a non-compact four-cycle via Ouyang's embedding before performing the angular $\theta_{1,2}$ integrals therein as was done in \cite{MQGP}, we will first perform the angular integral exactly and then take the UV limit of the resultant (incomplete) elliptic integrals, in this section.

\item
In the MQGP limit of (\ref{limits_Dasguptaetal-ii}), after integrating out the the warped squashed resolved warped deformed conifold, one gets an approximate black hole $AdS_5$. We further note the following points.
\begin{enumerate}
\item
We choose the finite AdS boundary at $r=r_0$, which  corresponds to the boundary common to the IR and the inner UV-IR interpolating regions.

 \item
 The scale $\Lambda$ corresponding to gaugino condensation $\langle N_{ij} N_{kl}\rangle\epsilon^{ik}\epsilon^{jl} = \left[\frac{2 \Lambda^{3M+1}}{\lambda^{M-1}}\right]^{\frac{1}{M}}$ ($N_{ij}\equiv A_i B_j$ where $A_i, B_j, i,j=1,2$ are defined in the sentence above (\ref{W flavors})) and hence the deformation parameter of the deformed conifold (${\rm det} N = \left(\frac{\Lambda^{3M+1}}{\lambda^{M-1}}\right)^{\frac{1}{M}}$) arises in the deep IR ($r<r_0$) where the $SU(M)$ gauge coupling after the end of the Seiberg duality cascade, diverges.

   \item
   {\it Unlike the ``hard wall models" which follow a bottom-up approach and hence are toy models, in our top-down approach, the gauge field $A_t(r)$ corresponding to a non-zero chemical potential (in the presence of which we calculate $T_c$ in Section {\bf 3}), is obtained from its EOM from the DBI action on the world volume of flavor $D7$-branes where the DBI action is constructed from pull-backs of type IIB metric and NS-NS $B$ field of [1].}
\end{enumerate}

\item
Using the sum of the five-dimensional Einstein-Hilbert and Gibbons-Hawking-York action and the $A_t(r)$ from step 1., the Hawking-Page transition or QCD deconfinement temperature $T_c$ is obtained.

\end{enumerate}

 We will assume $i\mu\in\mathbb{R}$ in Ouyang's embedding: $r^{\frac{3}{2}}e^{\frac{i}{2}(\psi-\phi_1-\phi_2)}\sin\frac{\theta_1}{2} \sin\frac{\theta_2}{2}=i|\mu|$,  which could be satisfied for $\psi=\phi_1+\phi_2+\pi$ and $r^{\frac{3}{2}}\sin\frac{\theta_1}{2} \sin\frac{\theta_2}{2}=|\mu|$. Using the same, one obtains the following metric for a space-time-filling wrapped $D7$-brane embedded in the resolved warped deformed conifold:
\begin{equation}
\label{i*g}
ds^2 = \frac{1}{\sqrt{h\left(r,\theta_2,\theta_1(r,\theta_2)\right)}}
\left(-g_1(r) dt^2+dx^2+dy^2+dz^2\right)+\sqrt{h\left(r,\theta_2,\theta_1(r,\theta_2)\right)}\Big[\frac{dr^2}{g_2(r)}+r^2 d{\cal M}_3^2\Big],
\end{equation}
where
{\small
\begin{eqnarray}
\label{eq:metric_D7}
& & d{\cal M}_3^2 = {h_1} \left({d\phi_2} (\cos ({\theta_2})+1)+{d\phi_1} \left(2-\frac{2 |\mu| ^2 \csc
   ^2\left(\frac{{\theta_2}}{2}\right)}{r^3}\right)\right)^2+ \nonumber\\
   & & {h_2} \left(\left(1-\left(1-\frac{2 |\mu| ^2
   \csc ^2\left(\frac{{\theta_2}}{2}\right)}{r^3}\right)^2\right) {d\phi_1}^2+\frac{|\mu| ^2 \left(\frac{3
   {dr}}{r}+{d\theta_2} \cot \left(\frac{{\theta_2}}{2}\right)\right)^2}{r^3 \left(\sin
   ^2\left(\frac{{\theta_2}}{2}\right)-\frac{|\mu| ^2}{r^3}\right)}\right)\nonumber\\
   & & +{h_5} \cos
   ({\phi_1}+{\phi_2}) \left(-\frac{{d\theta_2} |\mu|  \left(\frac{3 {dr}}{r}+{d\theta_2} \cot
   \left(\frac{{\theta_2}}{2}\right)\right)}{r^{3/2} \sqrt{\sin
   ^2\left(\frac{{\theta_2}}{2}\right)-\frac{|\mu| ^2}{r^3}}}-{d\phi_1} {d\phi_2} \sqrt{1-\left(1-\frac{2
   |\mu| ^2 \csc ^2\left(\frac{{\theta_2}}{2}\right)}{r^3}\right)^2} \sin ({\theta_2})\right)\nonumber\\
   & & +{h_5} \sin
   ({\phi_1}+{\phi_2}) \left(-\frac{|\mu|  \left(\frac{3 {dr}}{r}+{d\theta_2} \cot
   \left(\frac{{\theta_2}}{2}\right)\right) \sin ({\theta_2}) {d\phi_2}}{r^{3/2} \sqrt{\sin
   ^2\left(\frac{{\theta_2}}{2}\right)-\frac{|\mu| ^2}{r^3}}}+{d\phi_1}{d\phi_2} \sqrt{1-\left(1-\frac{2 |\mu| ^2 \csc
   ^2\left(\frac{{\theta_2}}{2}\right)}{r^3}\right)^2} \right)\nonumber\\
   & & +{h_4} \left({h_3}
   {d\theta_2}^2+{d\phi_2}^2 \sin ^2({\theta_2})\right).
\end{eqnarray}}
From (\ref{three-form-fluxes}), using the Ouyang embedding (implying $d\psi = d\phi_1 + d\phi_2,
d\theta_1 = - \tan\left(\frac{\theta_1}{2}\right)\left(3 \frac{dr}{r} + \cot\left(\frac{\theta_2}{2}\right)d\theta_2\right)$) \cite{MQGP}:
\begin{eqnarray}
\label{B_Ouyang}
& & B_2 = - \frac{3}{r}\tan\frac{\theta_1}{2}\left(B_{\theta_1\phi_1} + B_{\theta_1\psi}\right)dr\wedge d\phi_1 +
 \left[B_{\theta_2\phi_1}- \tan\frac{\theta_1}{2}\cot\frac{\theta_2}{2}\left(B_{\theta_1\phi_1} + B_{\theta_1\psi}\right)\right]d\theta_2\wedge d\phi_1
 \nonumber\\
 & & - \frac{3}{r}\tan\frac{\theta_1}{2}\left(B_{\theta_1\phi_2} + B_{\theta_1\psi}\right)dr\wedge d\phi_2
+\left[ B_{\theta_2\phi_2} - \tan\frac{\theta_1}{2}\cot\frac{\theta_2}{2}\left(B_{\theta_1\phi_2} + B_{\theta_1\psi}\right)\right]d\theta_2\wedge d\phi_2.
\end{eqnarray}
Hence, in the MQGP limit one obtains the following DBI action:
{
\begin{eqnarray}
\label{SDBI-arb-mu}
& & \hskip -0.45in S_{\rm DBI}=\int_{\mathbb{R}^{1,3}}\int_{r_h}^\infty dr \int_0^{2\pi}\int_0^{2\pi}d\phi_1d\phi_2\int_0^\pi d\theta_2\sqrt{det\left(i^*(g + B) + F\right)}\nonumber\\
& & \hskip -0.45in \sim N_f\int_{r_h}^\infty dr\int_{\theta_2=0}^{\pi}d\theta_2 \Biggl\{\left({F_{rt}}^2-1\right) \cot ^2\left(\frac{{\theta_2}}{2}\right) \csc ^4\left(\frac{{\theta_2}}{2}\right) \left(2 \left(5 |\mu| ^2-2 r^3\right) \cos
   ({\theta_2})+14 |\mu| ^2+3 r^3 \cos (2 {\theta_2})+r^3\right)\nonumber\\
   & &\hskip -0.45in \times \left(\left(8 |\mu| ^2-4 r^3\right) \cos ({\theta_2})+r^3 (\cos (2 {\theta_2})+3)\right)\Biggr\}^{\frac{1}{2}} + {\cal O}\left(\frac{1, h_5, \frac{a^2}{r^2}}{\sqrt{g_s N}}\right),
\end{eqnarray}
$i^*g$ denoting the pulled-back metric as given in (\ref{i*g}) and (\ref{eq:metric_D7}), and $i^*B$ denoting the pulled-back NS-NS $B$ as given in (\ref{B_Ouyang}).

In the MQGP limit, taking the large-$r$ limit after angular integration in (\ref{SDBI-arb-mu}), using the results of appendix {\bf A}, one obtains:
\begin{equation}
\label{S_aftertheta2UV}
S\sim\int_{r=r_h}^\infty dr\left[\sqrt{|\mu|}r^{\frac{9}{4}}\sqrt{1 - F_{rt}^2} + {\cal O}\left(r^{\frac{3}{2}},(1,h_5,\frac{a^2}{r^2})\left[\frac{1}{\sqrt{g_sN}},\frac{g_sM^2}{N}\right]\right)\right].
\end{equation}

With $e^{-\phi}\approx \frac{1}{g_s} - \frac{N_f}{2\pi} ln\mu$ in the MQGP limit, one obtains:
\begin{eqnarray}
\label{At}
& & A_t = r \ _2F_1\left(\frac{2}{9}, \frac{1}{2}, \frac{11}{9}, -\frac{r^{\frac{9}{2}} \left(\frac{1}{g_s} - \frac{N_f ln \mu}{2 \pi}\right)^2}{C^2}\right)\approx \frac{72 \pi ^3 C^3 {g_s}^3 \left(\frac{1}{r}\right)^{23/4} \Gamma
   \left(\frac{11}{9}\right)}{23 \Gamma \left(\frac{2}{9}\right) ({g_s}
   N_f \log (\mu )-2 \pi )^3} -\frac{36 \pi  C {g_s}
   \left(\frac{1}{r}\right)^{5/4} \Gamma \left(\frac{11}{9}\right)}{5 \Gamma
   \left(\frac{2}{9}\right) ({g_s} N_f \log (\mu )-2 \pi )}\nonumber\\
   & & +\frac{2^{4/9}
   \Gamma \left(\frac{5}{18}\right) \Gamma \left(\frac{11}{9}\right) (C
   {g_s})^{4/9}}{\sqrt[18]{\pi } ({g_s} N_f \log (\mu )-2 \pi )^{4/9}} \equiv \gamma_1 - \frac{\gamma_2}{r^{\frac{5}{4}}} + \frac{\gamma_3}{r^{\frac{23}{4}}}.
   \end{eqnarray}
   Now, (\ref{At}) implies:
   \begin{eqnarray}
   \label{muC}
   & & \mu_C = \int_{r_h}^\infty F_{rt}dr\nonumber\\
   & & = \frac{2^{4/9} \left(C g_s\right)^{\frac{4}{9}} \Gamma \left(\frac{5}{18}\right) \Gamma \left(\frac{11}{9}\right) }{\sqrt[18]{\pi } \left(2 \pi -g_s {Nf} \log (\mu )\right)^{\frac{4}{9}}}-{r_h} \,
   _2F_1\left(\frac{2}{9},\frac{1}{2};\frac{11}{9};-\frac{{r_h}^{9/2} (g_s {Nf} \log (\mu )-2 \pi )^2}{4 C^2 g_s^2 \pi ^2}\right).
   \end{eqnarray}

 Choosing a $\gamma$: $\int_{r_h}^{r_\Lambda}\sqrt{g}\left(A_t - \gamma\right)^2\sim\int_{r_h}^{r_\Lambda}r^3\left(A_t - \gamma\right)^2<\infty$, i.e.,
 \begin{equation}
 \frac{8}{11} {\gamma_2} {r_\Lambda}^{11/4} (\gamma -{\gamma_1})+\frac{1}{4}
   {r_\Lambda}^4 (\gamma -{\gamma_1})^2+\frac{2}{3} {\gamma_2}^2
   {r_\Lambda}^{3/2} = 0,
 \end{equation}
 this is solved for:
 \begin{equation}
 \gamma = \frac{{\gamma_3}}{r_\Lambda^{23/4}}+\frac{1}{33} {\gamma_2}
   \left(-\frac{33}{r_\Lambda^{5/4}}+\frac{2 \left(24+5 i
   \sqrt{6}\right)}{{r_\Lambda}^{5/4}}\right).
 \end{equation}
 Utilizing that dimensionally $[C]=[r^{\frac{9}{4}}]$, this implies that one can impose a Dirichlet boundary condition at a cut-off $r_0: A_t(r_0) - \gamma=0$ where the cut-off is given by:
 \begin{equation}
 \label{cut off}
 \frac{C g_s \pi}{r_0^{\frac{9}{4}}\left( - 2 \pi + g_s N_f ln \mu\right)} = \pm\sqrt{\frac{23}{10}}.
 \end{equation}
 As $e^{-\phi}\approx \frac{1}{g_s} - \frac{N_f ln\mu}{2\pi}>0$ we choose the minus sign in   (\ref{cut off}).
 Writing $C\equiv m_\rho^{\frac{9}{4}}$ on dimensional grounds, where $m_\rho$ provides the mass scale of the lightest vector boson, one obtains:
 \begin{equation}
 \label{mrho}
 m_\rho = \frac{\left(\frac{23}{10}\right)^{2/9} {r_0} \left(\frac{2 \pi -{g_s} {N_f} \log
   (|\mu|)}{{g_s}}\right)^{4/9}}{\pi ^{4/9}}.
 \end{equation}
 If $m_\rho=760$ MeV the cut-off $r_0$ in units of $MeV$, from (\ref{mrho}), is given by:
\begin{equation}
\label{r0}
 r_0 = \frac{760 \left(\frac{10}{23}\right)^{2/9} \pi ^{4/9}}{\left(\frac{2 \pi -{g_s} {N_f}   \log (|\mu| )}{{g_s} \kappa }\right)^{4/9}}.
 \end{equation}

Our next task would be to establish a relationship between the QCD deconfinement temperature and $r_0$, incorporating thereby the effects of non-zero baryon chemical potential and charge density, and in the process working out the dependence of $T_c$ on $N_f$.
We consider the Einstein-Hilbert (EH) action along with the Gibbons-Hawking York surface term of the form
\begin{equation}
I=-\frac{1}{2\kappa^2}\int_{M} d^5x \sqrt{g}\left(R+\frac{12 }{L^2}\right)-\frac{1}{\kappa^2}\int_{\partial M} d^4x \sqrt{g_B}K.
  \end{equation}
  where $g_B$ is the metric at the boundary and $K$ is the extrinsic curvature of the boundary.
  Now, the cut-off thermal AdS  metric is given as:
  \begin{equation}
  ds^2=\frac{r^2}{L^2}(-dt^2+dx^2+dy^2+dz^2)+\frac{L^2}{r^2}dr^2.
  \end{equation}
  The radial coordinate $r$ varies from the IR cut-off at $r=r_0$ to the boundary at $r=\infty$. The  AdS-black hole/brane metric is given as:
  \begin{equation}
  ds^2=\frac{r^2}{L^2}(-g(r)dt^2+dx^2+dy^2+dz^2)+\frac{L^2}{g(r)r^2}dr^2
  \end{equation}
  where $g(r)=1-\frac{r_h^4}{r^4} + {\cal O}\left(\frac{g_s M^2}{N}\right)$. The Hawking temperature is given by $T_h=\frac{r_h}{\pi L^2}$ \cite{MQGP}. In the black hole case the periodicity of $t$ is given as $0\leq t\leq \frac{\pi L^2}{r_h}$, while in thermal AdS it is not constrained. In each case, we have $R=-(20/L^2)$ and hence the on shell EH action is given as
  \begin{equation}
  I_{M}=\frac{4}{L^2\kappa^2}\int d^5x \sqrt{g}.
   \end{equation}
The GHY surface term can be written as:
 \begin{equation}
  I_{\partial M}=-\frac{1}{\kappa^2}\int_{\partial M} d^4x~ \partial_n\sqrt{g_B}.
   \end{equation}
   where $n$ is defined as the unit normal to the boundary.

  Now for the regularity of the action at the boundary for both the solution, we integrate up to a UV cut-off $r=r_{\Lambda}$ but will take the limit of $r_{\Lambda}\rightarrow\infty$ at the end. The regularized action  for thermal AdS background is given by:
   \begin{equation}
  V_1=\frac{4}{\kappa^2 L^5}\int^{\beta}_0 dt\int_{r_0}^{r_\Lambda} dr~ r^3-\frac{4}{\kappa^2 L^5}\int^{\beta}_0 dt\left(\sqrt{g(r)}r^4\right)
  \end{equation}
  For the black hole in AdS, the same is given by
  \begin{equation}
  V_2=\frac{4}{\kappa^2 L^5}\int^{\frac{\pi L^2}{r_h}}_0 dt\int_{max(r_0,r_h)}^{r_\Lambda} dr~ r^3-\frac{2}{\kappa^2 L^5}\int^{\frac{\pi L^2}{r_h}}_0 dt~~ r^4(1+g(r)).
  \end{equation}
  Comparing the two energy densities at $r=r_\Lambda$ and using $\beta=(\pi L^2/r_h)\sqrt{g_1(r_\Lambda)}$ we get:
 \begin{eqnarray}
\Delta V & = & \lim_{r_\Lambda\rightarrow\infty}\Biggl(V_2(r_\Lambda)-V_1(r_\Lambda)\Biggr)
\nonumber\\
& = &\frac{\pi}{L^3\kappa^2  r_h}\frac{r_h^4}{2}\hspace{35pt} r_0>r_h\nonumber\\
& = & \frac{\pi}{L^3\kappa^2  r_h}\left(r_0^4-\frac{r_h^4}{2}\right)\hspace{25pt} r_0<r_h.
\end{eqnarray}
 The Hawking-Page phase transition occurs when $\Delta V$ is equal to zero giving $r_h=2^{1/4}r_0$
 which gives the transition temperature
 \begin{equation}
 \label{Tc}
T_c = 2^{\frac{1}{4}}r_0/L^2\pi.
 \end{equation}
The result of (\ref{Tc}) also appears in \cite{Herzog-Tc} but unlike \cite{Herzog-Tc}, we also incorporate the GHY surface term and show that the result is unchanged. So, from (\ref{r0}) and (\ref{Tc}), one obtains:
  \begin{equation}
 \label{Nf_1}
 N_f = \frac{\frac{46 \pi }{{g_s}} \pm \frac{288800 2^{9/16} 5^{3/4} \sqrt[4]{19} \sqrt{23}
   \sqrt[8]{{g_s} N}}{\pi ^{19/8} {g_s}^{5/4} N^{5/4} {T_c}^{9/4}}}{23 \log (|\mu|)};
 \end{equation}
 we choose the plus sign as, in accordance with lattice calculations, $T_c$ must decrease with $N_f$ \cite{dTcoverdNfnegative}.
 In the MQGP limit taking $g_s=0.8$   in (\ref{Nf_1}), one obtains:
 \begin{equation}
 \label{Nf_2}
 N_f=\frac{7.85398 + \frac{2.94676}{{T_c}^{9/4}}}{\log (|\mu|)}.
 \end{equation}
  Hence, {\bf for   $\mu = 13.7 i, N_f = 3$, one obtains the QCD deconfinement temperature $T_c=175-190$ MeV, consistent with lattice calculations \cite{Lattice_Tc} and the correct number of light quark flavors}.

 Now, dimensionally, $[\mu]=[r^{\frac{3}{2}}]$ and using the  AdS/CFT dictionary, hence mass dimensions of 3/2. Curiously, if in the mass term (\ref{W masses}), one were to set $\sqrt{|\mu|} = m_q^{\frac{3}{4}}$, one would obtain, {\bf in units of $MeV$, $m_q\approx 5.6$ - exactly the mass scale of the first generation light quarks!}

 The thermodynamical stability conditions are governed by inequalities imposed on certain thermodynamical quantities such as $\Delta S<0, \Delta E>0 ~{\rm and} ~\Delta H>0$ (which measure deviations from
equilibrium values implied). Considering that $\Delta E(S,V,N)$ and $\partial^2 E(S,V,N) >0$  and expanding $\partial^2 E(S,V,N)$ around equilibrium values of $(S_0, V_0, N_0)$ leads to three conditions  $C_v>0, \left.\frac{\partial \mu_C}{\partial T}\right|_{N_f}<0, \left.\frac{\partial\mu_C}{\partial N_f}\right|_{T}>0$ for the system to be in stable thermodynamic equilibrium at constant value of S, V and N \cite{Bruno}. From (\ref{muC}), one sees that for $g_s=0.8,N_f=3,\mu=13.7i$:
   \begin{eqnarray}
   \label{th-stab}
   & &  \left.\frac{\partial \mu_C}{\partial T}\right|_{N_f}= - \frac{\partial S}{\partial N_f}\Biggr|_T = \pi \sqrt{4\pi g_sN}\left.\frac{\partial \mu_C}{\partial r_h}\right|_{N_f}=\pi\sqrt{4\pi g_sN}\left(-\frac{1}{\sqrt{\frac{{r_h}^{9/2} (g_s {Nf} \log (|\mu| )-2 \pi )^2}{4 \pi ^2 C^2 g_s^2}+1}}\right)<0;\nonumber\\
   & &  \left.\frac{\partial \mu_C}{\partial N_f}\right|_{T}= \frac{4\ 2^{4/9} \Gamma \left(\frac{5}{18}\right) \Gamma \left(\frac{11}{9}\right) \log (|\mu| )}{9 \sqrt[18]{\pi } C \left(\frac{({g_s} {Nf} \log (|\mu| )-2
   \pi )^2}{C^2 {g_s}^2}\right)^{13/18}}\nonumber\\
   & &  -\frac{4 {g_s} {r_h} \log (|\mu| ) \left(\frac{1}{\sqrt{\frac{{r_h}^{9/2} (g_s {Nf} \log (|\mu| )-2
   \pi )^2}{4 \pi ^2 C^2 g_s^2}+1}}-\, _2F_1\left(\frac{2}{9},\frac{1}{2};\frac{11}{9};-\frac{{r_h}^{9/2} (g_s {Nf} \log (|\mu| )-2 \pi )^2}{4
   C^2 {g_s}^2 \pi ^2}\right)\right)}{9 ({g_s} {N_f} \log (|\mu| )-2 \pi )}>0,\nonumber\\
   & &
\end{eqnarray}
which demonstrates the thermodynamical stability of the type IIB background of \cite{metrics}.

Hence, ensuring thermodynamical stability and with the lightest vector meson mass as an input,  for an appropriate imaginary Ouyang embedding parameter, {\bf it is possible to obtain the QCD deconfinement temperature consistent with lattice results for the right number of light quark flavors}, in the MQGP limit from the type IIB background of \cite{metrics} {\bf in such a way that the modulus of the Ouyang  embedding parameter gives the correct first generation quark mass scale!}

 As alluded to in Section {\bf 1}, the above is expected to be related to the fact that the underlying type IIB background possesses $SU(3)$ structure and not $SU(3)$ holonomy, eventually translated into the non-K\"{a}hlerity of the non-extremal resolved warped deformed conifold. The reason is the following. The starting point of the calculations of this section involves  pull-backs of the ten-dimensional non-extremal resolved warped deformed conifold metric and the NS-NS $B$ on to the $D7$-branes' world volume in the evaluation of the DBI action. Assuming a constant axion-dilaton modulus and disregarding the contribution from the RR-sector for simplification of explanation, schematically the equations of motion will consist of $R_{mn}\sim ( H)^2_{mn}$\footnote{The RR-sector field strengths $F_{p=1,3,5\ {\rm or}\ 2,4}$ will contribute, schematically, as $\left(F_p\right)^2_{mn}$.} implying violation of Ricci-flatness of the ten-dimensional background. This can be recast into the language of contorsions (Section {\bf 5}) wherein the NS-NS field strength $H$ plays the role of contorsion such that the covariant spinorial derivatives, apart from a spin-connection, necessarily require the inclusion of $H$ in the metric-compatible connection. It turns out after evaluation of the $SU(3)$-structure torsion classes (\cite{transport-coefficients} and Section {\bf 5}) that the non-extremal resolved warped deformed conifold is non-K\"{a}her. This is encoded, e.g., in the relationship between between the mass of the lightest vector meson (appearing through an integration constant in the solution to the $A_t(r)$'s EOM obtained from the aforementioned DBI action; $A_t(r)$ being determined from the variation of the DBI action constructed from pull-backs of the background metric and NS-NS $B$) and $T_c$.

\section{$N_f=2$ Gauge Field Fluctuations}

Within the framework of linear response theory, the Einstein's relation according to which the ratio of the DC electrical conductivity and charge susceptibility yields the diffusion constant, must be satisfied. Using the $U(1)$ background of Sec. {\bf 4}, we explore this issue and see if the same imposes any non-trivial constraints on any of the parameters. The main result of this section is that imposing the Einstein's relation requires the Ouyang embedding parameter corresponding to the holomorphic embedding of $N_f$ $D7$-branes in the non-extremal resolved warped deformed conifold, to have a specific dependence on the horizon radius $r_h$.

We first discuss the EOMs and their solutions for non-abelian gauge field fluctuations for $N_f=2$ about the background calculated in Sec. {\bf 3} using the formalism of \cite{Erdmenger_et_al}. Using the revised background field strength of {\bf 3}, we also obtain the EOM and its solution for the $U(1)$ gauge field. We then calculate the DC electrical conductivity and the charge susceptibility, and comment on the Einstein relation relating their ratio to the diffusion constant.

Considering a chemical potential with $SU(2)$ flavor structure the general action is given by:
\begin{equation}
S=-T_rT_{D7}\int d^8\xi \sqrt{\det(g+\hat{F})}
\end{equation}
where the group-theoretic factor $T_r=\frac{1}{2}$ for $SU(2)$ and the field strength tensor is given as:
\begin{equation}
\hat{F}_{\mu\nu}=\sigma^{a}(2\partial_{[{\mu}}\hat{A}^{a}_{\nu]}+\frac{r^2_h}{2\pi\alpha^{'}}f^{abc}\hat{A}^{b}_{\mu}\hat{A}^{c}_{\nu}),
\end{equation}
$\sigma^{a}$ are the Pauli matrices and $\hat{A}$ is given by
\begin{equation}
\hat{A}_{\mu}=\delta^{0}_{\mu}\tilde{A}_{0}+A_{\mu}
\end{equation}
with the $SU(2)$ background gauge field
\begin{equation}
\tilde{A}^{3}_{0}\sigma^{3}=\tilde{A}_{0}\left(
                                           \begin{array}{cc}
                                             1 & 0 \\
                                             0 & -1 \\
                                           \end{array}
                                         \right).
\end{equation}
Now collecting the induced metric $g$ and the background field tensor $\tilde{F}$ as another background tensor $G=g+\tilde{F}$ we get equation of motion for gauge field fluctuation $A^{a}_{\mu}$ on $D7$-brane from the action quadratic in the same gauge fluctuation as in \cite{Erdmenger_et_al}:
\begin{equation}
\partial_\kappa[\sqrt{{\rm det}\ G}(G^{\nu\kappa}G^{\sigma\mu}-G^{\nu\sigma}G^{\kappa\mu})\widehat{F^a_{\mu\nu}}]=\sqrt{{\rm det}\ G}\frac{r^2_h}{2\pi{\alpha^\prime}}
\tilde{A}^3_0f^{ab3}(G^{\nu t}G^{\sigma\mu}-G^{\nu\sigma}G^{t\mu})\widehat{F^b_{\mu\nu}}.
\end{equation}
This simplifies to yield:
\begin{eqnarray}
\label{eom}
& & -2\partial_u[\sqrt{{\rm det}\ G}(G^{uu}G^{yy})(2\partial_u A^a_y)]-2\partial_t[\sqrt{{\rm det}\ G}G^{yy}G^{tt}(2\partial_tA^a_y)+\sqrt{{\rm det}\ G}G^{yy}G^{tt}f^{ab3}\tilde{A}^3_0\frac{r^2_h}{2\pi{\alpha^\prime}} A^b_y]\nonumber\\
& & = -2\sqrt{{\rm det}\ G}\frac{r^2_h}{2\pi{\alpha^\prime}}\tilde{A}^3_0f^{ab3}G^{yy}G^{tt}(2\partial_t A^b_y)-2\sqrt{{\rm det}\ G}\frac{r^2_h}{2\pi{\alpha^\prime}}\tilde{A}^3_0G^{yy}G^{tt}f^{ab3}f^{bc3}
\tilde{A}^3_0\frac{r^2_h}{2\pi{\alpha^\prime}}A^c_y.
\end{eqnarray}
Now, choosing the momentum four-vector in $\mathbb{R}^{1,3}$ as $q^\mu = (w, q, 0, 0)$, and with a slight abuse of notation, writing $A_\mu^a(x,u)=\int d^4q e^{- i w t + i q x}A^a_\mu(q,u)$, the simplification of (\ref{eom}) and rewriting in terms of the gauge-invariant variables or electric field components $E^a_T = \omega A^a_y, a=1,2,3$ as well as a further simplification using $X \equiv E^1 + i E^2, Y \equiv E^1 - i E^2$, in the $q=0$-limit, their solutions up to linear order in $w$, are presented in Appendix {\bf B}.

In the same appendix, for the purpose of evaluation of DC electrical conductivity, the on-shell action too is worked out.  As shown in \cite{Erdmenger_et_al}, the on-shell action is given by:
\begin{equation}
S_{\rm on-shell}\sim T_r T_{D7}\int d^4x\sqrt{{\rm det}\ G}\left.\left(G^{\nu u}G^{\nu'\mu}-G^{\nu \nu'}G^{u \mu}\right)A^a_{\nu'}\widehat{F^a_{\mu\nu}}\right|_{u=0}.
\end{equation}
Working in the gauge $A^a_u=0$, in appendix {\bf B}, the following on-shell action's integrand is worked out:
\begin{eqnarray}
\label{DBI-EOM}
& & \sqrt{{\rm det}\ G}\left[\frac{4G^{uu}G^{xx}(G^{ut}G^{ut}-G^{uu}G^{tt})}{q^2(G^{uu}G^{xx})+w^2(G^{tt}G^{uu}-G^{ut}G^{ut})}E^a_x(\partial_uE^a_x)-\frac{4}{w^2}G^{uu}G^{\alpha \alpha}E^a_{\alpha}(\partial_uE^a_{\alpha})+...\right]_{u=0}\nonumber\\
& & = 4\left(\frac{r_h u(u^4-1)}{w^2(\frac{r_h}{u})^{3/4}\sqrt{\frac{r^4_h\sqrt{\frac{r_h}{u}}}{r^4_h\sqrt{\frac{r_h}{u}}+c^2e^{2\phi}u^4}}}E^a_x(\partial_uE^a_x) + \frac{r_h u(u^4-1)}{w^2(\frac{r_h}{u})^{3/4}\sqrt{\frac{r^4_h\sqrt{\frac{r_h}{u}}}{r^4_h\sqrt{\frac{r_h}{u}}+c^2e^{2\phi}u^4}}}E^a_{\alpha}(\partial_uE^a_{\alpha})+...
\right)_{u=0}\nonumber\\
& & \sim \left.\frac{r_h^{\frac{1}{4}}u^{\frac{7}{4}}}{w^2}\left(E^a_x(\partial_uE^a_x) + E^a_{\alpha}(\partial_uE^a_{\alpha})\right)+....\right|_{u\rightarrow0},
\end{eqnarray}
where the dots include the flavor anti-symmetric terms.

Defining the longitudinal electric field as $E_{x}(q,u)= E_{0}(q) \frac{E_{q}(u)}{E_{q}(u=0)}$, the flux factor as defined in \cite{[10]} in the zero momentum limit, using (\ref{F}) and (\ref{DBI-EOM}) will hence be given as:
\begin{eqnarray}
{\cal F}(q,u)=  -\frac{ e^{-\phi(u)}  r_h^{\frac{1}{4}}u^{\frac{7}{4}} }{w^2 }  \frac{E_{-q}(u)\partial_{u}E_{q}(u)}{ E_{-q}(u=0)E_{q}(u=0)},
\end{eqnarray}
and  the retarded Green's function for $E_{x}$, using the prescription of \cite{[10]}, will be given by: ${\cal G}(q,u)= -2 {\cal F}(q,u)$. The retarded Green function for $A_{x}$ is $w^2$ times above expression and for $q=0$, it gives
\begin{eqnarray}
{\cal G}_{x x} = \left.2 e^{-\phi(u)}   r_h^{\frac{1}{4}}u^{\frac{7}{4}} \frac{ \partial_{u}E_{q}(u)}{ E_{q}(u)}\right|_{u=0}.
\end{eqnarray}
The spectral functions in zero momentum limit will be given as:
\begin{equation}
\label{correlator_sigma}
{\cal X}_{x x}(w,q=0)= -2 Im {\cal G}_{x x}(w,0) = { e^{-\phi(u)}  r_h^{\frac{1}{4}} }Im\left[u^{\frac{7}{4}} \frac{ \partial_{u}E_{q}(u)}{ E_{q}(u)}\right]_{u=0}.
\end{equation}
The DC conductivity, using the discussion of sub-section {\bf 2.3} (b), is given by the following expression \cite{Mateos}, \cite{[10]}:
\begin{equation}
\label{conductivity}
\sigma = \lim_{w\rightarrow0}\frac{{\cal X}_{x x}(w,q=0)}{w}=\lim_{u\rightarrow0,w\rightarrow0}\frac{r_h^{\frac{1}{4}}u^{\frac{7}{4}}\Im m\left(\frac{E'(u)}{E(u)}\right)}{w}.
\end{equation}

The final result for the DC conductivity $\sigma$ is given as under:
\begin{equation}
\label{sigma-DC}
\sigma=\frac{r_h^{\frac{1}{4}}}{\pi T}\Im m\left(\frac{c_2\left(\frac{i}{16}(-)^{\frac{3}{4}}c_1 + \frac{\gamma_0}{4}c_2\right) - c_3\frac{c_1\gamma_0}{4}}{c_2^2}\right)\sim\left(g_sN\right)^{\frac{1}{8}}T^{-\frac{3}{4}}\frac{c_1}{c_2}.
\end{equation}
Interestingly, {\bf this mimics a one-dimensional interacting system - Luttinger liquid - on a lattice for appropriately tuned Luttinger parameter \cite{fraccond-Luttinger-1d}}.
 \footnote{One of us (AM) wishes to thank S. Mukerjee for pointing out this fact as well as \cite{fraccond-Luttinger-1d}.}

Another physically relevant quantity  is the charge susceptibility $\chi$, which is thermodynamically defined as response of the charge  density to the change in chemical potential, is given by the following expression  \cite{J. Mas et al [2008]}:
\begin{eqnarray}
\label{chi-a}
& & \chi=\left.\frac{\partial n_q}{\partial \mu_C}\right|_{T},
\end{eqnarray}
 where $n_q = \frac{\delta S_{DBI}}{\delta F_{rt}}$, and the chemical potential $\mu_C$ is defined as
 $\mu_C=\int_{r_h}^{r_B} { F_{rt}} dr$. The charge density will be given as:
\begin{equation}
\label{nq-ii}
n_q = \frac{\delta S_{\rm DBI}}{\delta F_{rt}} \sim \frac{F_{rt}\sqrt{|\mu|}r^{\frac{9}{4}}}{\sqrt{1-F_{rt}^2}},
\end{equation}
and using (\ref{chi-a}), one gets the following charge susceptibility:
\begin{eqnarray}
\label{chi-ii}
& & \frac{1}{\chi} = \int_{r_h}^\infty dr \frac{dF_{rt}}{dn_q}  = \int_{r_h}^\infty dr \frac{r^{\frac{9}{2}}}{\sqrt{|\mu|}\left( \frac{C^2}{\left(\frac{1}{g_s} - \frac{N_f}{2\pi}\log|\mu|\right)^2} + r^{\frac{9}{2}}\right)^{\frac{3}{2}}}\nonumber\\
& & = \frac{1}{45 \sqrt{\mu
   } {r_h}^{5/4} \left(  \frac{C^2}{\left(\frac{1}{g_s} - \frac{N_f}{2\pi}\log|\mu|\right)^2}+{r_h}^{9/2}\right)}\Biggl\{414 {r_h}^{9/2}\ _2F_1 \left(-\frac{1}{2},\frac{5}{18};\frac{23}{18};-\frac{
    \frac{C^2}{\left(\frac{1}{g_s} - \frac{N_f}{2\pi}\log|\mu|\right)^2}}{{r_h}^{9/2}}\right)\nonumber\\
    & & +\left(4  \frac{C^2}{\left(\frac{1}{g_s} - \frac{N_f}{2\pi}\log|\mu|\right)^2}-5 {r_h}^{9/2}\right)
  \  _2F_1 \left(\frac{5}{18},\frac{1}{2};\frac{23}{18};-\frac{  \frac{C^2}{\left(\frac{1}{g_s} - \frac{N_f}{2\pi}\log|\mu|\right)^2}}{{r_h}^{9/2}}\right)\Biggr\}\nonumber\\
   & & = \frac{4}{5 \sqrt{|\mu| }\left(4\pi g_sN\right)^{\frac{5}{8}} {T}^{5/4}} + {\cal O}\left(\frac{1}{\left( g_sN\right)^{\frac{23}{8}}}\right).
\end{eqnarray}

Hence, the charge susceptibility is given by:
\begin{equation}
\chi\sim\sqrt{|\mu| }\left( g_sN\right)^{\frac{5}{8}} {T}^{5/4}.
\end{equation}
Given that one is in the regime of linear response theory, one expects the Einstein's relation: $\frac{\sigma}{\chi}=D\sim\frac{1}{T}$, to hold\footnote{One of us (AM) thanks V.B.Shenoy and S. Mukerjee for clarifications on this point.}. However, a naive application yields $\frac{\sigma}{\chi}\sim \frac{c_1}{c_2}\frac{1}
{\sqrt{|\mu|g_sN}}\frac{1}{T^2}$. One expects the Ouyang embedding parameter to be related to the deformation parameter if there were supersymmetry. In the MQGP limit, there is approximate supersymmetry. The resolution parameter possesses an $r_h$-dependence.
{\bf If one assumes that $|\mu|\sim\frac{1}{r_h^2}$ (in $\alpha^\prime=1$-units), then the Einstein's relation is preserved.}

The fact that the Ouyang embedding parameter turns out to be dependent on the horizon radius is reminiscent of the fact that the resolution parameter too turns out to be dependent on the horizon radius \cite{K. Dasgupta  et al [2012]}, and serves as an important constraint while studying Ouyang embeddings. Further, the 1+1-dimensional subspace singled out in the plane wave basis of the Fourier modes of the gauge field fluctuations, via the evaluation of the electrical conductivity, provides an important prediction that the theory mimicks a 1+1-dimensional Luttinger liquid for appropriately tuned interaction parameter. These comprise the second set of significant and new results of our paper.

\section{$SU(3)/G_2$ Structure Torsion Classes  of the type IIA mirror/$M$-Theory Uplift}

As argued in subsection {\bf 2.3}, the M-theory uplift of the type IIB holographic dual \cite{metrics} of thermal QCD with fundamental quarks is expected to possess a $G_2$ structure, but the explict construction of the same has thur far, been missing in the literature. We will present, locally, an explict $SU(3)$ structure of the SYZ type IIA mirror in {\bf 5.1} (along with demonstration of approximate supersymmetry in terms of constraints on the torsion classes upon comparison with \cite{Butti et al [2004]}) and an explicit $G_2$ structure of its local uplift to M-theory in the MQGP limit of (\ref{limits_Dasguptaetal-ii}) in {\bf 5.2}. This will comprise the final set of significant and new results of this paper.

Flux compactifications involving the NS-NS flux, typically require the internal six-dimensional geometry's departure from
K\"{a}hlerity and even from being a complex manifold \cite{Louis_et_al},\cite{torsion}. The results of sections {\bf 3} and {\bf 4} arise due to the non-K\"{a}hlerity and non-conformality of the type IIB background of \cite{metrics}. In this Math-oriented section, we will be quantifying this departure from K\"{a}hlerity of the delocalized type IIA mirror of the resolved deformed conifold of \cite{metrics}. Further, we will also be quantifying the departure of the seven-fold relevant to \cite{MQGP}'s local M-theory uplift  from being a $G_2$-holonomy manifold due to non-zero $G_4$-fluxes. To be more specific, in this section, in the MQGP limit, we (i) work out the $SU(3)$ structure torsion classes of the local type IIA mirror's six-fold and (ii) work out a local $G_2$-structure and $G_2$ structure torsion classes.

Utilizing the results of appendix {\bf C}, the five dimensional $(r,\theta_1,\theta_2,\phi_1,\phi_2,\psi)$ type IIA metric's large-$N$ small-$\theta$ expansion in the UV can be summarized as:
\begin{eqnarray}
\label{D5metric}
& & \hskip -1.1in\sqrt{g_sN}\left(
\begin{array}{ccccc}
 1 & 0 & -\frac{\sqrt[3]{3} {g_s}^{3/2} M {N_f} \log (r)}{\sqrt{2} N^{3/10} \pi ^{5/4} {\theta_1}^4} & \frac{9 3^{5/6} \sqrt{{g_s}} M {\theta_1}
   \log (r)}{N^{7/10} \sqrt[4]{\pi }} & \frac{3 \sqrt[3]{3} {g_s}^{3/2} M {N_f} r \log (r)}{2 \sqrt{2} N^{7/20} \pi ^{5/4} {\theta_1}^3} \\
 0 & 1 & -\frac{27 3^{5/6} a^2 {g_s}^{3/2} M r^2 {\theta_1} \log (r)}{N^{3/5} \sqrt[4]{\pi }} & \frac{\sqrt{2} \sqrt[4]{\pi }}{3^{2/3} N^{9/20}} & -\frac{3
   \sqrt[3]{3} {g_s}^{3/2} M {N_f} \log (r)}{32 \sqrt{2} N^{3/20} \pi ^{5/4} {\theta_1}^3} \\
 -\frac{\sqrt[3]{3} {g_s}^{3/2} M {N_f} \log (r)}{\sqrt{2} N^{3/10} \pi ^{5/4} {\theta_1}^4} & -\frac{27 3^{5/6} a^2 {g_s}^{3/2} M r^2
   {\theta_1} \log (r)}{N^{3/5} \sqrt[4]{\pi }} & 3^{2/3} {\theta_1}^2 & \frac{2 \sqrt{2}}{3 \sqrt[6]{3} \sqrt[5]{N} {\theta_1}} & -\frac{4
   \sqrt[5]{N}}{9 \sqrt[3]{3} {\theta_1}^3} \\
 \frac{9 3^{5/6} \sqrt{{g_s}} M {\theta_1} \log (r)}{N^{7/10} \sqrt[4]{\pi }} & \frac{\sqrt{2} \sqrt[4]{\pi }}{3^{2/3} N^{9/20}} & \frac{2 \sqrt{2}}{3
   \sqrt[6]{3} \sqrt[5]{N} {\theta_1}} & \frac{3^{2/3} {\theta_1}^2}{\sqrt[5]{N}} & -\frac{\sqrt{2}}{\sqrt[6]{3}} \\
 \frac{3 \sqrt[3]{3} {g_s}^{3/2} M {N_f} r \log (r)}{2 \sqrt{2} N^{7/20} \pi ^{5/4} {\theta_1}^3} & -\frac{3 \sqrt[3]{3} {g_s}^{3/2} M {N_f}
   \log (r)}{32 \sqrt{2} N^{3/20} \pi ^{5/4} {\theta_1}^3} & -\frac{4 \sqrt[5]{N}}{9 \sqrt[3]{3} {\theta_1}^3} & -\frac{\sqrt{2}}{\sqrt[6]{3}} & \frac{2
   \sqrt[5]{N}}{3 \sqrt[3]{3} {\theta_1}^2}
\end{array}
\right).\nonumber\\
& &
\end{eqnarray}
Assuming taking the large$-N$ limit before taking the UV limit, the above is approximated by:
\begin{eqnarray}
\label{D5ii}
& & \left(
\begin{array}{ccccc}
 \sqrt{{g_s} N} & 0 & 0 & 0 & 0 \\
 0 & \sqrt{{g_s} N} & 0 & 0 & 0 \\
 0 & 0 & 3^{2/3} \sqrt{{g_s} N} {\theta_1}^2 & 0 & -\frac{4 \sqrt[5]{N} \sqrt{{g_s} N}}{9 \sqrt[3]{3} {\theta_1}^3} \\
 0 & 0 & 0 & 0 & -\frac{\sqrt{2} \sqrt{{g_s} N}}{\sqrt[6]{3}} \\
 0 & 0 & -\frac{4 \sqrt[5]{N} \sqrt{{g_s} N}}{9 \sqrt[3]{3} {\theta_1}^3} & -\frac{\sqrt{2} \sqrt{{g_s} N}}{\sqrt[6]{3}} & \frac{2 \sqrt[5]{N}
   \sqrt{{g_s} N}}{3 \sqrt[3]{3} {\theta_1}^2}
\end{array}
\right).
\end{eqnarray}
We will consider the following more general (and therefore partly phenomenological) three-dimensional metric in $(\phi_1,\phi_2,\psi)$-space:
\begin{equation}
\label{d=3}
\left(
\begin{array}{ccc}
 {g_{11}} \sqrt{{g_s} N} {\theta_1}^2 & 0 & -\frac{{g_{13}} \sqrt[5]{N} \sqrt{{g_s} N}}{{\theta_1}^3} \\
 0 & 0 & -{g_{23}} \sqrt{{g_s} N} \\
 -\frac{{g_{13}} \sqrt[5]{N} \sqrt{{g_s} N}}{{\theta_1}^3} & -{g_{23}} \sqrt{{g_s} N} & \frac{{g_{33}} \sqrt[5]{N} \sqrt{{g_s}
   N}}{{\theta_1}^2}
\end{array}
\right).
\end{equation}

The three eigenvalues of (\ref{d=3})  and the normalized eigenvectors are worked out in appendix {\bf C}. From the same, the sechsbeins that would diagonalize the type IIA mirror metric in the large-$N$ small-$\theta_1$ limit  up to ${\cal O}(\theta_1^2)$, are given by
${\cal M}^{-1}\left(\begin{array}{c} d\phi_1 \\ d\phi_2 \\ d\psi\end{array}\right)$ (${\cal M}$ being the modal matrix as given in (\ref{M})):
\begin{eqnarray}
\label{inv_frames}
& & \hskip -0.2in e^1 = \left( g_s N\right)^{\frac{1}{4}}\frac{1}{r\sqrt{1 - \frac{r_h^4}{r^4}}}dr\nonumber\\
& &  \hskip -0.2in  e^2 = \left(g_s N\right)^{\frac{1}{4}} d\theta_1\nonumber\\
& &  \hskip -0.2in e^3 =\left(g_s N\right)^{\frac{1}{4}} d\theta_2\nonumber\\
& &  \hskip -0.2in e^4 =\sqrt{\frac{{g_{13}} \sqrt{{g_s}} N^{7/10}}{{\theta_1}^3}+\frac{0.5 {g_{33}} \sqrt{{g_s}} N^{7/10}}{{\theta_1}^2}}\left[{d\phi_1} {g_{23}} \left(\frac{-\frac{{g_{13}}^3}{\sqrt{2}}+0.18 {g_{13}}^2 {g_{33}} {\theta_1}-0.02 {g_{13}} {g_{33}}^2
   {\theta_1}^2}{{g_{13}}^3 {g_{23}}}\right)\right.\nonumber\\
& &  \hskip -0.2in \left. +{d\phi_2} {g_{33}}^3 \sqrt[5]{N} \left(\frac{0.05 {g_{13}}^3-0.04 {g_{13}}^2 {g_{33}}
   {\theta_1}+0.008 {g_{13}} {g_{33}}^2 {\theta_1}^2}{{g_{13}}^5 {g_{23}}}\right)\right.\nonumber\\
   & &\left. +{d\psi} {g_{13}}^2 {g_{23}}
   \left(\frac{\frac{{g_{13}}^3}{\sqrt{2}}+0.18 {g_{13}}^2 {g_{33}} {\theta_1}-0.07 {g_{13}} {g_{33}}^2 {\theta_1}^2}{{g_{13}}^5 {g_{23}}}\right)\right]\nonumber\\
   & &  \hskip -0.2in \equiv g_s^{\frac{1}{4}}N^{\frac{7}{20}}\Lambda_4(\theta_1)\left(\alpha_{41}d\phi_1 + \alpha_{42}d\phi_2 + \alpha_{43}d\psi\right)\nonumber\\
   & &  \hskip -0.2in e^5 = \sqrt{\frac{0.5 {g_{33}} \sqrt{{g_s}} N^{7/10}}{{\theta_1}^2}-\frac{{g_{13}} \sqrt{{g_s}} N^{7/10}}{{\theta_1}^3}}\left[ {d\phi_1}  {g_{23}} \left(\frac{\frac{{g_{13}}^3}{\sqrt{2}}+0.18 {g_{13}}^2 {g_{33}} {\theta_1}+0.02 {g_{13}} {g_{33}}^2
   {\theta_1}^2}{{g_{13}}^3 {g_{23}}}\right)\right.\nonumber\\
   & & \left. +{d\phi_2} {g_{33}}^3 \sqrt[5]{N} \left(\frac{0.05 {g_{13}}^3+0.04 {g_{13}}^2 {g_{33}}
   {\theta_1}+0.008 {g_{13}} {g_{33}}^2 {\theta_1}^2}{{g_{13}}^5 {g_{23}}}\right)\right.\nonumber\\
   & &  \hskip -0.2in\left. +{d\psi} {g_{13}}^2 {g_{23}}
   \left(\frac{\frac{{g_{13}}^3}{\sqrt{2}}-0.18 {g_{13}}^2 {g_{33}} {\theta_1}-0.07 {g_{13}} {g_{33}}^2 {\theta_1}^2}{{g_{13}}^5 {g_{23}}}\right)\right]\nonumber\\
   & &  \hskip -0.2in\approx g_s^{\frac{1}{4}}N^{\frac{7}{20}}\Lambda_5(\theta_2)\left(-\alpha_{41}d\phi_1 + \alpha_{42}d\phi_2 + \alpha_{43}d\psi\right)\nonumber\\
    & &  \hskip -0.2in e^6 = \sqrt{\frac{0.074 {g_{33}}^3 \sqrt{{g_s}} N^{7/10}}{{g_{13}}^2}} \left[ {g_{33}}^3 \left(\frac{0.07{g_{23}}{g_{13}} {\theta_1}^3   {d\phi_1} + {d\phi_2} \left(-0.07
   {g_{13}}^2-0.04 {g_{33}}^2 {\theta_1}^2\right) - 0.012 {g_{23}} {\theta_1}^3{g_{33}}
   {\theta_1} {d\psi}  }{{g_{13}}^4 {g_{23}}}\right)\right]\nonumber\\
   & & \hskip -0.2in \equiv g_s^{\frac{1}{4}}N^{\frac{7}{20}}\left(\alpha_{61}d\phi_1 + \alpha_{62}d\phi_2 + \alpha_{63}d\psi\right).
   \end{eqnarray}
For ensuring a non-singular nature of these sechsbeins and their orthonormality, we will demand that as $\theta_1\rightarrow0$ as $\alpha_\theta\epsilon^{\frac{5}{2}}$ with $\epsilon\stackrel{<}{\sim}1$ and $\alpha_\theta\sim N^{-\frac{1}{5}}\ll  1$,
\begin{equation}
\label{gij_constr}
g_{13}\sim\alpha_\theta^3;\  g_{33}\sim\alpha_\theta^2;\ g_{11}\sim \alpha_\theta^{-3}.
\end{equation}
It is crucial to verify the orthonormality of the (inverse) frames obtained in (\ref{inv_frames}), and it will turn out that this will require $g_{23}$ to be large.  We need to verify: $G^{IIA}_{\mu\nu} = e^a_\mu e^b_\nu\eta_{ab}$. This is verified in Appendix {\bf C}.

Defining $E^1 = e^1 + i e^6,\ E^2 = e^2 + i e^3, E^3 = e^4 + i e^5$, one can write the following fundamental two-form $J$ and the holomorphic three form $\Omega$ as $J=\frac{i}{2}\Biggl(E^1\wedge \bar{E}^1 + E^2\wedge\bar{E}^2 + E^3\wedge \bar{E}^3\Biggr)$ and $\Omega = E^1\wedge E^2\wedge E^3$. From (\ref{inv_frames}) and using
(\ref{gij_constr}),  one sees that near $\theta_1=0$:
\begin{eqnarray}
\label{des}
& & d^{1,2,3}=0\nonumber\\
& & de^{4,5} \sim g_s^{\frac{1}{4}}N^{\frac{7}{20}}\frac{d\theta_1}{\theta_1}\wedge e^{4,5}\frac{\lambda_{4,5}}{\Lambda_{4,5}^2}\sim N^{\frac{1}{10}} e^2\wedge e^{4,5}\frac{\lambda_{4,5}}{\theta_1\Lambda_{4,5}^2}\nonumber\\
& & de^6 \sim g_s^{\frac{1}{4}}N^{\frac{7}{20}}\frac{d\theta_1}{\theta_1}\wedge\left(3\alpha_{61}d\phi_1 + 2\alpha_{62}d\phi_2 + 4\alpha_{63}d\psi\right),
\end{eqnarray}
implying:
\begin{eqnarray}
\label{dEs}
& & dE^1\sim g_s^{-\frac{1}{4}}N^{-\frac{1}{4}}\frac{\left(E^2 + \bar{E}^2\right)}{\theta_1}\wedge\left[\left(E^1 - \bar{E}^1\right)\chi_1 + E^3\chi_2 + \bar{E}^3\chi_3\right],\ \Re e\chi_1=0,\ \chi_{2,3}\in\mathbb{C}\nonumber\\
& & dE^2=0\nonumber\\
& & dE^3\sim g_s^{-\frac{1}{4}}N^{\frac{1}{10}}\frac{\left(E^2 + \bar{E}^2\right)}{\theta_1}\wedge\left[\frac{\lambda_4}{2\Lambda_4^2}\left(E^3 + \bar{E}^3\right) + \frac{\lambda_5}{2\Lambda_5^2}\left(E^3 - \bar{E}^3\right)\right].
\end{eqnarray}

 We will next work out the $SU(3)$-structure torsion classes of the delocalized type IIA mirror and the $G_2$-structure torsion classes of the M-theory uplift in the MQGP limit. For the paper to be self-contained, we have given a self-contained introduction to $G$-structures as well as, specifically, $SU(3)$-structure and $G_2$-structure in appendices C and D, respectively.

\subsection{$SU(3)$-Structure Torsion Classes of Type IIA Mirror}

We will now quantify the deviation of the type IIB resolved warped deformed  conifold as well as its delocalized SYZ type IIA mirror  from being a complex manifold and/or K\"{a}hler by evaluating the $SU(3)$ structure torsion classes.

\noindent To quantify the deviation from K\"{a}hlerity of the resolved warped deformed conifold background of \cite{metrics}, we looked at the  five $SU(3)$ structure torsion classes in \cite{transport-coefficients}. Use was made of the observation that the  resolved warped deformed conifold can be written in the form of the \cite{Papadopoulos-Tseytlin [2001]}
 ansatz  in the string frame:
\begin{eqnarray*}
\label{metricD10}
ds^2 =  h^{-1/2}  ds^2_{\mathbb{R}^{1,3}}
 + e^x ds_\mathcal{M}^2  = h^{-1/2}  dx^2_{1,3} + \sum_{i=1}^6 G_i^2\ ,
\end{eqnarray*}
where \cite{Butti et al [2004], M.K. Bena+I.Klebanov [2008]}:
\begin{eqnarray*}
\label{Gforms}
& & G_1 \equiv e^{x(\tau)+g(\tau)/2}\,e_1,\nonumber\\
& & G_2 \equiv {\cal A}\,e^{(x(\tau)+g(\tau))/2}\,e_2 + {\cal B}(\tau)\,e^{(x(\tau)-g(\tau))/2}\,(\epsilon_2-a e_2)\ ,\nonumber\\
& & G_3 \equiv e^{(x(\tau)-g(\tau))/2}\,(\epsilon_1-a e_1)\ , \nonumber\\
& & G_4 \equiv {\cal B}(\tau)\,e^{(x(\tau)+g(\tau))/2}\,e_2 - {\cal A}\,e^{(x(\tau)-g(\tau))/2}\,(\epsilon_2-a e_2)\ , \nonumber\\
& & G_5 \equiv e^{x(\tau)/2}\,v^{-1/2}(\tau)d\tau\ , \nonumber\\
& & G_6 \equiv e^{x(\tau)/2}\,v^{-1/2}(\tau)(d\psi + \cos\theta_2d\phi_2 + \cos\theta_1d\phi_1),
\end{eqnarray*}
wherein ${\cal A}\equiv\frac{\cosh\tau+a(\tau)}{\sinh\tau},  {\cal B}(\tau)\equiv \frac{e^{g(\tau)}}{\sinh\tau}$. The $e_i$s are one-forms on ${S}^2$ and the $\epsilon_i$s a set of one-forms on ${S}^3$. As $r\sim e^{\frac{\tau}{3}}$,
in the MQGP limit, the metric  matches the RWDC metric with the identifications:
\begin{eqnarray*}
\label{identifics}
& & \frac{e^x(\tau)}{v(\tau)}\sim\frac{\sqrt{4\pi g_sN}}{9}\left(1 + {\cal O}(r_h^2e^{-\frac{2\tau}{3}})\right);\nonumber\\
 & & v(\tau)\sim\frac{3}{2}\left[1 + {\cal O}\left(\left\{\frac{g_sM^2}{N}a_{\rm res}^2,r_h^2\right\}e^{-\frac{2\tau}{3}}\right)\right];\nonumber\\
& &  e^{x(\tau)}\sim\frac{\sqrt{4\pi g_sN}}{6}\left[1 + {\cal O}\left(\frac{g_sM^2}{N}a_{\rm res}^2e^{-\frac{2\tau}{3}}\right)\right];\nonumber\\
& & g(\tau)\sim - 2 e^{-2\tau};\ a(\tau)\sim-2e^{-\tau}.
\end{eqnarray*}
 In the UV, ${\cal A}\sim1$ and ${\cal B}(\tau)\sim e^{-\tau}$ the five torsion classes were evaluated in \cite{transport-coefficients}. In the MQGP limit one sees that :
\begin{eqnarray}
\label{W12345}
& & \hskip -0.7in W_1 \sim \frac{e^{-3\tau}}{\sqrt{4\pi g_s N}}\ll  1\ ({\rm in\ the\ UV});\nonumber\\
& &\hskip -0.7in W_2 \sim \left(4\pi g_sN\right)^{\frac{1}{4}}e^{-3\tau}\left(d\tau\wedge e_\psi + e_1\wedge e_2 + \epsilon_1\wedge e_2\right)\ll  1\ ({\rm in\ the\ UV});\nonumber\\
& &\hskip -0.7in W_3 \sim \left.\sqrt{4\pi g_sN}\left(32\sqrt{\frac{2}{3}}e^{-3\tau}\left(e_1\wedge \epsilon_1 + e_2\wedge \epsilon_2\right)\wedge e_\psi + 2\sqrt{\frac{2}{3}}e_2\wedge \epsilon_1\wedge d\tau e^{-\tau} + 32e_1\wedge \epsilon_2\wedge d\tau e^{-3\tau} \right)\right|_{\theta_1\sim0;\ {\rm UV}}\ll  1;\nonumber\\
& & \hskip -0.7in W_4  \sim - \frac{2}{3}e^{-g(\tau)}d\tau = 2 W^3_4 = 2 W^{\bar{3}}_4;\nonumber\\
& & \hskip -0.7in W_5^{(\bar 3)} \sim -\frac{1}{2}\left(d\tau - i e_\psi\right),
\end{eqnarray}
implying that in the UV and near $\theta_i=0$, $T\in W_4\oplus W_5$ such that $\frac{2}{3}\Re e W^{\bar{3}}_5 = W^{\bar{3}}_4 = - \frac{1}{3}d\tau $ implying supersymmetry is preserved locally \cite{Butti et al [2004]}. This, in addition to (\ref{sLag-conditions}), provides a non-trivial justification for the application of SYZ mirror construction.  Obviously, in the strict $r\rightarrow\infty$ limit, one obtains a Calabi-Yau three-fold in which $W_{1,2,3,4,5}=0$.

We will now be addressing the issue of approximate supersymmetry of the delocalized type  type IIA mirror, by explicitly evaluating the $SU(3)$ structure torsion classes, locally, for the same. However, before doing so, let us get back to the issue of the $G_3$-fluxes being approximately of the (2,1)-type, as mentioned in sub-section {\bf 2.2}. For this purpose, we will closely be following \cite{Ouyang}. In \cite{Ouyang}, a basis of one-forms consisting of the following holomorphic forms and their complex conjugates, was constructed:
\begin{eqnarray}
\lambda &=& 3\frac{dr}{r} + i e_\psi, \\
\sigma_1 &=& \cot\frac{\theta_1}{2}(d\theta_1 -i \sin\theta_1 d\phi_1), \\
\sigma_2 &=& \cot\frac{\theta_2}{2}(d\theta_2 -i \sin\theta_2 d\phi_2),
\label{oneforms}
\end{eqnarray}
where $e_\psi\equiv d\psi+\cos\theta_1 d\phi_1 +\cos\theta_2 d\phi_2$ is the one-form associated with the $U(1)$ fiber of $T^{1,1}$. In the following, we will also be using a convenient shorthand notation introduced in \cite{Ouyang}:
\begin{equation}
\label{def-Omega}
\Omega_{ij} \equiv d\theta_i \wedge \sin\theta_j d\phi_j.
\end{equation}
We see that (\ref{oneforms}) and (\ref{def-Omega}) together imply: $d\sigma_{1,2}=i\Omega_{11,22}$. Using (\ref{oneforms}), the following basis of imaginary self-dual (2,1) forms were constructed for the conifold:
\begin{eqnarray}
\label{2-1forms}
\eta_1 &=& \lambda \wedge \omega_2 \\
\eta_2 &=& {\frac12} \lambda \wedge (\sigma_1\wedge{\bar{\sigma}_{\bar{2}}} -\sigma_2 \wedge {\bar{\sigma}_{\bar{1}}}) \nonumber \\
&=& \cot(\theta_1/2)\cot(\theta_2/2) \lambda \wedge (d\theta_1 \wedge d\theta_2 +\sin(\theta_1) d\phi_1 \wedge \sin(\theta_2) d\phi_2 )\\
\eta_3 &=&\left(\frac{dr}{r} \wedge e_\psi + {\frac12} \Omega_{22} \right) \wedge \sigma_1 = \left(\frac{i}{6}\lambda\wedge{\bar\lambda} -  \frac{i}{2}d\sigma_2\right)\wedge \sigma_1, \\
\eta_4 &=&\left(\frac{i}{6}\lambda\wedge{\bar\lambda} -  \frac{i}{2}d\sigma_1\right)\wedge \sigma_2, \\
\eta_5 &=& \bar{\lambda} \wedge \sigma_1 \wedge \sigma_2 \nonumber \\
&=& \bar{\lambda} \wedge (d\theta_1 \wedge d\theta_2 -\sin(\theta_1) d\phi_1 \wedge \sin(\theta_2) d\phi_2 - i (\Omega_{12}-\Omega_{21})).
\end{eqnarray}
In the $r\gg a, ({\rm deformation\ parameter})^{\frac{2}{3}}$-limit of the asymmetry factors in (\ref{three-form-fluxes}) - justified by working in the UV - using the results of  \cite{Ouyang}, one obtains:
\begin{eqnarray}
\label{G_3}
& & G_3= \frac{2M}{i} \left[\left(1+\frac{3g_s N_f}{2\pi}\log r\right)\eta_1 +
\frac{3g_s N_f}{8\pi}\left(\eta_4-\eta_3\right) \right]\nonumber\\
 & & + {\cal O}\left(\left(g_sN_f\right)^2; \left(\frac{a^2}{r^2}, \frac{a^2\log r}{r}, \frac{a^2\log r}{r^2}, \frac{a^2\log r}{r^3}\right); \left(\frac{{\rm deformation\ parameter }^2}{r^3} \right)\right),
\end{eqnarray}
where the ${\cal O}\left(\left(g_sN_f\right)^2\right)$ terms are:
\begin{eqnarray}
\label{gsNfsquared G_3}
& & \hskip -0.3in\frac{3M\left(g_sN_f\right)^2}{4r}\left(-\frac{3}{4\pi}\log r - \frac{1}{2\pi}\log\left(\sin\frac{\theta_1}{2}\sin\frac{\theta_2}{2}\right)\right)\left(1 + \frac{9}{4\pi}\log r + \frac{1}{2\pi}\log\left(\sin\frac{\theta_1}{2}\sin\frac{\theta_2}{2}\right)\right) dr\wedge\left(\Omega_{11} - \Omega_{22}\right).\nonumber\\
& &
\end{eqnarray}
In other words, $G_3$ is of the (2,1) type in the UV and near $\theta_1\sim\frac{1}{N^{\frac{1}{5}}}, \theta_2\sim\frac{1}{N^{\frac{3}{10}}}$. This makes the discussion below equation (\ref{T}), more concrete. This, interestingly is related to our result of (\ref{W12345}) wherein
it was shown that the type IIB $SU(3)$-structure torsion classes are given by $T\in W_4\oplus W_5: \frac{2}{3}\Re e W^{\bar{3}}_5 = W^{\bar{3}}_4$ (column ``(B)", Table 2 of \cite{Butti et al [2004]}).

Let us now calculate the type IIA $SU(3)$ structure torsion classes and see if, locally, supersymmetry continues to be preserved. Using (\ref{dEs}) one obtains:
\begin{eqnarray}
\label{dJ}
& & dJ \sim g_s^{-\frac{1}{4}}N^{\frac{1}{10}}\frac{\left(E^2 + \bar{E}^2\right)}{\theta_1}\wedge\ E^3\wedge\bar{E}^3
\left(\frac{\lambda_4}{\Lambda_4^2} + \frac{\lambda_5}{\Lambda_5^2}\right) \nonumber\\
& & + \frac{g_s^{\frac{1}{4}}N^{-\frac{1}{4}}}{\theta_1}\left(E^2 + \bar{E}^2\right)\wedge\left[2\chi_1 E^1\wedge\bar{E}^1 + \left(\chi_2E^3\wedge\bar{E}^1 + c.c.\right) + \left(\chi_3\bar{E}^3\wedge E^1 + c.c.\right)\right].
\end{eqnarray}
From (\ref{dJ}), we see:
\begin{equation}
\label{W3}
W_3\leftrightarrow \left[dJ\right]^{(2,1)}_0\sim \frac{g_s^{\frac{1}{4}}N^{-\frac{1}{4}}}{\theta_1}E^2\wedge\left[\left(\chi_2E^3\wedge\bar{E}^1 + c.c.\right) + \left(\chi_3\bar{E}^3\wedge E^1 + c.c.\right)\right],
\end{equation}
(where the subscript 0 implies picking out the primitive component or in other words $J\wedge \left[dJ\right]^{(2,1)}_0=0$), i.e. $W_3$ is suppressed in the large-N MQGP limit. Similarly,
\begin{equation}
\label{W4}
W_4 = \frac{1}{2}J \lrcorner dJ=\alpha_{W_4}\frac{N^{\frac{1}{10}}}{g_s^{\frac{1}{4}}\theta_1}\left(E^2 + \bar{E}^2\right),
\end{equation}
$\alpha_{W_4}$ being a constant. Also,
\begin{equation}
\label{W1}
W_1\leftrightarrow \left[dJ\right]^{(3,0)}=0.
\end{equation}

In the large-$N$ MQGP limit,
\begin{equation}
\label{dOmega}
d\Omega\sim \frac{N^{\frac{1}{10}}}{g_s^{\frac{1}{4}}\theta_1}\bar{E}^2\wedge E^1\wedge E^2\left[\left(\frac{\lambda_4}{2\Lambda_4^2} + \frac{\lambda_5}{2\Lambda_5^2}\right)E^3 +
\left(\frac{\lambda_4}{2\Lambda_4^2} - \frac{\lambda_5}{2\Lambda_5^2}\right)\bar{E}^3\right],
\end{equation}
implying
\begin{equation}
\label{W2}
W_2\leftrightarrow \left[d\Omega\right]^{(2,2)}_0\sim \frac{N^{\frac{1}{10}}}{g_s^{\frac{1}{4}}\theta_1}\bar{E}^2\wedge E^1\wedge E^2\wedge\bar{E}^3\left(\frac{\lambda_4}{2\Lambda_4^2} - \frac{\lambda_5}{2\Lambda_5^2}\right).
\end{equation}
We see that this deviation  from the local type IIA mirror being a complex manifold can be fine tuned away if, e.g., we
consider that in the $\theta_1\rightarrow0$-limit, instead of just (\ref{gij_constr}), one has:
\begin{equation}
\label{fine-tuning-2}
g_{33}\sim\frac{1}{10}\alpha_\theta^2,\ {\rm implying}\ \lambda_4\approx\lambda_5;\ \Lambda_4\approx\Lambda_5.
\end{equation}
 Also, writing $\Omega = \Omega_+ + i \Omega_-$:
 \begin{eqnarray}
 \label{dOmega+}
 & & d\Omega_+\sim\frac{N^{\frac{1}{10}}}{g_s^{\frac{1}{4}}\theta_1}E^2\wedge\bar{E}^2\Biggl[\left(\frac{\lambda_4}{2\Lambda_4^2} + \frac{\lambda_5}{2\Lambda_5^2}\right)E^1\wedge E^3 - c.c.\nonumber\\
 & & + \left(\frac{\lambda_4}{2\Lambda_4^2} - \frac{\lambda_5}{2\Lambda_5^2}\right)E^1\wedge\bar{E}^2 -c.c.\Biggr],
 \end{eqnarray}
 implying
\begin{equation}
\label{W5}
W_5=\frac{1}{2}\Omega_+\lrcorner d\Omega_+ = \alpha_{W_5}\frac{N^{\frac{1}{10}}}{g_s^{\frac{1}{4}}\theta_1}\left(E^2\left[\frac{\lambda_4}{2\Lambda_4^2} + \frac{\lambda_5}{2\Lambda_5^2}\right] + c.c.\right),
\end{equation}
where $\alpha_{W_5}$ is a constant.

 Using (\ref{fine-tuning-2}),(\ref{W4}) and (\ref{W5}), for example by demanding: $\alpha_{W_5}\frac{\lambda_4}{\Lambda_4^2}=\left(2\ {\rm or}\ \frac{2}{3}\right)\alpha_{W_4}$, analogous respectively to \cite{Dasgupta_et_al_Heterotic_small_instantons} or the Klebanov-Strassler-like background \cite{Butti et al [2004]}, locally, one obtains supersymmetry after the delocalized SYZ mirror symmetry.

So, in the MQGP limit, locally in the UV:
$T_{SU(3)}^{\rm IIB}\in W_4\oplus W_5: \frac{2}{3}W^{\bar{3}}_5=W^{\bar{3}}_4$ ({\bf implying supersymmetry} \cite{Butti et al [2004]})$\stackrel{\rm delocalized\ SYZ\ mirror}{\longrightarrow}T_{SU(3)}^{\rm IIA}\in W_2\oplus W_4\oplus W_5$, i.e., the large-$N$-suppression of the resolved warped deformed conifold(appearing in \cite{metrics})'s deviation from being complex, is lost in taking the mirror. After a fine tuning (\ref{fine-tuning-2}),
\begin{equation}
T_{SU(3)}\in W_4\oplus W_5: W_4\sim\Re e W_5,
\end{equation}
{\bf implying supersymmetry} of the delocalized SYZ mirror \cite{Butti et al [2004]}. Hence, the fine tuned type IIA mirror, locally, is approximately complex and supersymmetric in the MQGP limit.

The explicit torsion class chasing under SYZ mirror construction of the type IIB holographic dual of \cite{metrics} and demonstration of approximate supersymmetry (in the MQGP limit) {\bf provides one of the few examples in the literature} of:
\begin{itemize}
 \item
seeing explicitly what happens to supersymmetry under application of SYZ mirror symmetry to holographic string duals consisting of non-extremal resolved warped deformed conifold backgrounds in the language of $SU(3)$ structure torsion classes

\item
explicit construction of the $SU(3)$ structure of type IIA SYZ mirror of the type IIB fluxed non-extremal resolved warped deformed conifolds, and seeing that despite non-K\"{a}hlerity, there is still approximate supersymmetry in the MQGP limit.

\end{itemize}

\subsection{$G_2$-Structure Torsion Classes of M-Theory Uplift}

We now evaluate the $G_2$-structure torsion classes specially to see the possibility of generating, locally, seven-folds of $G_2$-holonomy despite having four-form fluxes $G_4$.  As discussed in appendix D,
the tensor $T_A^{\ M}$, like the space W, possesses 49 components and hence fully defines $\nabla \varphi $. In general $T_{AB}$ cab be split into torsion components
as
\begin{equation}
T=T _{1}g+T _{7}\lrcorner \varphi +T _{14}+T _{27}
\label{torsioncomps}
\end{equation}%
where $T _{1}$ is a function and gives the $\mathbf{1}$ component of $T$
. We also have $T _{7}$, which is a $1$-form and hence gives the $\mathbf{
7}$ component, and, $T _{14}\in \Lambda _{14}^{2}$ gives the $\mathbf{14}$
component. Further, $T _{27}$ is traceless symmetric, and gives the $\mathbf{27}$
component. Writing $T_i$ as $W_i$, we can split $W$ as
\begin{equation}
W=W_{1}\oplus W_{7}\oplus W_{14}\oplus W_{27}.  \label{Wsplit}
\end{equation}

From \cite{G2-Structure}, we see that a $G_2$ structure can be defined as:
\begin{equation}
\label{G_2_i}
\varphi_0 = \frac{1}{3!}{f}_{ABC}e^{ABC} = e^{-\phi^{IIA}}{f}_{abc}e^{abc} + e^{-\frac{2\phi^{IIA}}{3}}J\wedge e^{x_{10}},
\end{equation}
where $A,B,C=1,...,6,10; a,b,c,=1,...,6$ and ${f}_{ABC}$ are the structure constants of the imaginary octonions.
Now, substituting the non-zero $f_{abc}$ \cite{Chiossi+Salamon}, one obtains:
\begin{equation}
\varphi_0 = \frac{1}{g_s}\left(e^{135} - e^{146} - e^{236} - e^{245}\right) + \frac{1}{g_s^{\frac{2}{3}}}\left(e^{127} + e^{347} +
e^{567}\right),
\end{equation}
implying:
\begin{equation}
*_7\varphi_0=\frac{1}{g_s}\left(e^{1367} + e^{1457} + e^{2357} + e^{2467}\right) + \frac{1}{g_s^{\frac{2}{3}}}\left(e^{3456} + e^{1256} + e^{1234}\right).
\end{equation}
Hence, using (\ref{des}), one obtains:
\begin{eqnarray}
\label{dPsi}
& & d\varphi_0 \sim\frac{g_s^{-\frac{1}{4}}N^{-\frac{1}{4}}}{\theta_1}\Biggl[\frac{1}{g_s}\left(\frac{\lambda_5}{\Lambda_5}e^{1325} +\frac{\lambda_4}{\Lambda_4}e^{1246} - \left[\frac{\gamma_{62}}{\Lambda_5}e^{1425} + \gamma_{63}e^{1426}\right]
\right)\nonumber\\
& & + \frac{1}{g_s^{\frac{2}{3}}}\left(\frac{\lambda_5}{\Lambda_5}e^{2567} -\frac{\lambda_4}{\Lambda_4}e^{3247} - \left[\frac{\gamma_{61}}{\Lambda_4}e^{5247} + \gamma_{63}e^{5267}\right]
\right)\Biggr],
\end{eqnarray}
and
\begin{eqnarray}
\label{d*Psi}
& & d*_7\varphi_0 \sim\frac{1}{\left(g_sN\right)^{\frac{1}{4}}\theta_1}\Biggl( \frac{1}{g_s}\Biggl[\frac{\gamma_{61}}{\Lambda_4} e^{13247} + \frac{\gamma_{62}}{\Lambda_5}e^{13257} + \frac{\lambda_4}{\Lambda_4}e^{12457} + \frac{\lambda_5}{\Lambda_5}e^{14257}\Biggr]\nonumber\\
& & + \frac{1}{g_s^{\frac{2}{3}}}\Biggl[\frac{\lambda_4}{\Lambda_4} + \frac{\lambda_5}{\Lambda_5} + \gamma_{63}\Biggr]e^{32456}\Biggr).
\end{eqnarray}
One can show \cite{Chiossi+Salamon}:
\begin{eqnarray}
& & d\varphi_0 = 4 W_1 *_7\varphi_0 - 3 W_7\wedge\varphi_0 - *_7 W_{27}\nonumber\\
& & d*_7\varphi_0 = - 4 W_7\wedge *_7\varphi_0 - 2 *_7W_{14},
\end{eqnarray}
where $W_{27}$ corresponds to the symmetric traceless rank-two tensor $h_{AB}$ contracted with the ${\varphi_0}_{ABC}$ of (\ref{G_2_i}) to give a rank-three $\chi_{ABC}$ valued in $W_{27}$ via $\chi_{ABC}=h_{[A}^{d}\varphi _{BC]D}$, and $W_{14}$ corresponds to the anti-symmetric rank $\omega_{AB}$ satisying $\omega\lrcorner\varphi_0=0$.
One therefore sees that the non-zero $G_2$ structure torsion classes are given by:
\begin{eqnarray}
\label{nonzeroG2structure}
& & W_{27} = - *_7d\varphi_0 \sim - \frac{1}{\left(g_sN\right)^{\frac{1}{4}}\theta_1}\left(\frac{1}{g_s}\left[\frac{\lambda_5}{\Lambda_5}e^{467} - \frac{\lambda_4}{\Lambda_4}e^{357} - \left\{-\frac{\gamma_{62}}{\Lambda_5}e^{367} + \gamma_{63}e^{357}\right\}\right]\right.\nonumber\\
& & \left. + \frac{1}{g_s^{\frac{2}{3}}}\left[\frac{\lambda_5}{\Lambda_5}e^{134} + \frac{\lambda_4}{\Lambda_4}e^{156} - \left\{\frac{\gamma_{61}}{\Lambda_4}e^{136} - \gamma_{63}e^{134}\right\}\right]\right),\nonumber\\
& & W_{14} = -\frac{1}{2}*_7d*_7\varphi_0 \sim \frac{1}{\left(g_sN\right)^{\frac{1}{4}}\theta_1}\Biggl[\frac{1}{g_s}\left(-\frac{\gamma_{61}}{\Lambda_4}e^{56} + \frac{\gamma_{62}}{\Lambda_5}e^{46} + \left\{\frac{\lambda_4}{\Lambda_4}-\frac{\lambda_5}{\Lambda_5}\right\}e^{36}\right)\nonumber\\
& & + \frac{e^{17}}{g_s^{\frac{2}{3}}}\left(-\frac{\lambda_4}{\Lambda_4} + \frac{\lambda_5}{\Lambda_5} - \gamma_{63}\right)\Biggr].
\end{eqnarray}
Hence, $T_{G_2}\in W_{14}\oplus W_{27}$.
 However, in the $\theta_1\rightarrow0$-limit in which the above expressions have been worked out, assuming as we have that $\theta_1\rightarrow\frac{1}{N^{\frac{1}{5}}}$, the non-zero $G_2$ torsion classes are large-$N$ suppressed. If all torsion classes of a $G$ structure become trivial the manifold is supposed to possess a holonomy given by $G$.
  So, the MQGP limit accelerates the approach of the seven-fold relevant to the eleven-dimensional uplift, locally, to being a $G_2$-holonomy manifold.

To our knowledge, {\bf for the first time, an explicit $G_2$ structure of the local M-theory uplift of string theory duals of thermal QCD at finite gauge coupling, has been constructed} in this subsection and comprises the last significant and new result of this paper.

\section{Summary and Significance of Results Obtained}

Systems like QGP are expected to be strongly coupled. In fact, as mentioned in the Introduction, apart from having a large t'Hooft coupling, it is believed that {\it the same must also be characterized by finite gauge coupling \cite{Natsuume-sQGP}}. It is hence important to have a framework in the spirit of gauge-gravity duality, to be able to address this regime in string theory. {\it Finite gauge coupling would under this duality translate to finite string coupling hence necessitating addressing the same from $M$ theory perspective.} As explained in detail in {\bf 2.2}, the MQGP limit is particularly suited for holographic studies of strongly coupled large-$N$ thermal QCD due to the calculational simplifications effected by the same, e.g., as regard the effective number of $D3/D5/D7$ branes, ten-dimensional warp factors, fluxes, etc. thereby simplifying construction of the type IIA mirror via the Strominger-Yau-Zaslow construction from a triple T-dual, and the eventual uplift of this type IIA mirror to M-theory.

Continuing the line of reasoning of our previous efforts \cite{MQGP,transport-coefficients}, there are two sets of issues discussed in this paper.  We broadly classify them as Physics and Math  issues. The former arise as a consequence of the latter in the sense that the inherent non-K\"{a}hlerity of the parent type IIB background of \cite{metrics} and its delocalized SYZ type IIA mirror is responsible for the Physics and it is hence natural to quantify and classify these mathematical characteristics.

\begin{enumerate}
\item{\bf Physics Issues and Significance of Results Obtained}

\noindent There are two Physics-related issues discussed in this paper.
\begin{itemize}
\item
{\bf New results}: First, we evaluated the DBI action of the $N_f$ flavor $D7$ branes in the presence of a $U(1)$ gauge field (assuming it to have only a non-zero temporal component with only a radial dependence, corresponding to a baryon chemical potential) by {\it first} evaluating in the MQGP limit, the angular integrals exactly and then taking the UV limit of the (incomplete) elliptic integrals so obtained. Demanding square integrability of the aforementioned $U(1)$ gauge field and using the Dirichlet boundary condition at an IR cut-off and demanding a mass parameter appearing in the solution to be related to the mass of the lightest known vector meson mass, we related the mass of the lightest vector meson to the IR cut-off. The computation of the QCD deconfinement transition temperature or equivalently the critical temperature $T_c$ corresponding to the first order Hawking-Page phase transition between a thermal AdS and an black AdS backgrounds, is then carried out from five-dimensional Einstein-Hilbert and Gibbons-Hawking-York actions (having integrated out the five-dimensional compact directions of the type IIB background of \cite{metrics} in the MQGP limit).  Hence, from a top-down approach using the type IIB holographic dual of \cite{metrics} in the presence of a finite chemical potential and the MQGP limit, to the best of  our knowledge {\it it has been shown for the first time that}:
\begin{itemize}
\item
 it is possible to obtain the QCD deconfinement temperature  consistent with lattice results for $N_f$ equal to three, ensuring at the same time the thermodynamical stability of the type $IIB$ background;
 \item
  the Ouyang embedding parameter required to be dialed in to reproduce $T_c$ is happily exactly what also reproduces the mass scale of the first generation (light) quarks;
 \item
 $T_c$ decreases with increase in $N_f$ in accordance with lattice computations.
\end{itemize}
{\bf Significance}: Being able to reproduce the confinement-deconfinement temperature compatible with lattice results, serves as a non-trivial check for a proposed holographic dual of large-$N$ thermal QCD. In this respect, the result of section {\bf 3} is very significant as it is able to successfully incorporate in a self-consistent way, a lattice-compatible $T_c$ for the right number of light quark flavors and light quark masses, thermodynamical stability, the right lightest vector mass for the number of quark colors $N_c$ given in in the IR (relevant to a low value of $T_c$) by $M$ which can be tuned to equal 3 (as one ends up with an $SUN(M)$ gauge theory at finite temperature in the IR at the end of the Seiberg duality cascade).

\item
{\bf New Results}: Using the aforementioned $U(1)$ background, we then looked at both $U(1)$ and $SU(2)$ (for $N_f=2$) gauge fluctuations. By looking at two-point correlation functions of either the former or the diagonal sector of the latter, we calculated the DC electrical conductivity and the temperature dependence of the same (well above $T_c$), and found:
\begin{itemize}
\item
 demanding the Einstein relation (ratio of electrical conductivity and charge susceptibility to equal the diffusion constant) to be satisfied within linear perturbation theory, requires a non-trival dependence of the Ouyang embedding paramter on the horizon radius;

 \item
 a prediction that the temperature dependence of the DC electrical conductivity above $T_c$, curiously mimics a one-dimensional Luttinger liquid with an appropriately tuned interaction parameter.

\end{itemize}

{\bf New insight}: The following two fold significance of this set of results provides new insights into the Physics of large-$N$ thermal QCD at finite gauge coupling.
 \begin{itemize}

  \item
  Given that one is working within linear perturbation/response theory, one expects the Einstein relation relating the ratio of the DC electrical conductivity and charge susceptibility to the diffusion constant, to hold. This necessitates taking the Ouyang embedding parameter, analogous to the resolution parameter \cite{K. Dasgupta  et al [2012]}, to be dependent on the horizon radius with a specific form of dependence. Thus far, this realization was missing in the literature.

 \item
  The temperature dependence at temperatures above $T_c$, i.e., the deconfined phase curiously  mimics a one-dimensional Luttinger liquid for a specific choice of the Luttinger parameter. The one-dimensional identification could be due to the $(t,x)$ singled out in the plane-wave basis of the Fourier modes of the gauge field fluctuations upon the choice of the dual $q^\mu=(w,q,0,0)$.

\end{itemize}

\end{itemize}

\item{\bf Math Issues and Significance of Results Obtained}
\begin{itemize}
\item
{\bf New results}: In \cite{transport-coefficients}, we saw that the five $SU(3)$ structure torsion classes, in the MQGP limit, satisfied (schematically):
$T_{SU(3)}^{\rm IIB}\in W_1 \oplus W_2 \oplus W_3 \oplus W_4 \oplus W_5 \sim \frac{e^{-3\tau}}{\sqrt{g_s N}} \oplus \left(g_s N\right)^{\frac{1}{4}} e^{-3\tau}\oplus \sqrt{g_s N}e^{-3\tau}\oplus -\frac{2}{3} \oplus -\frac{1}{2}$ such that $\frac{2}{3}W^{\bar{3}}_5=W^{\bar{3}}_4$ in the UV, implying a Klebanov-Strassler-like supersymmetry \cite{Butti et al [2004]}. Locally, the type IIA torsion classes after a fine tuning of the delocalized SYZ type IIA mirror metric, are: $T_{SU(3)}^{\rm IIA}\in W_2 \oplus W_3 \oplus W_4 \oplus W_5 \sim \gamma_2g_s^{-\frac{1}{4}} N^{\frac{3}{10}} \oplus g_s^{-\frac{1}{4}}N^{-\frac{1}{20}} \oplus g_s^{-\frac{1}{4}} N^{\frac{3}{10}} \oplus g_s^{-\frac{1}{4}} N^{\frac{3}{10}}\approx W_2\oplus W_4\oplus W_5$.  Further, $W_4\sim \Re e W_5$ indicative of supersymmetry after constructing the delocalized SYZ mirror.

{\bf Significance}: Apart from quantifying the departure from $SU(3)$ holonomy due to intrinsic contorsion supplied by the NS-NS three-form $H$, via the evaluation of the $SU(3)$ structure torsion classes, to our knowledge for the first time in the context of holographic thermal QCD {\bf at finite gauge coupling}: \begin{itemize}
 \item
 the existence of approximate supersymmetry of the type IIB holographic dual of \cite{metrics} in the MQGP limit near the coordinate branch $\theta_1=\theta_2=0$ is demonstrated, which apart from the existence of a special Lagrangian three-cycle (as shown in \cite{transport-coefficients} and sub-section {\bf 2.3}) is essential for construction of the local SYZ type IIA mirror;

\item
    it is demonstrated that the large-$N$ suppression of the deviation of the type IIB resolved warped deformed conifold from being a complex manifold, is lost on being duality-chased to type IIA - it is also shown that one further fine tuning  $\gamma_2=0$ in $W_2^{\rm IIA}$ can ensure that the local type IIA mirror is complex;

 \item
  for the local type IIA $SU(3)$ mirror,  the possibility of surviving approximate supersymmetry is demonstrated which is essential from the point of view of the end result of application of the SYZ mirror prescription.

    \end{itemize}

\item
{\bf New result}: We work out a local $G_2$ structure wherein the torsion classes are: $T_{G_2}\in W_2^{14} \oplus W_3^{27} \sim \frac{1}{\left(g_sN\right)^{\frac{1}{4}} }\oplus \frac{1}{\left(g_sN\right)^{\frac{1}{4}}}$. Hence, the approach of the seven-fold, locally, to having a $G_2$ holonomy ($W_1^{G_2}=W_2^{G_2}=W_3^{G_2}=W_4^{G_2}=0$)  is accelerated in the MQGP limit.

{\bf Significance}: As stated in subsection {\bf 2.3}, the global uplift to M-theory of the type IIB background of \cite{metrics} is expected to involve a seven-fold of $G_2$ structure (not $G_2$-holonomy due to non-zero $G_4$). It is hence extremely important to be able to see this, at least locally. It is in this sense that the results of {\bf 5.2} are of great significance as one explicitly sees, for the first time, in the context of  holographic thermal QCD {\bf at finite gauge coupling}, though locally, the aforementioned $G_2$ structure having worked out the non-trivial $G_2$-structure torsion classes.

\end{itemize}

\end{enumerate}
\section*{Acknowledgements}

We would like to thank M. Dhuria for participation in the earlier stages of the project. One of us (KS) is supported by a senior research fellowship (SRF) from the Ministry of Human Resource and Development (MHRD). One of us (AM) would like to thank
King's College (specially G. Papadopoulos), Chalmers University of Technology (specially G. Ferretti), Purdue University (specially M. Kruczenski) and Brown University for hospitality where part of this work was completed, as well as K. Dasgupta for several useful clarifications. He would also like to thank T. Das, P. Majumdar and specially S. Mukerjee, V. Shenoy for very useful condensed matter Physics related clarifications and references. AM was partly supported by the CSIR, Govt. of India, project number CSR-656-PHY. His work is also partly funded under the Associates Program of the Abdus Salam ICTP and he would also like to thank the Instituto de Fisica Teorica (IFT UAM-CSIC) in Madrid for its support via the Centro de Excelencia Severo Ochoa Program under Grant SEV-2012-0249, during the workshop `eNLarge Horizons' where part of the work was completed. Some results of this paper were presented by AM at seminars at King's College, Purdue and Brown universities and at conferences: eNLarge Horizons (IFT UAM-CSIC), PASCOS 2015(ICTP) and MG14 (University of Rome, La Sapienza), and the workshop ``Applications of AdS/CFT to QCD
and condensed matter physics" (as part of the thematic semester on ``AdS/CFT, Holography, Integrability") at Centre de Recherches Mathematiques (CRM) of Montreal.

\appendix

\section{Ouyang's holomorphic embedding of $N_f$ $D7$-branes in a (predominantly) resolved conifold}
\setcounter{equation}{0} \seceqaa

Let us now discuss Ouyang's holomorphic embedding of $N_f$ $D7$-branes in a (predominantly) resolved conifold (based on \cite{Ouyang}). The conifold, expressed as a quadric in $\mathbb{CP}^3[2]$: $z_1 z_2 = z_3 z_4$ can be mapped to $\mathbb{CP}^1\times\mathbb{CP}^1$ via Segr\'{e}-embedding: $\mathbb{CP}^1(A_1,A_2)\times\mathbb{CP}^1(B_1,B_2)
\hookrightarrow\mathbb{CP}^3(z_1=A_1B_1,z_2=A_2B_2,z_3=A_1B_2,z_4=A_2B_1)$. Hence, the holomorphic embedding of $D7$-branes $z_1=0$ would correspond to two branches $A_1=0$ and $B_1=0$. Given that there is an $SU(N_f)$ flavor symmetry with each of the two branches, one generates an $SU(N_f)\times SU(N_f)$ symmetry. Cancelation of gauge anomalies requires addition of two flavors of opposite chirality with each of the two branches - following the notation of \cite{Ouyang}, let us denote the same by: $q/\tilde{q}$ transforming as $(N_f,1)/(1,N_f)$ under $SU(N_f)\times SU(N_f)$ and $(N+M,1)/(\overline{N+M},1)$ under $SU(M+N)\times SU(N)$, and $Q/\tilde{Q}$ transforming as $(\bar{N}_f,1)/(1,\bar{N}_f)$, under $SU(N_f)\times SU(N_f)$ and transforming as $(1,N)/(1,\overline{N})$  under $SU(M+N)\times SU(N)$.
With $A_i, B_j$ transforming respectively as $(N+M,\overline{N})$ and $(\overline{N+M},N)$ the color-invariant and flavor-invariant superpotential will be
\begin{equation}
\label{W flavors}
W_{\rm flavors} = \lambda_{\tilde{q}A_1Q}\tilde{q} A_1 Q + \lambda_{\tilde{Q}B_1q}\tilde{Q}B_1q.
\end{equation}
 As $A_i, B_i$ are dimension-$\frac{3}{4}$ fields \cite{kw}, \cite{ks}, \cite{Ouyang}, taking $q_i,\tilde{q}_j,Q_k,\tilde{Q}_k$ to be having the same dimension, they will hence be dimension-$\frac{9}{8}$. The mass terms in the superpotential breaking the $SU(N_f)\times SU(N_f)$ symmetry to the diagonal $SU(N_f)$ in \cite{Ouyang} are:
 \begin{equation}
 \label{W masses}
 W_{\rm masses} = \sqrt{\mu}q\tilde{q} + \sqrt{\mu}Q\tilde{Q}.
 \end{equation}
  Then following \cite{Ouyang}, rewrite the total superpotential as:
\begin{eqnarray}
\label{Wtotal}
& & W_{\rm flavors} + W_{\rm masses} = \lambda_{\tilde{q}A_1Q}\tilde{q} A_1 Q + \lambda_{\tilde{Q}B_1q}\tilde{Q}B_1q + \lambda_{\tilde{q}A_1Q}\sqrt{\mu}q\tilde{q} + \lambda_{\tilde{Q}B_1q}\sqrt{\mu}\tilde{Q}Q \nonumber\\
& &  = \left(\begin{array}{cc}
\tilde{q} & \tilde{Q}
\end{array} \right)\left(\begin{array}{cc}
\lambda_{\tilde{q}A_1Q}\sqrt{\mu} & \lambda_{\tilde{q}A_1Q}A_1 \\
\lambda_{\tilde{Q}B_1q}B_1 & \lambda_{\tilde{Q}B_1q}\sqrt{\mu}
\end{array}
\right)\left(\begin{array}{c}
q\\
Q
\end{array}\right).
\end{eqnarray}
Hence, the 3-7 strings become massless when the $D3$-branes and $D7$-branes intersect. This corresponds to null eigenvalues of the mass matrix $\left(\begin{array}{cc}
\lambda_{hA_1qQ}\sqrt{\mu} & \lambda_{hA_1qQ}A_1 \\
\lambda_{\tilde{q}_1B_1\tilde{Q}}B_1 & \lambda_{\tilde{q}_1B_1\tilde{Q}}\sqrt{\mu}
\end{array}
\right)$, i.e., the Ouyang embedding equation $z_1=\mu$ or:
\begin{equation}
\label{Ouyang-rc}
z_1=\left ( 9 a^2 r^4 + r^6 \right ) ^{1/4} e^{\imath/2(\psi-\phi_1-\phi_2)}\,\sin\frac{\theta_1}{2}\,\sin\frac{\theta_2}{2}=\mu.
\end{equation}
Using $D7$-branes monodromy arguments for Ouyang embedding, assuming a very small $|\mu|$ - as will turn out to be the case in Sec. {\bf 3} -  $\tau\sim\frac{N_f}{2\pi\imath} \log z$, close to $D7$-branes, implies in the IR:
\begin{eqnarray}
\label{dilaton-IR}
& & e^{-\Phi} = \frac{1}{g_s} -\frac{N_f}{8\pi} ~{\rm log} \left(r^6 + 9a^2 r^4\right) -
\frac{N_f}{2\pi} {\rm log} \left({\rm sin}~\frac{\theta_1}{2} ~ {\rm sin}~\frac{\theta_2}{2}\right);
\end{eqnarray}
the first term on the right hand side of (\ref{dilaton-IR}) has been included to yield the correct value of the dilaton for $N_f=0$ \cite{Ouyang}.

The values for the axion $C_0$ and the five form $F_5$ are given by \cite{Ouyang}:
\begin{eqnarray*}
\label{axfive}
&&C_0 ~ = ~ \frac{N_f}{4\pi} (\psi - \phi_1 - \phi_2) [{\rm since}\ \int_{S^1}dC_0=N_f],\nonumber\\
&& F_5 ~ = ~ \frac{1}{g_s} \left[ d^4 x \wedge d h^{-1} + \ast(d^4 x \wedge dh^{-1})\right].
\end{eqnarray*}

\section{Constancy of the axion-dilaton modulus in the UV}
\setcounter{equation}{0} \seceqbb

In this appendix, we discuss the UV constancy of the axion-dilaton modulus from two perspectives: (a) from F-theory and (b) locally using the background of \cite{metrics}.

\noindent $\bullet$ {\it From F-theory}:

Let us make some remarks about the axion-dilaton modulus in the UV away from the $N_f$ $D7$-branes by looking at the modular $j$ function for finite $g_s$ from a Weierstrass variety worth of a generic F-theory uplift (though such a global uplift of the type IIB background of \cite{metrics} is not explicitly known).  Let us assume that  in $\left\{z_{i\neq1}=1\right\}$-patch, the F-theory Weierstrass variety:
\begin{equation}
\label{Weierstrass}
y^2 = x^3 + f(\left\{z_{i\neq1}=1,z_1\right\})x + g(\left\{z_{i\neq1}=1,z_1\right\})
\end{equation}
is written in the abovementioned coordinate patch as:
\begin{equation}
\label{Weierstrass-1}
y^2 = x^3 + F(z_1) x + G(z_1)
\end{equation}
where
\begin{equation}
\label{Weierstrass-2}
F(z_1)= f_0\prod_{i=1}^8(z_1 - {\cal Z}_i),\
\Delta(z)=\Delta_0\prod_{j=1}^{24}(z_1 - Z_j), \left\{{\cal Z}_i\right\}\neq \left\{Z_i\right\}.
\end{equation}
Of course this does not imply that the global F-theory uplift involves an elliptically fibered K3. {\it This discussion, in the same spirit as a similar discussion in \cite{IR-UV-desc_Dasgupta_etal} (but for small string coupling), is to give a plausibility argument that if one works at finite string coupling, then the holomorphic Ouyang embedding would automatically guarantee a constant axion-dilaton modulus and hence conformality in the UV.}

For finite $g_s$, one should in principle consider the entire infinite series for the $j$-function:
 \begin{equation}
 \label{j-i}
 j(\tau) = \frac{1}{q} + 744 + 19,688 q + 21,493,760 q^2 + ..... = \frac{4\left(24 F(z_1)\right)^3}{27 G^2(z_1) + 4 F^3(z_1)}
 \end{equation}
 $(q\equiv e^{2i\pi\tau})$.  So, truncating (\ref{j-i}) at the first term yields  \footnote{There is a small typo in
 \cite{IR-UV-desc_Dasgupta_etal}} for large $z_1$:
 \begin{equation}
 \label{j-ii}
  \tau = \frac{i}{g_s} + \frac{i}{2\pi}\log\left(\frac{55,296 f_0^3}{\Delta_0}\right)
 + \sum_{n=1}^\infty\frac{1}{n z_1^n}\left(\sum_{i=1}^{24}Z_i^n - 3\sum_{i=1}^8{\cal Z}_i^n\right).
 \end{equation}
 Truncating the series (\ref{j-i}) at ${\cal O}(q^2)$, one obtains for large $z_1$:
 \begin{equation}
 \label{j-iii}
\hskip -0.3in \tau = \frac{i}{g_s} + -\frac{i \log \left(-\frac{\sqrt{-\sqrt{\frac{21233664 b^6}{a^2}-\frac{571392 b^3}{a}-1625}+\frac{4608 b^3}{a}-62}}{3\sqrt{3646}}\right)}{\pi }  + \frac{\frac{2,304 i f_0^3\left(\sum_{i=1}^{24}Z_i - 3\sum_{i=1}^8{\cal Z}_i\right)}{\pi  a  \sqrt{\frac{21233664 b^6}{a^2}-\frac{571392 b^3}{a}-1625}}}{z_1} + {\cal O}\left(\frac{1}{z_1^2}\right).
 \end{equation}
 So, in general one expects:
 \begin{equation}
 \label{j-iv}
\tau = \frac{i}{g_s} + \frac{i{\cal F}(f_0,\Delta_0)}{\pi} + \sum_{m=1}^\infty\frac{C_n(\theta_{1,2},\phi_{1,2},\psi;f_0,\Delta_0) + i D_n(\theta_{1,2},\phi_{1,2},\psi;f_0,\Delta_0)}{r^{\frac{3}{2}n}}
\end{equation}
 in the Ouyang embedding, implying $\beta\rightarrow0$ or conformality as $\Lambda$(energy scale)$\equiv r\rightarrow\infty$ (the UV).

\noindent $\bullet$ {\it Locally, from (\ref{dilaton-IR})}:

Near the $\theta_{1,2}=0$-branch in  the originally IR-valued $e^{-\Phi}$ written out in (\ref{dilaton-IR}), choosing $\gamma_\theta$ and $\gamma_r$ in such a way that in the UV: $\frac{3 N_f}{4\pi}\gamma_r = \frac{N_f\gamma_\theta}{\pi}$, then $e^{-\Phi}$ in the UV would approach a constant implying a vanishing $\beta$ or conformality in the UV. So, the $\theta_{1,2}=0$-branch mimics the required axion-dilaton behavior in the UV.

\section{Details of Exact Angular Integration in the DBI Action and Its UV Limit}
\setcounter{equation}{0} \seceqcc

The $\theta_2$ integral in the DBI action of (\ref{SDBI-arb-mu}), is expressed in terms of elliptic integral of the first kind $F(\phi;\mu)\equiv \int_0^{\phi}\frac{d\theta}{\sqrt{1 - \mu \sin^2\theta}}$ as well as incomplete integral of the first kind $\Pi(\nu;\phi|\mu)\equiv\int_0^\phi\frac{d\theta}{\left(1 - \nu\sin^2\theta\right)\sqrt{1 - \mu\sin^2\theta}}$:
\begin{eqnarray}
\label{elliptic-fns}
& & \hskip -0.2in F\Biggl(\sin ^{-1}\left.\left(\sqrt{\frac{\frac{|\mu| }{\sqrt{|\mu| ^2-r^3}}-\frac{-7 |\mu| ^2+\sqrt{25 |\mu| ^4-104 |\mu| ^2 r^3+16 r^6}+4 r^3}{2 \left(|\mu| ^2+2
   r^3\right)}}{\frac{|\mu| }{\sqrt{|\mu| ^2-r^3}}+\frac{-7 |\mu| ^2+\sqrt{25 |\mu| ^4-104 |\mu| ^2 r^3+16 r^6}+4 r^3}{2 \left(|\mu| ^2+2
   r^3\right)}}}\right)\right|\nonumber\\
   & & \frac{\left(\frac{|\mu| }{\sqrt{|\mu| ^2-r^3}}-\frac{-7 |\mu| ^2-\sqrt{25 |\mu| ^4-104 |\mu| ^2 r^3+16 r^6}+4 r^3}{2 \left(|\mu| ^2+2 r^3\right)}\right)
   \left(-\frac{|\mu| }{\sqrt{|\mu| ^2-r^3}}-\frac{-7 |\mu| ^2+\sqrt{25 |\mu| ^4-104 |\mu| ^2 r^3+16 r^6}+4 r^3}{2 \left(|\mu| ^2+2 r^3\right)}\right)}{\left(-\frac{|\mu|
   }{\sqrt{|\mu| ^2-r^3}}-\frac{-7 |\mu| ^2-\sqrt{25 |\mu| ^4-104 |\mu| ^2 r^3+16 r^6}+4 r^3}{2 \left(|\mu| ^2+2 r^3\right)}\right) \left(\frac{|\mu| }{\sqrt{|\mu|
   ^2-r^3}}-\frac{-7 |\mu| ^2+\sqrt{25 |\mu| ^4-104 |\mu| ^2 r^3+16 r^6}+4 r^3}{2 \left(|\mu| ^2+2 r^3\right)}\right)}\Biggr);\nonumber\\
& & \hskip -0.2in F\Biggl(\sin ^{-1}\left.\left(\sqrt{\frac{-2 |\mu| ^3-7 |\mu| ^2 \sqrt{|\mu| ^2-r^3}+4 r^3 \sqrt{|\mu| ^2-r^3}+\sqrt{|\mu| ^2-r^3} \sqrt{25 |\mu| ^4-104 |\mu| ^2 r^3+16 r^6}-4 |\mu|
   r^3}{-2 |\mu| ^3+7 |\mu| ^2 \sqrt{|\mu| ^2-r^3}-4 r^3 \sqrt{|\mu| ^2-r^3}-\sqrt{|\mu| ^2-r^3} \sqrt{25 |\mu| ^4-104 |\mu| ^2 r^3+16 r^6}-4 |\mu|  r^3}}\right)\right|\nonumber\\
   & & \frac{\left(-2 |\mu|
   ^3+7 |\mu| ^2 \sqrt{|\mu| ^2-r^3}-4 r^3 \sqrt{|\mu| ^2-r^3}-\sqrt{|\mu| ^2-r^3} \sqrt{25 |\mu| ^4-104 |\mu| ^2 r^3+16 r^6}-4 |\mu|  r^3\right)}{\left(2 |\mu| ^3+7 |\mu| ^2 \sqrt{|\mu| ^2-r^3}-4
   r^3 \sqrt{|\mu| ^2-r^3}-\sqrt{|\mu| ^2-r^3} \sqrt{25 |\mu| ^4-104 |\mu| ^2 r^3+16 r^6}+4 |\mu|  r^3\right) }\nonumber\\
   & & \times\frac{ \left(2 |\mu| ^3+7 |\mu| ^2
   \sqrt{|\mu| ^2-r^3}-4 r^3 \sqrt{|\mu| ^2-r^3}+\sqrt{|\mu| ^2-r^3} \sqrt{25 |\mu| ^4-104 |\mu| ^2 r^3+16 r^6}+4 |\mu|  r^3\right)}{\left(-2 |\mu| ^3+7 |\mu| ^2 \sqrt{|\mu| ^2-r^3}-4 r^3 \sqrt{|\mu|
   ^2-r^3}+\sqrt{|\mu| ^2-r^3} \sqrt{25 |\mu| ^4-104 |\mu| ^2 r^3+16 r^6}-4 |\mu|  r^3\right)}\Biggr);
   \nonumber\\
   & & \hskip -0.2in \Pi \Biggl(\frac{\left(\frac{|\mu| }{\sqrt{|\mu| ^2-r^3}}+1\right) \left(\frac{|\mu| }{\sqrt{|\mu| ^2-r^3}}+\frac{-7 |\mu| ^2+\sqrt{25 |\mu| ^4-104 |\mu| ^2 r^3+16 r^6}+4 r^3}{2
   \left(|\mu| ^2+2 r^3\right)}\right)}{\left(1-\frac{|\mu| }{\sqrt{|\mu| ^2-r^3}}\right) \left(\frac{-7 |\mu| ^2+\sqrt{25 |\mu| ^4-104 |\mu| ^2 r^3+16 r^6}+4 r^3}{2 \left(|\mu|
   ^2+2 r^3\right)}-\frac{|\mu| }{\sqrt{|\mu| ^2-r^3}}\right)};\nonumber\\
   & & \hskip -0.2in \sin ^{-1}\left.\left(\sqrt{\frac{-2 |\mu| ^3-7 |\mu| ^2 \sqrt{|\mu| ^2-r^3}+4 r^3 \sqrt{|\mu| ^2-r^3}+\sqrt{|\mu|
   ^2-r^3} \sqrt{25 |\mu| ^4-104 |\mu| ^2 r^3+16 r^6}-4 |\mu|  r^3}{-2 |\mu| ^3+7 |\mu| ^2 \sqrt{|\mu| ^2-r^3}-4 r^3 \sqrt{|\mu| ^2-r^3}-\sqrt{|\mu| ^2-r^3} \sqrt{25 |\mu| ^4-104 |\mu|
   ^2 r^3+16 r^6}-4 |\mu|  r^3}}\right)\right|\nonumber\\
   & & \hskip -0.2in \frac{\left(-2 |\mu|
   ^3+7 |\mu| ^2 \sqrt{|\mu| ^2-r^3}-4 r^3 \sqrt{|\mu| ^2-r^3}-\sqrt{|\mu| ^2-r^3} \sqrt{25 |\mu| ^4-104 |\mu| ^2 r^3+16 r^6}-4 |\mu|  r^3\right)}{\left(2 |\mu| ^3+7 |\mu| ^2 \sqrt{|\mu| ^2-r^3}-4
   r^3 \sqrt{|\mu| ^2-r^3}-\sqrt{|\mu| ^2-r^3} \sqrt{25 |\mu| ^4-104 |\mu| ^2 r^3+16 r^6}+4 |\mu|  r^3\right) }\nonumber\\
   & & \hskip -0.2in \times\frac{ \left(2 |\mu| ^3+7 |\mu| ^2
   \sqrt{|\mu| ^2-r^3}-4 r^3 \sqrt{|\mu| ^2-r^3}+\sqrt{|\mu| ^2-r^3} \sqrt{25 |\mu| ^4-104 |\mu| ^2 r^3+16 r^6}+4 |\mu|  r^3\right)}{\left(-2 |\mu| ^3+7 |\mu| ^2 \sqrt{|\mu| ^2-r^3}-4 r^3 \sqrt{|\mu|
   ^2-r^3}+\sqrt{|\mu| ^2-r^3} \sqrt{25 |\mu| ^4-104 |\mu| ^2 r^3+16 r^6}-4 |\mu|  r^3\right)}\Biggr).\nonumber\\
   & &
\end{eqnarray}
In the large-$r$ limit of (\ref{elliptic-fns}) after angular integrations, the finite radial integrand of (\ref{SDBI-arb-mu}) is given by:
\begin{eqnarray*}
& & -\frac{1}{72 \sqrt{2} |\mu| ^3 r^6}\Biggl[\sqrt{\left({F_{rt}}^2-1\right) |\mu| ^4} \Biggl(-64 i |\mu|  \sqrt{i |\mu|  \left(\frac{1}{r}\right)^{3/2}} F\left.\left(\frac{1}{4} \left(\frac{\sqrt{2} |\mu|
   }{r^{3/2}}-4 i \sinh ^{-1}(1)\right)\right|4 i |\mu|  \left(\frac{1}{r}\right)^{3/2}-1\right) r^9\nonumber\\
   & & +32 i \sqrt{i |\mu|  \left(\frac{1}{r}\right)^{3/2}} \sqrt{|\mu| ^2-r^3}
   F\left.\left(\frac{1}{4} \left(\frac{\sqrt{2} |\mu|
   }{r^{3/2}}-4 i \sinh ^{-1}(1)\right)\right|4 i |\mu|  \left(\frac{1}{r}\right)^{3/2}-1\right) r^9\nonumber\\
   & & +64 i |\mu|  \sqrt{i |\mu|
   \left(\frac{1}{r}\right)^{3/2}} \Pi \left(i |\mu|  \left(\frac{1}{r}\right)^{3/2}-1;i \sinh ^{-1}(1)-\frac{|\mu| }{2 \sqrt{2} r^{3/2}}|4 i |\mu|
   \left(\frac{1}{r}\right)^{3/2}-1\right) r^9\nonumber\\
   & & -64 i |\mu|  \sqrt{i |\mu|  \left(\frac{1}{r}\right)^{3/2}} \Pi \left(1-3 i |\mu|  \left(\frac{1}{r}\right)^{3/2};i \sinh
   ^{-1}(1)-\frac{|\mu| }{2 \sqrt{2} r^{3/2}}|4 i |\mu|  \left(\frac{1}{r}\right)^{3/2}-1\right) r^9\nonumber\\
   & & -16 |\mu| ^2 \sqrt{i |\mu|  \left(\frac{1}{r}\right)^{3/2}}
   F\left.\left(\frac{1}{4} \left(\frac{\sqrt{2} |\mu|
   }{r^{3/2}}-4 i \sinh ^{-1}(1)\right)\right|4 i |\mu|  \left(\frac{1}{r}\right)^{3/2}-1\right)r^{15/2}\nonumber\\
   & & +136 |\mu|  \sqrt{i |\mu|
   \left(\frac{1}{r}\right)^{3/2}} \sqrt{|\mu| ^2-r^3} F\left(\frac{1}{4} \left(\frac{\sqrt{2} |\mu| }{r^{3/2}}-4 i \sinh ^{-1}(1)\right)|4 i |\mu|
   \left(\frac{1}{r}\right)^{3/2}-1\right) r^{15/2}\nonumber\\
   & & +16 |\mu| ^2 \sqrt{i |\mu|  \left(\frac{1}{r}\right)^{3/2}} \Pi \left(i |\mu|  \left(\frac{1}{r}\right)^{3/2}-1;i \sinh
   ^{-1}(1)-\frac{|\mu| }{2 \sqrt{2} r^{3/2}}|4 i |\mu|  \left(\frac{1}{r}\right)^{3/2}-1\right) r^{15/2}\nonumber\\
   & & -16 |\mu| ^2 \sqrt{i |\mu|  \left(\frac{1}{r}\right)^{3/2}} \Pi
   \left(1-3 i |\mu|  \left(\frac{1}{r}\right)^{3/2};i \sinh ^{-1}(1)-\frac{|\mu| }{2 \sqrt{2} r^{3/2}}|4 i |\mu|  \left(\frac{1}{r}\right)^{3/2}-1\right) r^{15/2}\nonumber\\
   & & +32 |\mu|
   ^3 r^6+104 i |\mu| ^3 \sqrt{i |\mu|  \left(\frac{1}{r}\right)^{3/2}} F\left.\left(\frac{1}{4} \left(\frac{\sqrt{2} |\mu|
   }{r^{3/2}}-4 i \sinh ^{-1}(1)\right)\right|4 i |\mu|  \left(\frac{1}{r}\right)^{3/2}-1\right) r^6\nonumber\\
   & & +108 i |\mu| ^2 \sqrt{i |\mu|  \left(\frac{1}{r}\right)^{3/2}} \sqrt{|\mu| ^2-r^3} F\left.\left(\frac{1}{4} \left(\frac{\sqrt{2} |\mu|
   }{r^{3/2}}-4 i \sinh ^{-1}(1)\right)\right|4 i |\mu|  \left(\frac{1}{r}\right)^{3/2}-1\right) r^6\nonumber\\
   & & +280 i |\mu| ^3 \sqrt{i |\mu|  \left(\frac{1}{r}\right)^{3/2}} \Pi
   \left(i |\mu|  \left(\frac{1}{r}\right)^{3/2}-1;i \sinh ^{-1}(1)-\frac{|\mu| }{2 \sqrt{2} r^{3/2}}|4 i |\mu|  \left(\frac{1}{r}\right)^{3/2}-1\right) r^6\nonumber\\
   & & -280 i |\mu| ^3
   \sqrt{i |\mu|  \left(\frac{1}{r}\right)^{3/2}} \Pi \left(1-3 i |\mu|  \left(\frac{1}{r}\right)^{3/2};i \sinh ^{-1}(1)-\frac{|\mu| }{2 \sqrt{2} r^{3/2}}|4 i |\mu|
   \left(\frac{1}{r}\right)^{3/2}-1\right) r^6\nonumber\\
   & & +288 i |\mu| ^3 \sqrt{i |\mu|  \left(\frac{1}{r}\right)^{3/2}} \Pi \left(1-\frac{i |\mu| }{r^{3/2}};i \sinh
   ^{-1}(1)-\frac{|\mu| }{2 \sqrt{2} r^{3/2}}|4 i |\mu|  \left(\frac{1}{r}\right)^{3/2}-1\right) r^6\nonumber\\
   & & +32 \sqrt{6} |\mu| ^2 \sqrt{|\mu| ^2-r^3} r^6+28 |\mu| ^4 \sqrt{i |\mu|
   \left(\frac{1}{r}\right)^{3/2}} F\left.\left(\frac{1}{4} \left(\frac{\sqrt{2} |\mu|
   }{r^{3/2}}-4 i \sinh ^{-1}(1)\right)\right|4 i |\mu|  \left(\frac{1}{r}\right)^{3/2}-1\right)
   r^{9/2}\nonumber\\
   & & +114 |\mu| ^3 \sqrt{i |\mu|  \left(\frac{1}{r}\right)^{3/2}} \sqrt{|\mu| ^2-r^3} F\left.\left(\frac{1}{4} \left(\frac{\sqrt{2} |\mu|
   }{r^{3/2}}-4 i \sinh ^{-1}(1)\right)\right|4 i |\mu|  \left(\frac{1}{r}\right)^{3/2}-1\right)r^{9/2}\nonumber\\
    \end{eqnarray*}
   \begin{eqnarray}
   \label{integrand_large_r}
 & & +68 |\mu| ^4 \sqrt{i |\mu|  \left(\frac{1}{r}\right)^{3/2}} \Pi \left(i |\mu|
   \left(\frac{1}{r}\right)^{3/2}-1;i \sinh ^{-1}(1)-\frac{|\mu| }{2 \sqrt{2} r^{3/2}}|4 i |\mu|  \left(\frac{1}{r}\right)^{3/2}-1\right) r^{9/2}\nonumber\\
   & & -68 |\mu| ^4 \sqrt{i |\mu|
   \left(\frac{1}{r}\right)^{3/2}} \Pi \left(1-3 i |\mu|  \left(\frac{1}{r}\right)^{3/2};i \sinh ^{-1}(1)-\frac{|\mu| }{2 \sqrt{2} r^{3/2}}|4 i |\mu|
   \left(\frac{1}{r}\right)^{3/2}-1\right) r^{9/2}\nonumber\\
   & & +72 |\mu| ^4 \sqrt{i |\mu|  \left(\frac{1}{r}\right)^{3/2}} \Pi \left(1-\frac{i |\mu| }{r^{3/2}};i \sinh
   ^{-1}(1)-\frac{|\mu| }{2 \sqrt{2} r^{3/2}}|4 i |\mu|  \left(\frac{1}{r}\right)^{3/2}-1\right) r^{9/2}\nonumber\\
   & & +\sqrt{i |\mu|  \left(\frac{1}{r}\right)^{3/2}} \sqrt{|\mu| ^2-r^3}
   \left(-32 i r^6-8 |\mu|  r^{9/2}+52 i |\mu| ^2 r^3+14 |\mu| ^3 r^{3/2}+7 i |\mu| ^4\right)\nonumber\\
   & & \times F\left.\left(\frac{1}{4} \left(\frac{\sqrt{2} |\mu|
   }{r^{3/2}}-4 i \sinh ^{-1}(1)\right)\right|4 i |\mu|  \left(\frac{1}{r}\right)^{3/2}-1\right) r^3\nonumber\\
   & & +14 i |\mu| ^5 \sqrt{i |\mu|  \left(\frac{1}{r}\right)^{3/2}} F\left.\left(\frac{1}{4} \left(\frac{\sqrt{2} |\mu|
   }{r^{3/2}}-4 i \sinh ^{-1}(1)\right)\right|4 i |\mu|  \left(\frac{1}{r}\right)^{3/2}-1\right) r^3\nonumber\\
   & & +i |\mu| ^4 \sqrt{i |\mu|
   \left(\frac{1}{r}\right)^{3/2}} \sqrt{|\mu| ^2-r^3} F\left.\left(\frac{1}{4} \left(\frac{\sqrt{2} |\mu|
   }{r^{3/2}}-4 i \sinh ^{-1}(1)\right)\right|4 i |\mu|  \left(\frac{1}{r}\right)^{3/2}-1\right) r^3\nonumber\\
   & & +34 i |\mu| ^5 \sqrt{i |\mu|  \left(\frac{1}{r}\right)^{3/2}} \Pi \left(i |\mu|  \left(\frac{1}{r}\right)^{3/2}-1;i \sinh
   ^{-1}(1)-\frac{|\mu| }{2 \sqrt{2} r^{3/2}}|4 i |\mu|  \left(\frac{1}{r}\right)^{3/2}-1\right) r^3\nonumber\\
   & & -34 i |\mu| ^5 \sqrt{i |\mu|  \left(\frac{1}{r}\right)^{3/2}} \Pi
   \left(1-3 i |\mu|  \left(\frac{1}{r}\right)^{3/2};i \sinh ^{-1}(1)-\frac{|\mu| }{2 \sqrt{2} r^{3/2}}|4 i |\mu|  \left(\frac{1}{r}\right)^{3/2}-1\right) r^3\nonumber\\
   & & +36 i |\mu| ^5
   \sqrt{i |\mu|  \left(\frac{1}{r}\right)^{3/2}} \Pi \left(1-\frac{i |\mu| }{r^{3/2}};i \sinh ^{-1}(1)-\frac{|\mu| }{2 \sqrt{2} r^{3/2}}|4 i |\mu|
   \left(\frac{1}{r}\right)^{3/2}-1\right) r^3\nonumber\\
   & & +8 |\mu| ^5 \sqrt{i |\mu|  \left(\frac{1}{r}\right)^{3/2}} \sqrt{|\mu| ^2-r^3} F\left.\left(\frac{1}{4} \left(\frac{\sqrt{2} |\mu|
   }{r^{3/2}}-4 i \sinh ^{-1}(1)\right)\right|4 i |\mu|  \left(\frac{1}{r}\right)^{3/2}-1\right) r^{3/2}\nonumber\\
   & & -2 |\mu|  \sqrt{i |\mu|  \left(\frac{1}{r}\right)^{3/2}} \sqrt{|\mu|
   ^2-r^3} \left(32 r^{15/2}+8 i |\mu|  r^6+24 |\mu| ^2 r^{9/2}+6 i |\mu| ^3 r^3+4 |\mu| ^4 r^{3/2}+i |\mu| ^5\right)\nonumber\\
   & & \times E\left.\left(\frac{1}{4} \left(\frac{\sqrt{2} |\mu|
   }{r^{3/2}}-4 i \sinh ^{-1}(1)\right)\right|4 i |\mu|  \left(\frac{1}{r}\right)^{3/2}-1\right)\Biggr)\Biggr]\nonumber\\
   & & \approx \sqrt{|\mu|}\sqrt{1-F_{rt}^2}r^{\frac{9}{4}} + {\cal O}(r^{\frac{3}{2}}).
\end{eqnarray}

\section{$N_f=2$ Gauge Field Fluctuations' EOMs, Solution and On-Shell Action}
\setcounter{equation}{0} \seceqdd

Choosing the momentum four-vector in $\mathbb{R}^{1,3}$ as $q^\mu = (w, q, 0, 0)$, and  writing $A_\mu^a(x,u)=\int d^4q e^{- i w t + i q x}A^a_\mu(q,u)$, it follows in a straightforward way from (\ref{eom}) that in momentum space:
\begin{eqnarray}
& &  A_y^a{''}+\frac{\partial_u(\sqrt{{\rm det}\ G}G^{uu}G^{yy})}{\sqrt{{\rm det}\ G}G^{uu}G^{yy})}A_y^a{'}-w^2\frac{G^{tt}}{G^{uu}}A^a_y-\frac{(iw)}{2}\frac{G^{tt}}{G^{uu}}\frac{r^2_h}{2\pi{\alpha^\prime}}\tilde{A}^3_0f^{ab3}A^b_y\nonumber\\
& & +(iw)\frac{G^{tt}}{G^{uu}}\frac{r^2_h}{2\pi{\alpha^\prime}}\tilde{A}^3_0f^{ab3}A^b_y-\frac{1}{2}(\frac{r^2_h}{2\pi{\alpha^\prime}}\tilde{A}^3_0)^2
\frac{G^{tt}}{G^{uu}}f^{ab3}f^{bc3}A^c_y=0.
\end{eqnarray}
Set $q=0$, $A^a_y=\frac{1}{\omega}E^a_T$ implying:
\begin{eqnarray}
& & E_T^a{''}+\frac{\partial_u(\sqrt{{\rm det}\ G}G^{uu}G^{yy})}{\sqrt{{\rm det}\ G}G^{uu}G^{yy})}E_T^a{'}-w^2\frac{G^{tt}}{G^{uu}}E^a_T+\frac{iw}{2}\frac{G^{tt}}{G^{uu}}\frac{r^2_h}{2\pi{\alpha^\prime}}
\tilde{A}^3_0f^{ab3}E^b_T-\frac{1}{2}(\frac{r^2_h}{2\pi{\alpha^\prime}}\tilde{A}^3_0)^2\frac{G^{tt}}{G^{uu}}f^{ab3}f^{bc3}
E^c_T=0.\nonumber\\
& &
\end{eqnarray}
Now take $a=1$ and $b=2$:
\begin{eqnarray}
\label{E^1}
 E_T^1{''}+\frac{\partial_u(\sqrt{{\rm det}\ G}G^{uu}G^{yy})}{\sqrt{{\rm det}\ G}G^{uu}G^{yy})}E_T^1{'}-\frac{G^{tt}}{G^{uu}}[w^2-\frac{1}{2}(\frac{r^2_h}{2\pi{\alpha^\prime}}\tilde{A}^3_0)^2]E^1_T+
 \frac{iw}{2}\frac{G^{tt}}{G^{uu}}\frac{r^2_h}{2\pi{\alpha^\prime}}\tilde{A}^3_0f^{ab3}E^2_T=0.
\end{eqnarray}
Take $a=2,b=1$:
\begin{eqnarray}
\label{E^2}
 E_T^2{''}+\frac{\partial_u(\sqrt{{\rm det}\ G}G^{uu}G^{yy})}{\sqrt{{\rm det}\ G}G^{uu}G^{yy})}E_T^2{'}-\frac{G^{tt}}{G^{uu}}[w^2-\frac{1}{2}(\frac{r^2_h}{2\pi{\alpha^\prime}}
 \tilde{A}^3_0)^2]E^2_T-\frac{iw}{2}\frac{G^{tt}}{G^{uu}}\frac{r^2_h}{2\pi{\alpha^\prime}}\tilde{A}^3_0f^{ab3}E^1_T=0.
\end{eqnarray}
Take $a=3$:
\begin{eqnarray}
\label{E^3}
E_T^3{''}+\frac{\partial_u(\sqrt{{\rm det}\ G}G^{uu}G^{yy})}{\sqrt{{\rm det}\ G}G^{uu}G^{yy})}E_T^3{'}-w^2\frac{G^{tt}}{G^{uu}}E_T^3=0.
\end{eqnarray}

Defining $X=E^1+ i E^2$, $Y=E^1- i  E^2$, $\overline{A^3_0}\equiv\frac{r_h}{2\pi\alpha^\prime}\tilde{A}^3_0$, the $SU(2)$  equations of motion (\ref{E^1}) - (\ref{E^3}) can be rewritten as:
  \begin{eqnarray}
  \label{SU2_EOMs}
& &\hskip -1in \frac{\partial^2 X}{\partial u^2}+\frac{[16C^6 e^{6\phi}\sqrt{\frac{r_h}{u}} u^{14}(2u^4-1)+6C^2e^{2\phi}r_h^{9}\sqrt{\frac{r_h}{u}}u^{5}(13u^4-5)+r^{14}_h(23u^4-7)+3C^4e^{4\phi}r_h^5u^9(29u^4-13)]}{4(u^4-1)\sqrt{\frac{r_h}{u}}(r_h^4\sqrt{\frac{r_h}{u}}+C^2 e^{2\phi}u^{5})^3}\frac{\partial X}{\partial u} \nonumber\\
& & \hskip -1in+\frac{1}{\pi^2T^2(u^4-1)^2}(w-\overline{A^3_0})^2 X= 0\nonumber\\
& & \hskip -1in \frac{\partial^2 Y}{\partial u^2}+\frac{[16C^6e^{6\phi}\sqrt{\frac{r_h}{u}} u^{14}(2u^4-1)+6C^2e^{2\phi}r_h^{9}\sqrt{\frac{r_h}{u}}u^{5}(13u^4-5)+r^{14}_h(23u^4-7)+3C^4e^{4\phi}r_h^5u^9(29u^4-13)]}{4(u^4-1)\sqrt{\frac{r_h}{u}}(r_h^4\sqrt{\frac{r_h}{u}}+C^2e^{2\phi} u^{5})^3}\frac{\partial Y}{\partial u}\nonumber\\
& & \hskip -1in +\frac{1}{\pi^2T^2(u^4-1)^2}(w+\overline{A^3_0})^2 Y = 0;\nonumber\\
 & &\hskip -1in \frac{\partial^2 E^3}{\partial u^2}+\frac{[16C^6 e^(6\phi)\sqrt{\frac{r_h}{u}} u^{14}(2u^4-1)+6C^2e^{2\phi}r_h^{9}\sqrt{\frac{r_h}{u}}u^{5}(13u^4-5)+r^{14}_h(23u^4-7)+3C^4 e^{4\phi}r_h^5u^9(29u^4-13)]}{4(u^4-1)\sqrt{\frac{r_h}{u}}(r_h^4\sqrt{\frac{r_h}{u}}+C^2e^{2\phi} u^{5})^3}\frac{\partial E^3}{\partial u}\nonumber\\
& & \hskip -1in +\frac{w^2}{\pi^2T^2(u^4-1)^2} E^3 = 0.
 \end{eqnarray}
Similar to $E^3(u)$, the $U(1)$ EOM corresponding to gauge-invariant $E(u)$ is:
  \begin{eqnarray}
  \label{U1_EOM}
& & \hskip -0.9in  \frac{\partial^2 E}{\partial u^2}+\frac{[16C^6 e^{6\phi}\sqrt{\frac{r_h}{u}} u^{14}(2u^4-1)+6C^2e^{2\phi}r_h^{9}\sqrt{\frac{r_h}{u}}u^{5}(13u^4-5)+r^{14}_h(23u^4-7)+3C^4e^{4\phi}r_h^5u^9(29u^4-13)]}{4(u^4-1)\sqrt{\frac{r_h}{u}}(r_h^4\sqrt{\frac{r_h}{u}}+C^2 e^{2\phi}u^{5})^3}\frac{\partial E}{\partial u}\nonumber\\
& & \hskip-0.9in +\left[\frac{w^2}{\pi^2T^2(u^4-1)^2}\right]E = 0.
  \end{eqnarray}

The solution to EOM of either $E^3(u)$ or $E(u)$ is worked out as follows.  Defining
\begin{eqnarray}
\label{Sigma}
& & \hskip -0.8in \Sigma(u)\equiv  \frac{[16C^6 e^{6\phi}\sqrt{\frac{r_h}{u}} u^{14}(2u^4-1)+6C^2e^{2\phi}r_h^{9}\sqrt{\frac{r_h}{u}}u^{5}(13u^4-5)+r^{14}_h(23u^4-7)+3C^4e^{4\phi}r_h^5u^9(29u^4-13)]}{4(u^4-1)\sqrt{\frac{r_h}{u}}(r_h^4\sqrt{\frac{r_h}{u}}+C^2 e^{2\phi}u^{5})^3},\nonumber\\
& &
\end{eqnarray}
the EOM for $Z(u)\equiv E^3(u)$ or $E(u)$ can be written as:
\begin{equation}
\label{Z}
(u-1)^2\frac{d^2Z(u)}{du^2} + \frac{(u-1)\Sigma(u)}{(u+1)(u^2+1)}\frac{dZ(u)}{du} + \frac{w_3^2 Z(u)}{(u+1)^2(u^2+1)^2} = 0.
\end{equation}
One realizes that $u=1$ is a regular singular point with solutions to the indicial equation given by: $\pm i\frac{w_3}{4}$ and we choose the minus sign for incoming-wave solutions:
$Z(u)=(1-u)^{-\frac{iw_3}{4}}{\cal Z}(u)$. Using a perturbative ansatz:
\begin{equation}
\label{pert}
{\cal Z}(u) = {\cal Z}^{(0)}(u) + w_3 {\cal Z}^{(1)}(u) + {\cal O}(w_3^2),
\end{equation}
one finds (\ref{Z}) splits up into the following system of differential equations:
\begin{eqnarray}
\label{systemdiffeqs}
& & (u-1)^2 \frac{d^2{\cal Z}^{(0)}}{du^2}  + \frac{(u-1)\Sigma(u)}{(u+1)(u^2+1)}\frac{d{\cal Z}^{(0)}}{du} = 0;\nonumber\\
& & (u-1)^2\frac{d{\cal Z}^{(1)}}{du^2} + \frac{(u-1)\Sigma(u)}{(u+1)(u^2+1)}\frac{d{\cal Z}^{(1)}}{du} = \frac{i}{4}\left\{-1 + \frac{\Sigma(u)}{4(u+1)(u^2+1)}\right\}{\cal Z}^{(0)}(u) + \frac{i}{2}(u-1)\frac{d{\cal Z}^{(0)}}{du},\nonumber\\
\end{eqnarray}
 with the following solutions to (\ref{systemdiffeqs}):
\begin{eqnarray}
\label{sol1}
& & {\cal Z}^{(0)}(u) = \frac{2 c_1 \left(-21 \sqrt[4]{1-2 u} u \ _2F_1\left(\frac{1}{4},\frac{1}{4};\frac{5}{4};2 u\right)+6 u^2+u-2\right)}{3 u^{3/4} \sqrt[4]{2 u-1}}+c_2\nonumber\\
& & = \frac{4 (-1)^{3/4} c_1}{3 u^{3/4}}+c_2+14 (-1)^{3/4} c_1 \sqrt[4]{u}+{\cal O}\left(u^{5/4}\right)\nonumber\\
& & \equiv \frac{\alpha_0 c_1}{u^{\frac{3}{4}}} + c_2 + {\cal O}\left(u^{\frac{1}{4}}\right);\nonumber\\
& & {\cal Z}^{(1)}(u) = c_3+\frac{-14112 c_2 \sqrt[4]{1-2 u} u \ _2F_1\left(\frac{1}{4},\frac{1}{4};\frac{5}{4};2 u\right)+672 c_2 \left(6 u^2+u-2\right)+(68+68 i) \sqrt{2} c_1 \sqrt[4]{2 u-1}}{1008 u^{3/4}
   \sqrt[4]{2 u-1}}\nonumber\\
& & = \frac{\frac{4}{3} (-1)^{3/4} c_2+\frac{\left(\frac{17}{126}+\frac{17 i}{126}\right) c_1}{\sqrt{2}}}{u^{3/4}}+c_3+14 (-1)^{3/4} c_2
   \sqrt[4]{u} + {\cal O}\left(u^{5/4}\right)\nonumber\\
   & & \equiv \frac{\alpha_1 c_1 + \beta_1 c_2}{u^{\frac{3}{4}}} + c_3 + {\cal O}\left(u^{\frac{1}{4}}\right),
\end{eqnarray}
where $c_{1,2}\in\mathbb{R}$ and it is understood that $u\rightarrow0$ as $u\rightarrow\delta\rightarrow0$ and $c_{1,2,3}\rightarrow\delta^{\frac{3}{4}}:\frac{c_1}{c_2}$ is finite, to ensure finite gauge field perturbations ${\cal Z}^{(0),(1)}(u\rightarrow0)$ in (\ref{sol1}) and finite electrical conductivity (\ref{sigma-DC}). From (\ref{sol1}), we obtain the following:
\begin{eqnarray}
\label{sol2}
Z(u) &=& (1-u)^{-\frac{i w_3}{4}}{\cal Z}(u) = \frac{\alpha_0 c_1 + w_3\left[\alpha_1c_1 + \beta_1c_2\right]}{u^{\frac{3}{4}}}
+ c_2 + c_3 w_3 + c_1\gamma_0 u^{\frac{1}{4}} + \left(\frac{i}{4}\alpha_0c_1 + c_2\gamma_0\right)w_3u^{\frac{1}{4}} \nonumber\\
& & + \frac{i c_2}{4}w_3 u + ......;\nonumber\\
\frac{dZ(u)}{du} &=& \frac{1}{u^{\frac{7}{4}}}\left(-\frac{3\alpha_0c_1}{4} - \frac{\left(3\alpha_1c_1+ 3 \beta_1c_2\right)}{4}
w_3 + \frac{c_1\gamma_0}{4}u + \left(\frac{i}{16}\alpha_0c_1 + \frac{\gamma_0}{4}c_2\right)w_3u + ...\right).
\end{eqnarray}
We notice that the only distinction between the $SU(2)$ and $U(1)$ EOMs is the shift in the roots of the indicial equation corresponding to the horizon being a regular singular point; the incoming plane-wave root of the former (in $\alpha^\prime=\frac{1}{2\pi}$-units) is given by:
\begin{equation}
-\frac{i}{4}\left(w_3 + \overline{A^3_0}(u=1)\right) =  -\frac{i}{4}\left(w_3 + \left[\frac{2^{4/9} \Gamma \left(\frac{5}{18}\right) \Gamma \left(\frac{11}{9}\right)}{\sqrt[18]{\pi } \left(\frac{({g_s} {N_f} \log (\mu )-2
   \pi )^2}{C^2 {g_s}^2}\right)^{2/9}} - 1\right]r_h\right).
\end{equation}
We will not say more about this in this paper.

Let us work out the on-shell action to calculate the DC conductivity. For $\sigma=u$ the LHS of equation (\ref{eom}) simplifies to:
\begin{equation}
\partial_t\left[\sqrt{{\rm det}\ G}\left(2G^{tt}G^{uu}-2G^{ut}G^{ut}\right)\widehat{F^a_{ut}}\right]+\partial_x\left[\sqrt{{\rm det}\ G}\left(2G^{xx}G^{uu}\right)\widehat{F^a_{ux}}\right].
\end{equation}
Similarly the RHS simplifies to:
\begin{eqnarray}
& & \sqrt{{\rm det}\ G}\frac{r^2_h}{2\pi{\alpha^\prime}}
\tilde{A}^3_0f^{ab3}\left(G^{\nu t}G^{u\mu}-G^{\nu u}G^{t\mu}\right)\widehat{F^b_{\mu\nu}} = \sqrt{{\rm det}\ G}\frac{r^2_h}{2\pi{\alpha^\prime}}
\tilde{A}^3_0f^{ab3}\biggl[\left(G^{tt}G^{uu}-G^{tu}G^{tu}\right)\widehat{F^b_{ut}} \nonumber\\
& & +\left(G^{ut}G^{ut}-G^{uu}G^{tt}\right)\widehat{F^b_{tu}}\biggr]
= \sqrt{{\rm det}\ G}\frac{r^2_h}{2\pi{\alpha^\prime}}
\tilde{A}^3_0f^{ab3}\biggl[2G^{tt}G^{uu}-2G^{ut}G^{ut}\biggr]\widehat{F^b_{ut}}.
\end{eqnarray}
Now, working in the gauge $A^a_u=0$ which implies
\begin{eqnarray}
& & \partial_t\widehat{F^a_{ut}}=2\left(-iw\right)\partial_u A^a_t\nonumber\\
& & \partial_x\widehat{F^a_{ux}}=2\left(iq\right)\partial_u A^a_x,
\end{eqnarray}
we get the EOM:
\begin{eqnarray}
& & \left(-iw\right)\left(G^{tt}G^{uu}-G^{ut}G^{ut}\right)\partial_u A^a_t+\left(iq\right)\left(G^{xx}G^{uu}\right)\partial_u A^a_x = \frac{r^2_h}{2\pi{\alpha^\prime}}
\tilde{A}^3_0f^{ab3}\left(G^{tt}G^{uu}-G^{ut}G^{ut}\right)\partial_u A^b_t\nonumber\\
& & \Rightarrow\partial_uA^a_x=\frac{\left(G^{tt}G^{uu}-G^{ut}G^{ut}\right)}{\left(iq\right)\left(G^{xx}G^{uu}\right)}\biggl[\frac{r^2_h}{2\pi{\alpha^\prime}}
\tilde{A}^3_0f^{ab3}\left(\partial_uA^b_t\right)+iw\left(\partial_uA^a_t\right)\biggr].
\end{eqnarray}
Now, as shown in \cite{Erdmenger_et_al}, the on-shell action is given by:
\begin{equation}
\label{S-on-shell-1}
S_{\rm on-shell}\sim T_r T_{D7}\int d^4x\sqrt{{\rm det}\ G}\left.\left(G^{\nu u}G^{\nu'\mu}-G^{\nu \nu'}G^{u \mu}\right)A^a_{\nu'}\widehat{F^a_{\mu\nu}}\right|_{u=0},
\end{equation}
wherein:
\begin{eqnarray}
\label{S-on-shell-2}
& & \sqrt{{\rm det}\ G}\left(G^{\nu u}G^{\nu'\mu}-G^{\nu \nu'}G^{u \mu}\right)A^a_{\nu'}\widehat{F^a_{\mu\nu}}\nonumber\\
& & = \sqrt{{\rm det}\ G}\biggl[\left(G^{uu}G^{tt}-G^{ut}G^{ut}\right)A^a_{t}\widehat{F^a_{tu}}+\left(G^{tu}G^{tu}-G^{tt}G^{uu}\right)A^a_{t}
\widehat{F^a_{ut}} +\left(G^{uu}G^{xx}\right)A^a_x\widehat{F^a_{xu}}+\left(G^{tu}G^{xx}\right)A^a_x\widehat{F^a_{xt}}\nonumber\\
& & +\left(-G^{uu}G^{xx}\right)A^a_{x}\widehat{F^a_{ux}}+\left(-G^{ut}G^{xx}\right)A^a_{x}\widehat{F^a_{tx}}
+\left(G^{uu}G^{\alpha\alpha}\right)A^a_{\alpha}\widehat{F^a_{\alpha u}}+\left(G^{tu}G^{\alpha\alpha}\right)A^a_{\alpha}\widehat{F^a_{\alpha t}}\nonumber\\
& & +\left(-G^{uu}G^{\alpha\alpha}\right)A^a_{\alpha}\widehat{F^a_{u \alpha}}+\left(-G^{ut}G^{\alpha\alpha}\right)A^a_{\alpha}\widehat{F^a_{t \alpha}}\biggr]\nonumber\\
& & = \sqrt{{\rm det}\ G}\biggl[\left(2G^{ut}G^{ut}-2G^{uu}G^{tt}\right)A^a_{t}\widehat{F^a_{ut}}-\left(2G^{uu}G^{xx}\right)A^a_{x}\widehat{F^a_{ux}}-\left(2G^{uu}G^{\alpha\alpha}\right)A^a_{\alpha}\widehat{F^a_{u \alpha}}\biggr]\nonumber\\
& & =\sqrt{{\rm det}\ G}\biggl[4\left(G^{ut}G^{ut}-G^{uu}G^{tt}\right)A^a_{t}\left(\partial_uA^a_t\right)-4\left(G^{uu}G^{xx}\right)A^a_{x}\left(\partial_uA^a_x\right)
-4\left(G^{uu}G^{\alpha \alpha}\right)A^a_{\alpha}\left(\partial_u A^a_{\alpha}\right)\biggr].
\end{eqnarray}
In equation (\ref{S-on-shell-2}), the first term as an example can be simplified to:
\begin{eqnarray}
& & 4\sqrt{{\rm det}\ G}\left[\left(G^{ut}G^{ut}-G^{uu}G^{tt}\right)A^a_{t}\left(\partial_uA^a_t\right)\right]\nonumber\\
& & -\frac{A^a_x}{iq}\left(\left(G^{uu}G^{tt}-G^{ut}G^{ut}\right)\left[\frac{r^2_h}{2\pi{\alpha^\prime}}
\tilde{A}^3_0f^{ab3}\left(\partial_uA^b_t\right)+iw\left(\partial_uA^a_t\right)\right]-\left(G^{uu}G^{\alpha\alpha}\right)
A^a_{\alpha}\left(\partial_uA^a_{\alpha}\right)\right)
\nonumber\\
& & =4\sqrt{{\rm det}\ G}\left[\left(G^{ut}G^{ut}-G^{uu}G^{tt}\right)\left(\partial_uA^a_t\right)\left(A^a_{t}+\frac{w}{q}A^a_{x}\right)
+\left(G^{ut}G^{ut}-G^{uu}G^{tt}\right)\left(\frac{r^2_h}{2\pi{\alpha^\prime}}
\tilde{A}^3_0f^{ab3}\right)\left(\frac{A^a_{x}}{iq}\right)\left(\partial_uA^b_t\right)\right.\nonumber\\
& & \left. -\left(G^{uu}G^{\alpha\alpha}\right)A^a_{\alpha}\left(\partial_uA^a_{\alpha}\right)\right].
\end{eqnarray}
Let us work with the gauge-invariant electric field components $E^a_x=q A_t + w A^a_x$ and $E^a_{\alpha} = wA^a_{\alpha}, \alpha = y, z$. Differentiating we get
\begin{eqnarray}
\label{duEax}
& & \hskip -0.8in \partial_uE^a_x=q\partial_uA^a_t+wA^a_x\nonumber\\
& & \hskip -0.8in  = q\partial_uA^a_t+w\frac{w\left(G^{uu}G^{tt}-G^{ut}G^{ut}\right)}{\left(iq\right)\left(G^{uu}G^{xx}\right)}\left(\frac{r^2_h}{2\pi{\alpha^\prime}}
\tilde{A}^3_0f^{ab3}\partial_uA^b_t\right) + \frac{w^2}{q}\frac{\left(G^{uu}G^{tt}-G^{ut}G^{ut}\right)}{G^{xx}G^{uu}}\left(\partial_uA^a_t\right).
\end{eqnarray}
Now the terms in the on-shell action have to write in terms of $\partial_u E^a_x$. Assuming one will be interested in evaluation of flavor-diagonal two-point correlation functions for simplicity, we will disregard the flavor anti-symmetric terms and therefore obtain:
\begin{eqnarray}
\partial_uA^a_x=\frac{w}{q}\frac{\left(G^{tt}G^{uu}-G^{ut}G^{ut}\right)}{\left(G^{xx}G^{uu}\right)}\left(\partial_uA^a_t\right).
\end{eqnarray}
Substituting for $\partial_uA^a_x$, the action (\ref{S-on-shell-2}) then simplifies to:
\begin{eqnarray}
& & 4\sqrt{{\rm det}\ G}\left[\left(G^{ut}G^{ut}-G^{uu}G^{tt}\right)\left(A^a_{t}+\frac{w}{q}A^a_x\right)\left(\partial_uA^a_t\right)-4\left(G^{uu}G^{\alpha\alpha}\right)E^a_{\alpha}\left(\partial_uE^a_{\alpha}\right)\right]\nonumber\\
& & = \sqrt{{\rm det}\ G}\left[\frac{4}{q}\left(G^{ut}G^{ut}-G^{uu}G^{tt}\right)E^a_{x}\left(\partial_uA^a_t\right)-\frac{4}{w^2}\left(G^{uu}G^{\alpha\alpha}\right)
E^a_{\alpha}\left(\partial_uE^a_{\alpha}\right)\right].
\end{eqnarray}
Again disregarding the flavor-antisymmetric factor the expression for $\partial_uE^a_x$ in equation (\ref{duEax}), one gets: \begin{eqnarray}
& & \partial_uE^a_x=q\partial_uA^a_t+wA^a_x
=q\partial_uA^a_t+\frac{w^2}{q}\frac{\left(G^{uu}G^{tt}-G^{ut}G^{ut}\right)}{G^{xx}G^{uu}}(\partial_uA^a_t),
\end{eqnarray}
using which one obtains the following on-shell action's integrand:
\begin{eqnarray}
\label{DBI-EOM-ii}
& & \sqrt{{\rm det}\ G}\left[\frac{4G^{uu}G^{xx}(G^{ut}G^{ut}-G^{uu}G^{tt})}{q^2(G^{uu}G^{xx})+w^2(G^{tt}G^{uu}-G^{ut}G^{ut})}E^a_x(\partial_uE^a_x)-\frac{4}{w^2}G^{uu}G^{\alpha \alpha}E^a_{\alpha}(\partial_uE^a_{\alpha})+...\right]_{u=0}\nonumber\\
& & = 4\left(\frac{r_h u(u^4-1)}{w^2(\frac{r_h}{u})^{3/4}\sqrt{\frac{r^4_h\sqrt{\frac{r_h}{u}}}{r^4_h\sqrt{\frac{r_h}{u}}+c^2e^{2\phi}u^4}}}E^a_x(\partial_uE^a_x) + \frac{r_h u(u^4-1)}{w^2(\frac{r_h}{u})^{3/4}\sqrt{\frac{r^4_h\sqrt{\frac{r_h}{u}}}{r^4_h\sqrt{\frac{r_h}{u}}+c^2e^{2\phi}u^4}}}E^a_{\alpha}(\partial_uE^a_{\alpha})+...
\right)_{u=0}\nonumber\\
& & \sim \left.\frac{r_h^{\frac{1}{4}}u^{\frac{7}{4}}}{w^2}\left(E^a_x(\partial_uE^a_x) + E^a_{\alpha}(\partial_uE^a_{\alpha})\right)+....\right|_{u\rightarrow0},
\end{eqnarray}
where the dots include the flavor anti-symmetric terms.

\section{$SU(3)$ Structure Torsion Classes}
\setcounter{equation}{0}\seceqee

In this appendix, we will briefly review $SU(3)$ Structure Torsion classes. We will closely be following \cite{Louis_et_al}.

A $d$-dimensional Riemannian manifold ${\cal M}$,
has a $G$-structure if the structure group of the frame bundle  can be reduced
to $G \subset O(d)$.  A non-vanishing, globally defined
tensor or spinor $\xi$ is $G$-invariant if it is invariant under $G \subset O(d)$ rotations of the orthonormal frame. The existence of $\xi$ implies the existence of a $G$-structure.  
If the structure group of the frame
bundle is reduced to $G \subset O(d)$, the representation can be decomposed into irreducible
representations of $G$. For almost complex manifolds, this corresponds to the
decomposition under the $\left(P^\pm\right)_m^{\ n}\equiv\frac{1}{2}\left(\delta^m_n\pm i J_m^{\ n}\right)$ projections on to $\pm i$ eigenvalues of the almost complex structure. As there will usually be some tensor or spinor that will have a component in this decomposition which is invariant under $G$ implying the existence of a globally defined
non-vanishing $G$-invariant tensor or spinor. Now, two-forms are in the adjoint
representation $15$ of $SO(6)$ which decomposes under $U(3)$ as
$15 = 1 + 8 + (3 + \bar{3}) $.
Given a $U(3)$-structure, the singlet in the decomposition is the globally defined invariant two-form, which is precisely the fundamental two-form
$J$.
In the context of $SU(3)$ structure, there are two invariant tensors. First is the fundamental form $J$ as above. The second is the invariant
complex three-form . Three-forms are in the $20$ of $SO(6)$, giving two
singlets in the decomposition under $SU(3)$,
$15 = 1 + 8 + 3 + \bar{3}$ implying the existence of $J$ ,
$20 = 1 + 1 + 3 + \bar{3} + 6 + \bar{6}$ implying the existence of $\Omega  = \Omega^+ + i\Omega^- $.
There being no singlet in the decomposition of a five-form, one finds that
$J\wedge \Omega  = 0 $. Similarly, a six-form is a singlet of $SU(3)$, so we also must have that $J \wedge J \wedge J=3i/4 \Omega\wedge\bar{\Omega}$,
Conversely, a non-degenerate $J$ and  satisfying $J\wedge \Omega  = 0$ and $J \wedge J \wedge J=3i/4 \Omega\wedge\bar{\Omega}$ implies that ${\cal M}$ has $SU(3)$-structure.
We have the isomorphism $Spin(6) \cong SU(4)$ and the four-dimensional spinor representation
decomposes as $4 = 1 + 3$ implying the existence of $\eta$.
The singlet in the decomposition implies the existence of a globally defined
invariant spinor $\eta$. A metric and a globally defined
spinor $\eta$ implies that ${\cal M}$ has $SU(3)$-structure.

Now, One can define the Riemann curvature tensor $R_{mnp}^{\ q}$ and the torsion tensor $T_{mn}^{\ r}$  as follows:
$[\nabla^\prime_m,\nabla^\prime_n]V_p = - R_{mnp}^{\ q}V_q - 2T_{mn}^{\ r}\nabla_r^\prime V_p$ , where $V$ is an arbitrary vector field. The Levi-Civita connection is the unique torsionless
connection compatible with the metric and is given by the usual expression in terms
of Christoffel symbols $\Gamma_{mn}^{\ p} = \Gamma_{nm}^{\ p}$. Any metric-compatible connection can be written in terms of the Levi-Civita connection
$\nabla^{(T)} = \nabla + \kappa $,
where $\kappa_{mn}^{\ p}$ is the contorsion tensor. Metric compatibility implies
$\kappa_{mnp} = -\kappa_{mpn} $, where $\kappa_{mnp} = \kappa_{mn}^{\ r}g_{rp} $.
In general, the Levi-Civita connection does not preserve the $G$-invariant tensors (or spinor) $\xi$, i.e., $\nabla \xi\neq 0$. However, one can show that there always exists a connection $\nabla^{(T)}$ which is compatible with the $G$ structure so that $\nabla^{(T)}\xi = 0.$
On an almost Hermitian manifold one can always find $\nabla^{(T)}$ such that
$\nabla^{(T)}J = 0$. On a manifold with $SU(3)$-structure, it means we can always find $\nabla^{(T)}: \nabla^{(T)}J = 0,\ \nabla^{(T)}\Omega = 0$. Since the existence of $SU(3)$-structure is also
equivalent to the existence of an invariant spinor $\eta$, this is equivalent to the condition $\nabla^{(T)}\eta = 0$.
If $\kappa$ is the contorsion tensor corresponding to $\nabla^{(T)}$, then symmetries of $\kappa_{mnp}$ imply $\kappa\in\Lambda^1 \otimes \Lambda^2$ where$ \Lambda^n$ is the space of $n$-forms. Alternatively, since
$\Lambda^2 \cong so(d)$, $\kappa_{mn}^{\ p}$ can also be thought of as a one-form valued in the Lie-algebra $so(d)$, i.e., $\Lambda^1 \otimes so(d)$. Given the existence of a $G$-structure, we can decompose $so(d)$
into a part in the Lie algebra $g$ of $G \subset SO(d)$ and its orthogonal complement $g^\perp = so(d)/g$.  The contorsion $\kappa$ splits accordingly into
$\kappa = \kappa^0 + \kappa^g$, where $\kappa^0$ is the part in $\Lambda^1\otimes g^\perp$. Since an invariant tensor (or spinor) $\xi$ is fixed under $G$ rotations, the action of $g$ on $\xi$ vanishes and one has:
�$\nabla^{(T)}\xi = (\nabla + \kappa^0 + \kappa^g)\xi = (\nabla + \kappa^0) = 0$. Thus, any two $G$-compatible connections must differ by a piece proportional to $\kappa^g$ and they have a common term $\kappa^0 $ in $\Lambda^1\otimes g^\perp$ called the "intrinsic contorsion". Thus, the intrinsic contorsion/torsion,
is independent of the choice of $G$-compatible connection and is a measure
of the degree by which $\nabla\xi$ fails to vanish and as such is a measure solely of the $G$ structure. One can decompose $\kappa^0$ into irreducible $G$ representations providing a classification of $G$-structures in terms of which representations appear in the decomposition. In the special case when $\kappa^0$ vanishes so that $\nabla\xi = 0$, one says that the structure is ``torsion-free". For an almost Hermitian structure this is
equivalent to requiring that the manifold is complex and K\"{a}hler. In particular, it implies
that the holonomy of the Levi-Civita connection is contained in $G$. Let us consider the decomposition of $T^0$ in the case of $SU(3)$-structure. The relevant
representations are
$\Lambda^1\sim 3\oplus\bar{3}, g \sim 8, g^\perp\sim  1 \oplus 3 \oplus \bar{3}.$
Thus the intrinsic torsion, an element of $\Lambda^1\oplus su(3)^\perp$, can be decomposed into the following $SU(3)$ representations:
\begin{eqnarray}
& & \Lambda^1 \otimes su(3)^\perp = (3 \oplus \bar{3}) \otimes (1 \oplus 3 \oplus \bar{3)}
\nonumber\\
& & = (1 \oplus 1) \oplus (8 \oplus 8) \oplus (6 \oplus \bar{6}) \oplus (3 \oplus \bar{3}) \oplus (3 \oplus \bar{3})^\prime\equiv W_1\oplus W_2\oplus W_3\oplus W_4\oplus W_5.
\end{eqnarray}

The $SU(3)$ structure torsion classes \cite{torsion},\cite{uplift-IWASAWA} can be defined in terms of J, $ \Omega $, dJ, $ d{\Omega}$ and
the contraction operator  $\lrcorner : {\Lambda}^k T^{\star} \otimes {\Lambda}^n
T^{\star} \rightarrow {\Lambda}^{n-k} T^{\star}$,  $J$ being given by:
$$ J  =  e^1 \wedge e^2 + e^3 \wedge e^4 + e^5 \wedge e^6, $$
and
the (3,0)-form $ \Omega $ being given by
$$ \Omega  =  ( e^1 + ie^2) \wedge (e^3 +
ie^4) \wedge (e^5 + ie^6). $$
The torsion classes are defined in the following way:
\begin{itemize}
\item
$W_1 \leftrightarrow [dJ]^{(3,0)}$, given by real numbers
$W_1=W_1^+ + W_1^-$
with $ d {\Omega}_+ \wedge J = {\Omega}_+ \wedge dJ = W_1^+ J\wedge J\wedge J$
and $ d {\Omega}_- \wedge J = {\Omega}_- \wedge dJ = W_1^- J \wedge J \wedge J$;

\item
$W_2 \leftrightarrow [d \Omega]_0^{(2,2)}$ :
$(d{\Omega}_+)^{(2,2)}=W_1^+ J \wedge J + W_2^+ \wedge J$
and $(d{\Omega}_-)^{(2,2)}=W_1^- J \wedge J + W_2^- \wedge J$;

\item
 $W_3 \leftrightarrow [dJ]_0^{(2,1)}$ is defined
as $W_3=dJ^{(2,1)} -[J \wedge W_4]^{(2,1)}$;

\item
 $W_4 \leftrightarrow J \wedge dJ$ : $W_4 =\frac{1}{2} J\lrcorner dJ$;

 \item
 $W_5 \leftrightarrow
[d \Omega]_0^{(3,1)}$: $W_5 = \frac{1}{2} {\Omega}_+\lrcorner d{\Omega}_+$
(the subscript 0 indicative of the primitivity of the respective forms).
\end{itemize}

Depending on the classes of torsion one can obtain different types of manifolds, some of which are:
\begin{enumerate}
\item
(complex) special-hermitian  manifolds with $ W_1=W_2=W_4=W_5=0$ which
means that $ T \in W_3 $;

\item
(complex) K\"{a}hler  manifolds with $ W_1=W_2=W_3=W_4=0$ which means
$T\in W_5 $;

\item
(complex) balanced Manifolds with $W_1=W_2=W_4=0$ which
means $T\in W_3\oplus W_5$;

\item
(complex)
Calabi-Yau manifolds with $ W_1=W_2=W_3=W_4=W_5=0$ which means $T =0$.

\end{enumerate}

\section{$G_2$-Structure Torsion Classes}
\setcounter{equation}{0}\seceqff

In this appendix, we will give a brief description of seven-folds with $G_2$ structure borrowing extensively from \cite{Grigorian}.

If $V$ is a seven-dimensional real vector space, then a three-form $\varphi $ is said to be positive if it lies in the $GL\left( 7,
\mathbb{R}\right) $ orbit of $\varphi _{0}$, where $\varphi_0$ is a three-form on $\mathbb{R}^7$ which is preserved by $G_2$-subgroup of $GL(7,\mathbb{R})$.  The pair $\left( \varphi ,g\right)$ for a positive $3$-form $\varphi $ and corresponding metric $g$ constitute a $G_{2}$-structure.
The space of $p$-forms decompose as following irreps of $G_{2}$:
\begin{eqnarray}
\Lambda ^{1} &=&\Lambda _{7}^{1}  \label{l1decom} \nonumber\\
\Lambda ^{2} &=&\Lambda _{7}^{2}\oplus \Lambda _{14}^{2}  \label{l2decom} \nonumber\\
\Lambda ^{3} &=&\Lambda _{1}^{3}\oplus \Lambda _{7}^{3}\oplus \Lambda
_{27}^{3}  \label{l3decom} \nonumber\\
\Lambda ^{4} &=&\Lambda _{1}^{4}\oplus \Lambda _{7}^{4}\oplus \Lambda
_{27}^{4}  \label{l4decom} \nonumber\\
\Lambda ^{5} &=&\Lambda _{7}^{5}\oplus \Lambda _{14}^{5}  \label{l5decom} \nonumber\\
\Lambda ^{6} &=&\Lambda _{7}^{6}  \label{l6decom}
\end{eqnarray}
The subscripts denote the dimension of representation and components of same representation/dimensionality, are isomorphic to each other.
Let $M$ be a $7$-manifold with a $G_{2}$-structure $\left( \varphi ,g\right)
$. Then the components of spaces of $2$-, $3$-, $4$-, and $5$-forms are:
\begin{eqnarray*}
\Lambda _{7}^{2} &=&\left\{ \alpha \lrcorner \varphi {: }\alpha \in
\Lambda _{7}^{1}\right\} \\
\Lambda _{14}^{2} &=&\left\{ \omega \in \Lambda ^{2}{: }\left( \omega
_{AB}\right) \in \mathfrak{g}_{2}\right\} =\left\{ \omega \in \Lambda ^{2}%
{: }\omega \lrcorner \varphi =0\right\} \\
\Lambda _{1}^{3} &=&\left\{ f\varphi {: }f\in C^{\infty }\left(
M\right) \right\} \\
\Lambda _{7}^{3} &=&\left\{ \alpha \lrcorner \psi {: }\alpha \in
\Lambda _{7}^{1}\right\} \\
\Lambda _{27}^{3} &=&\left\{ \chi \in \Lambda ^{3}:
\chi_{ABC}=h_{[A}^{D}\varphi _{BC]D}{\rm for}\ h_{AB}~{\rm traceless,\ symmetric}\right\} \\
\Lambda _{1}^{4} &=&\left\{ f\psi {: }f\in C^{\infty }\left( M\right)
\right\} \\
\Lambda _{7}^{4} &=&\left\{ \alpha \wedge \varphi {: }\alpha \in
\Lambda _{7}^{1}\right\} \\
\Lambda _{27}^{4} &=&\left\{ \chi \in \Lambda ^{4}:\chi_{ABCD}=h_{[A}^{E}\psi _{BCD]E}{\rm for}\ h_{AB}~{\rm traceless,\ symmetric}
\right\} \\
\Lambda _{7}^{5} &=&\left\{ \alpha \wedge \psi {: }\alpha \in \Lambda
_{7}^{1}\right\} \\
\Lambda _{14}^{5} &=&\left\{ \omega \wedge \varphi {: }\omega \in
\Lambda _{14}^{2}{ }\right\}.
\end{eqnarray*}
 The metric $g$ defines a reduction of the frame bundle F to a principal $SO\left( 7\right) $-sub-bundle $Q$, that is, a sub-bundle of oriented orthonormal frames. Now, $g$ also defines a Levi-Civita connection $\nabla $ on the tangent bundle
$TM$, and hence on $F$. However, the $G_{2}$-invariant $3$-form $\varphi $
reduces the orthonormal bundle further to a principal $G_{2}$-subbundle $Q$.
The Levi-Civita connection can be pulled back to $Q$. On $Q$,  $\nabla $ can be uniquely decomposed as
\begin{equation}
\nabla =\bar{\nabla}+\mathcal{T}  \label{tors}
\end{equation}%
where $\bar{\nabla}$ is a $G_{2}$-compatible canonical connection on $P$, taking values in the sub-algebra $\mathfrak{g}_{2}\subset \mathfrak{so}%
\left( 7\right) $, while $\mathcal{T}$ is a $1$-form taking values in $%
\mathfrak{g}_{2}^{\perp }\subset \mathfrak{so}\left( 7\right) $; $\mathcal{T}$ is known as the intrinsic torsion of the
$G_{2}$-structure - the obstruction to the
Levi-Civita connection being $G_{2}$-compatible. Now $\mathfrak{so}%
\left( 7\right) $ splits under $G_{2}$ as
\begin{equation}
\mathfrak{so}\left( 7\right) \cong \Lambda ^{2}V\cong \Lambda _{7}^{2}\oplus
\Lambda _{14}^{2}.
\end{equation}
But $\Lambda _{14}^{2}\cong \mathfrak{g}_{2}$, so the orthogonal complement $\mathfrak{g%
}_{2}^{\perp }\cong \Lambda _{7}^{2}\cong V$. Hence $\mathcal{T}$ can be
represented by a tensor $T_{ab}$ which lies in $W\cong V\otimes V$. Now,
since $\varphi $ is $G_{2}$-invariant, it is $\bar{\nabla}$-parallel. So, the
torsion is determined by $\nabla \varphi $.

Following \cite{karigiannis-2007}, consider the $3$-form $\nabla _{X}\varphi
$ for some vector field $X$ from where:
\begin{equation}
\nabla _{X}\varphi \in \Lambda _{7}^{3}  \label{torsphi37}
\end{equation}%
and from Lemma 2.24 of \cite{karigiannis-2007}:
\begin{equation}
\nabla \varphi \in \Lambda _{7}^{1}\otimes \Lambda _{7}^{3}\cong W.
\label{torsphiW}
\end{equation}%
Due to the isomorphism between the $\Lambda^{a=1,...,5}_7$s, $\nabla \varphi $ lies in the same space as $T_{AB}$ and thus
completely determines it. Equation (\ref{torsphiW}) is equivalent to:
\begin{equation}
\nabla _{A}\varphi _{BCD}=T_{A}^{\ \ E}\psi _{EBCD}  \label{fulltorsion}
\end{equation}%
where $T_{AB}$ is the full torsion tensor. Equation (\ref{fulltorsion}) can be inverted to yield:
\begin{equation}
T_{A}^{\ M}=\frac{1}{24}\left( \nabla _{A}\varphi _{BCD}\right) \psi ^{MBCD}.
\label{tamphipsi}
\end{equation}%
The tensor $T_A^{\ M}$, like the space W, possesses 49 components and hence fully defines $\nabla \varphi $. In general $T_{AB}$ cab be split into torsion components
as
\begin{equation}
T=T _{1}g+T _{7}\lrcorner \varphi +T _{14}+T _{27}
\label{torsioncomps}
\end{equation}%
where $T _{1}$ is a function and gives the $\mathbf{1}$ component of $T$
. We also have $T _{7}$, which is a $1$-form and hence gives the $\mathbf{
7}$ component, and, $T _{14}\in \Lambda _{14}^{2}$ gives the $\mathbf{14}$
component. Further, $T _{27}$ is traceless symmetric, and gives the $\mathbf{27}$
component. Writing $T_i$ as $W_i$, we can split $W$ as
\begin{equation}
W=W_{1}\oplus W_{7}\oplus W_{14}\oplus W_{27}.  \label{Wsplit}
\end{equation}

\section{$G^{IIA}_{\theta_1\theta_2}=0$ in the UV and Details of Local Sechsbeins Relevant to Type IIA $SU(3)$ Structure Torsion Classes}
\setcounter{equation}{0} \seceqgg

In this appendix, after showing how to ensure $G^{IIA}_{\theta_1\theta_2}=0$ in the entire UV indicative of the possibility  that the local mirror of the warped deformed conifold is a warped resolved conifold, we provide details relevant to obtaining the type IIA sechsbeins for the directions $(r,\theta_1,\theta_2,\phi_1,\phi_2,\psi)$. These are relevant to section {\bf 5} where we discuss $SU(3)$-structure torsion classes of the delocalized type IIA mirror.

From \cite{MQGP}, the exact expression for $G^{IIA}_{\theta_1\theta_2}$ after a triple T duality is given by:
\begin{eqnarray}
\label{ExactIIAmetricth1th2_i}
& & \hskip -0.8in G^{IIA}_{\theta_1\theta_2} = \frac{1}{192 \pi ^{5/2} r^2 \left(3 \sin ^2({\theta_1})+2\right) \sqrt{\frac{{g_s}
   N}{r^4}}}\Biggl\{{g_s} \sin ^2({\theta_1}) \nonumber\\
   & & \hskip -0.8in \times\Biggl(\frac{128 \pi ^3 N {f_1}({\theta_1}) {f_2}({\theta_2}) \sin ({\theta_1}) \sin ^2({\theta_2}) (3
   {h_5} \csc ({\theta_1})+\csc ({\theta_2})) \left(3 \left(9 {h_5}^2-1\right) \sin ({\theta_1}) \sin ({\theta_2})-6 {h_5}-2 \csc
   ({\theta_1}) \sin ({\theta_2})\right)}{\sin ^2({\theta_1}) \left(3 \left(9 {h_5}^2-1\right) \sin ^2({\theta_2})-2\right)-12 {h_5} \sin
   ({\theta_1}) \sin ({\theta_2})-2 \sin ^2({\theta_2})}\nonumber\\
   & &\hskip -0.8in +576 \pi ^3 {h_5} N {f_1}({\theta_1}) {f_2}({\theta_2}) \sin ({\theta_1})
   \sin ({\theta_2})-384 \pi ^3 {h_5} N \left(3 \sin ^2({\theta_1})+2\right) \csc ^2({\theta_1}) ({f_1}({\theta_1}) {f_2}({\theta_2})
   \sin ({\theta_1}) \sin ({\theta_2})-1)
   \nonumber\\
   & & \hskip --0.8in  -243 {g_s}^2 M^2 {N_f} \log (r) \csc ^2({\theta_1}) \csc \left(\frac{{\theta_2}}{2}\right)\nonumber\\
   & & \hskip -0.8in \times\Biggl[2
   \sin ({\theta_1}) \left(2 \log (r) \left({g_s} {N_f} \log \left(\sin \left(\frac{{\theta_1}}{2}\right) \sin
   \left(\frac{{\theta_2}}{2}\right)\right)+2 \pi \right)+9 {g_s} {N_f} \log ^2(r)+{g_s} {N_f} \log \left(\sin
   \left(\frac{{\theta_1}}{2}\right) \sin \left(\frac{{\theta_2}}{2}\right)\right)\right) \nonumber\\
   & &\hskip -0.8in +{g_s} {N_f} \log (r) \csc
   \left(\frac{{\theta_1}}{2}\right)\Biggr]+\frac{1}{\sin ^2({\theta_1}) \left(3 \left(9 {h_5}^2-1\right) \sin ^2({\theta_2})-2\right)-12 {h_5} \sin
   ({\theta_1}) \sin ({\theta_2})-2 \sin ^2({\theta_2})}\nonumber\\
   & & \hskip -0.8in + \Biggl[81 {g_s} M^2 \sin ^2({\theta_2}) \Biggl\{(9 {h_5}-2 \csc ({\theta_1}) \csc ({\theta_2}))
   \Biggl(4 {g_s} {N_f} \log (r) \log \left(\sin \left(\frac{{\theta_1}}{2}\right) \sin \left(\frac{{\theta_2}}{2}\right)\right)+{g_s} {N_f}
   \log (r) \csc \left(\frac{{\theta_1}}{2}\right) \csc ({\theta_1})\nonumber\\
   & & \hskip -0.8in +18 {g_s} {N_f} \log ^2(r)+2 {g_s} {N_f} \log \left(\sin
   \left(\frac{{\theta_1}}{2}\right) \sin \left(\frac{{\theta_2}}{2}\right)\right)+8 \pi  \log (r)\Biggr)+{g_s} {N_f} \log (r) \csc
   \left(\frac{{\theta_1}}{2}\right) \left(2 \csc ^2({\theta_1})+3\right) \cot ({\theta_2})\Biggr\}\nonumber\\
   & & \hskip -0.8in \times \biggl[{g_s} {N_f} \log (r) \csc
   \left(\frac{{\theta_2}}{2}\right) \left(9 {h_5} \sin ({\theta_1}) \cos ({\theta_2})+\left(3 \sin ^2({\theta_1})+2\right) \csc
   ({\theta_2})\right)+2 \left(3 \sin ^2({\theta_1})+2\right)\nonumber\\
   & & \hskip -0.8in \left(2 \log (r) \left({g_s} {N_f} \log \left(\sin
   \left(\frac{{\theta_1}}{2}\right) \sin \left(\frac{{\theta_2}}{2}\right)\right)+2 \pi \right)+9 {g_s} {N_f} \log ^2(r)+{g_s} {N_f} \log
   \left(\sin \left(\frac{{\theta_1}}{2}\right) \sin \left(\frac{{\theta_2}}{2}\right)\right)\right)\nonumber\\
   & & \hskip -0.8in -2 {g_s} {N_f} \log (r) \cot ({\theta_2})
   \csc \left(\frac{{\theta_2}}{2}\right)\biggr]\Biggr]\Biggr)\Biggr\}
   \end{eqnarray}
Near $\theta_1=\theta_2=0$ and in the UV, (\ref{ExactIIAmetricth1th2_i}) simplifies to:
\begin{eqnarray}
\label{ExactIIAmetricth1th2_ii}
& & \hskip -0.8in G^{IIA}_{\theta_1\theta_2}=\frac{1}{96 \pi ^{5/2} {\theta_1} {\theta_2} \sqrt{{g_s} N}}
\Biggl\{{g_s} \Biggl[-3 \biggl(4 \pi ^3 {h_5} N {\theta_1} {\theta_2} {f_1}({\theta_1}) {f_2}({\theta_2}) (7 \sin ({\theta_1})+3 \sin
   (3 {\theta_1})) \sin ({\theta_2})\nonumber\\
   & & \hskip -0.8in+81 {g_s}^3 M^2 {N_f}^2 {\theta_1} \log ^2(r) \sin ({\theta_1}) \biggl\{9 \log (r)+2 \log \left[\sin
   \left(\frac{{\theta_1}}{2}\right) \sin \left(\frac{{\theta_2}}{2}\right)\right]\biggr\}+81 {g_s}^3 M^2 {N_f}^2 \log ^2(r)-64 \pi ^3 {h_5} N
   {\theta_1} {\theta_2}\biggr)\nonumber\\
   & & \hskip -0.8in +32 \pi ^3 N {f_1}({\theta_1}) {f_2}({\theta_2}) \left({\theta_1}^2+{\theta_2}^2\right) (3 {h_5} \sin
   ({\theta_1})+\sin ({\theta_2})) (3 {h_5} \sin ({\theta_2})+\sin ({\theta_1}))\nonumber\\
   & & \hskip -0.8in+162 {g_s}^3 M^2 {N_f}^2 \log (r)
   \left({\theta_1}^2+{\theta_2}^2\right) \left(9 \log (r)+2 \log \left[\sin \left(\frac{{\theta_1}}{2}\right) \sin
   \left(\frac{{\theta_2}}{2}\right)\right]\right)\nonumber\\
   & & \hskip -0.8in \times \left(2 \log (r) \log \left[\sin \left(\frac{{\theta_1}}{2}\right) \sin
   \left(\frac{{\theta_2}}{2}\right)\right]+9 \log ^2(r)+\log \left[\sin \left(\frac{{\theta_1}}{2}\right) \sin
   \left(\frac{{\theta_2}}{2}\right)\right]\right)\Biggr]\Biggr\}\end{eqnarray}
   The equation (\ref{ExactIIAmetricth1th2_ii}) yields:
  \begin{eqnarray}
  \label{GIIAth1th2-simpl}
& & \hskip -0.8in G^{IIA}_{\theta_1\theta_2}\sim\frac{g_s}{\sqrt{g_sN}}\Biggl\{ - 6\times 32\pi^3h_5N\left(f_1(\theta_1)f_2(\theta_2) \theta_1\theta_2 + 1\right)
+ g_s^3M^2N_f^2 \left[9(\log r)^3 + 4 \log\theta\right] + g_s^3M^2N_f^2 \frac{(\log r)^2}{\theta_1\theta_2} \nonumber\\
& & \hskip -0.8in+ N f_1(\theta_1)f_2(\theta_2)(3 h_5 \theta_1 + \theta_2)(3 h_5 \theta_2 + \theta_1) + g_s^3M^2N_f^2\left[9 (\log r)^2 + 4 (\log \theta)^2\right]\left[9(\log r)^2 + 4 \log r\log\theta + 2\log\theta\right] \Biggr\}.\nonumber\\
& &
\end{eqnarray}
Writing $f_i(\theta_i)\sim\cot\theta_i, i=1,2$, one sees that from (\ref{GIIAth1th2-simpl}), one obtains:
\begin{equation}
\frac{\theta_1^2+\theta_2^2}{\theta_1\theta_2}(3h_5\theta_1+\theta_2)(3h_5\theta_2+\theta_1)f_1(\theta_1)f_2(\theta_2)
\stackrel{\theta_2\ll  \theta_1}{\longrightarrow}h_5\left(\frac{\theta_1}{\theta_2}\right)^2,
\end{equation}
which if one assumes: $\theta_2=h_5^{\frac{\alpha\in(0,1)}{2}}\theta_1$ yields $h_5^{1-\alpha}\ll  1$.

So, near $\theta_1=\theta_2=0$ and in the UV, utilizing results for $G^{IIA}_{\theta_1\theta_1}$ and $G^{IIA}_{\theta_2\theta_2}$ of appendix A of \cite{MQGP}:
\begin{equation}
G^{IIA}_{\theta_i\theta_j}\sim\sqrt{g_sN}\left(f_i(\theta_i)f_j(\theta_j)\theta_i\theta_j + 1\right).
\end{equation}
By choosing: $f_1(\theta_1) = \pm\cot\theta_1, f_2(\theta_2) = \mp\cot\theta_2$, one ensures that $G^{IIA}_{\theta_1\theta_2}=0$ indicative of the possibility  that the local mirror of the warped deformed conifold is a warped resolved conifold $\forall r\in $ UV and not just $r=\sqrt{3}a$ as in \cite{MQGP}.

The other most dominant terms of the mirror type IIA metric of \cite{MQGP} are looked at in (\ref{xtheta2}) - (\ref{xz}).

\begin{eqnarray}
\label{xtheta1}
& & \hskip -0.8in (1)\ G^{IIA}_{x\theta_1}=\frac{1}{8 \sqrt{2} \pi
   ^{5/4} \sqrt{{g_s} N}}\Biggl\{\sqrt[3]{3} \sqrt[3]{\frac{1}{{g_s}}} {g_s}^{7/3} M {N_f} \log (r) \cot \left(\frac{{\theta_1}}{2}\right) \csc ({\theta_1}) \left(108 a^2
   \log (r)+r\right) \sqrt[4]{\frac{{g_s} N}{r^4}} \csc \left(\frac{{\theta_1}}{\sqrt[10]{N}}\right) \nonumber\\
   & & \hskip -0.8in \times \left(9 {h_5}+\left(3 \sqrt{6}-2 \cot
   ({\theta_1})\right) \cot \left(\frac{{\theta_1}}{\sqrt[10]{N}}\right)\right)\left(2 \cos ({\theta_1}) \cos
   \left(\frac{{\theta_1}}{\sqrt[10]{N}}\right)-9 {h_5} \sin ({\theta_1}) \sin \left(\frac{{\theta_1}}{\sqrt[10]{N}}\right)\right)\Biggr\}\nonumber\\
   & & \hskip -0.8in = \frac{\sqrt[3]{3} \sqrt[3]{\frac{1}{{g_s}}} {g_s}^{11/6} M {N_f} \sqrt[4]{\frac{{g_s}}{r^4}} \log (r) \cot \left(\frac{{\theta_1}}{2}\right)
   \left(3 \sqrt{6}-2 \cot ({\theta_1})\right) \cot ({\theta_1}) \left(108 a^2 \log (r)+r\right)}{4 \sqrt{2} \pi ^{5/4} \sqrt[20]{N} {\theta_1}^2} + {\cal O}\left(\frac{1}{N^{\frac{3}{20}}}\right),
\end{eqnarray}
implying the following most dominant term in the small-$\theta_{1,2}$ limit:
\begin{equation}
\label{Gphi1theta1}
G^{IIA}_{\phi_1\theta_1}\sim -\frac{\sqrt[3]{3} {g_s}^2 M \sqrt[5]{N} {N_f} \log (r)}{\sqrt{2} \pi ^{5/4} {\theta_1}^4}.
\end{equation}

\begin{eqnarray}
\label{xtheta2}
& & (2)\ G^{IIA}_{x\theta_2}=\frac{216 3^{5/6} a^2 \sqrt[3]{\frac{1}{{g_s}}} {g_s}^{7/3} M r^2 \log (r) \cos ^2({\theta_1}) \cot ({\theta_1}) \cot
   \left(\frac{{\theta_1}}{\sqrt[10]{N}}\right)}{\sqrt[4]{\pi } \sqrt[4]{{g_s} N} (\cos (2 {\theta_1})-5) \left(2 \cot
   ^2\left(\frac{{\theta_1}}{\sqrt[10]{N}}\right)+2 \cot ^2({\theta_1})+3\right)}\nonumber\\
   & & = \frac{108 3^{5/6} a^2 \sqrt[3]{\frac{1}{{g_s}}} {g_s}^{25/12} M r^2 {\theta_1} \log (r) \cos ^2({\theta_1}) \cot ({\theta_1})}{\sqrt[4]{\pi }
   N^{7/20} (\cos (2 {\theta_1})-5)} + {\cal O}\left(\frac{1}{N^{\frac{11}{20}}}\right),
\end{eqnarray}
implying the following most dominant term in the small-$\theta_{1,2}$ limit:
\begin{equation}
\label{Gphi1theta2}
G^{IIA}_{\phi_1\theta_2}\sim -\frac{27 3^{5/6} a^2 {g_s}^2 M r^2 {\theta_1} \log (r)}{\sqrt[4]{\pi } \sqrt[10]{N}}.
\end{equation}

\begin{eqnarray}
\label{ytheta2}
& & \hskip -0.8in (3) G^{IIA}_{y\theta_2} = -\frac{\sqrt[4]{\pi } \left(\frac{1}{{g_s}}\right)^{2/3} {g_s}^{2/3} \sqrt[4]{{g_s} N} (\cos (2 {\theta_1})-5) \sin
   \left(\frac{{\theta_1}}{\sqrt[10]{N}}\right) \cos ^2\left(\frac{{\theta_1}}{\sqrt[10]{N}}\right)}{\sqrt{2} 3^{2/3} \left(3 {h_5} \sin (2
   {\theta_1}) \sin \left(\frac{2 {\theta_1}}{\sqrt[10]{N}}\right)+3 \sin ^2({\theta_1}) \sin ^2\left(\frac{{\theta_1}}{\sqrt[10]{N}}\right)+2 \sin
   ^2({\theta_1}) \cos ^2\left(\frac{{\theta_1}}{\sqrt[10]{N}}\right)+2 \cos ^2({\theta_1}) \sin
   ^2\left(\frac{{\theta_1}}{\sqrt[10]{N}}\right)\right)}\nonumber\\
   & & \hskip -0.8in = \frac{\sqrt[4]{\pi } \left(\frac{1}{{g_s}}\right)^{2/3} {g_s}^{11/12} N^{3/20} (\cos (2 {\theta_1})-5)}{2 \sqrt{2} 3^{2/3} {\theta_1}} + {\cal O}\left(N^{\frac{1}{20}}\right),
\end{eqnarray}
implying the following most dominant term in the small-$\theta_{1,2}$ limit:
\begin{equation}
\label{phi2theta2}
G^{IIA}_{\phi_2\theta_2}\sim\frac{\sqrt{2} \sqrt[4]{\pi } \sqrt{{g_s}} \sqrt[20]{N}}{3^{2/3}}.
\end{equation}

\begin{eqnarray}
\label{ytheta1}
& &\hskip -0.8in (4) G_{y\theta_1} = \frac{9 3^{5/6} \sqrt[3]{\frac{1}{{g_s}}} {g_s}^{4/3} M r \log (r) \sin ({\theta_1}) \sqrt[4]{\frac{{g_s} N}{r^4}} \sin
   ^2\left(\frac{{\theta_1}}{\sqrt[10]{N}}\right) \left(2 \cos ({\theta_1}) \cot \left(\frac{{\theta_1}}{\sqrt[10]{N}}\right)-9 {h_5} \sin
   ({\theta_1})\right)}{\sqrt[4]{\pi } \sqrt{{g_s} N} \left(3 {h_5} \sin (2 {\theta_1}) \sin \left(\frac{2 {\theta_1}}{\sqrt[10]{N}}\right)+3
   \sin ^2({\theta_1}) \sin ^2\left(\frac{{\theta_1}}{\sqrt[10]{N}}\right)+2 \sin ^2({\theta_1}) \cos ^2\left(\frac{{\theta_1}}{\sqrt[10]{N}}\right)+2
   \cos ^2({\theta_1}) \sin ^2\left(\frac{{\theta_1}}{\sqrt[10]{N}}\right)\right)}\nonumber\\
   & & \hskip -0.8in = \frac{9 3^{5/6} \sqrt[3]{\frac{1}{{g_s}}}{g_s}^{13/12} M \log (r) \sin ({\theta_1})}{\sqrt[4]{\pi } N^{7/20} {\theta_1}} + {\cal O}\left(\frac{1}{N^{\frac{11}{20}}}\right),
\end{eqnarray}
implying the following most dominant term in the small-$\theta_{1,2}$ limit:
\begin{equation}
\label{phi2theta1}
G^{IIA}_{\phi_2\theta_1}\sim \frac{9 3^{5/6} {g_s} M \log (r) \sin ({\theta_1})}{\sqrt[4]{\pi } \sqrt[5]{N}}
\end{equation}

\begin{eqnarray}
\label{ztheta1}
& & \hskip -0.8in (5)\ G^{IIA}_{z\theta_1} = \frac{3 \sqrt[3]{3} {g_s}^2 M {N_f} r \log (r) \cot \left(\frac{{\theta_1}}{2}\right) \csc ^2({\theta_1}) \left(6 {h_5} \sin (2 {\theta_1})
   \cot \left(\frac{{\theta_1}}{\sqrt[10]{N}}\right)+\sin ^2({\theta_1}) \left(2 \cot ^2\left(\frac{{\theta_1}}{\sqrt[10]{N}}\right)+3\right)+2 \cos
   ^2({\theta_1})\right)}{8 \sqrt{2} \pi ^{5/4} \sqrt[4]{{g_s} N}}\nonumber\\
   & & = \frac{3 \sqrt[3]{3} {g_s}^{7/4} M {N_f} r \log (r) \cot \left(\frac{{\theta_1}}{2}\right)}{4 \sqrt{2} \pi ^{5/4} \sqrt[20]{N} {\theta_1}^2} + {\cal O}\left(\frac{1}{N^{\frac{1}{4}}}\right)
\end{eqnarray}
implying the following most dominant term in the small-$\theta_{1,2}$ limit:
\begin{equation}
\label{psitheta1}
G^{IIA}_{\psi\theta_1}\sim \frac{3 \sqrt[3]{3} {g_s}^2 M N^{3/20} {N_f} r \log (r)}{2 \sqrt{2} \pi ^{5/4} {\theta_1}^3}.
\end{equation}

\begin{eqnarray}
\label{ztheta2}
& & \hskip -0.8in (6)\ G^{IIA}_{z\theta_2} = \frac{1}{256 \sqrt{2} \pi ^{5/4} \sqrt[4]{{g_s} N} (\cos (2
   {\theta_1})-5)}\nonumber\\
   & & \hskip -0.8in \times\Biggl\{3 \sqrt[3]{3} {g_s}^2 M {N_f} \log (r) \csc ^2({\theta_1}) \csc ^3\left(\frac{{\theta_1}}{2 \sqrt[10]{N}}\right) \sec
   \left(\frac{{\theta_1}}{2 \sqrt[10]{N}}\right) \biggl((1-12 {h_5}) \cos \left(\left(2-\frac{2}{\sqrt[10]{N}}\right) {\theta_1}\right)\nonumber\\
   & & \hskip -0.8in+12 {h_5}
   \cos \left(2 \left(\frac{1}{\sqrt[10]{N}}+1\right) {\theta_1}\right)+\cos \left(2 \left(\frac{1}{\sqrt[10]{N}}+1\right) {\theta_1}\right)+6 \cos
   \left(\frac{2 {\theta_1}}{\sqrt[10]{N}}\right)+6 \cos (2 {\theta_1})-14\biggr)\nonumber\\
   & & \hskip -0.8in\times \left(4 \cos ({\theta_1}) \cos
   \left(\frac{{\theta_1}}{\sqrt[10]{N}}\right)-\sin ^2({\theta_1})+\cos ^2({\theta_1})-5\right)\Biggr\}\nonumber\\
   & &\hskip -0.8in = \frac{3 \sqrt[3]{3} {g_s}^{7/4} M \sqrt[20]{N} {N_f} \log (r) (6 \cos (2 {\theta_1})-7) \csc ^2({\theta_1}) \left(-\sin ^2({\theta_1})+\cos
   ^2({\theta_1})+4 \cos ({\theta_1})-5\right)}{32 \sqrt{2} \pi ^{5/4} {\theta_1}^3 (\cos (2 {\theta_1})-5)},
\end{eqnarray}
implying the following most dominant term in the small-$\theta_{1,2}$ limit:
\begin{equation}
\label{psitheta2}
G^{IIA}_{\psi\theta_2}\sim -\frac{3 \sqrt[3]{3} {g_s}^2 M N^{7/20} {N_f} \log (r)}{32 \sqrt{2} \pi ^{5/4} {\theta_1}^3}.
\end{equation}

\begin{eqnarray}
\label{xx}
& &  (7)\  G^{IIA}_{xx} = \frac{3^{2/3} \sin ^2({\theta_1}) \left(\cos \left(\frac{2 {\theta_1}}{\sqrt[10]{N}}\right)-5\right)}{\cos \left(\left(2-\frac{2}{\sqrt[10]{N}}\right)
   {\theta_1}\right)+\cos \left(2 \left(\frac{1}{\sqrt[10]{N}}+1\right) {\theta_1}\right)-2} = 3^{\frac{2}{3}} + {\cal O}\left(\frac{1}{N^{\frac{1}{5}}}\right),
\end{eqnarray}
implying the following most dominant term in the small-$\theta_{1,2}$ limit:
\begin{equation}
\label{phi1phi1}
G^{IIA}_{\phi_1\phi_1}\sim 3^{2/3} {\theta_1}^2 \sqrt{{g_s} N}.
\end{equation}

\begin{equation}
\label{phi2phi2}
(8)\ G^{IIA}_{\phi_2\phi_2} = \frac{3^{2/3} {\theta_1}^2 \sqrt{{g_s} N}}{\sqrt[5]{N}}.
\end{equation}

\begin{eqnarray}
\label{zz}
& & (9)\ G^{IIA}_{zz} = \frac{\left(\frac{1}{{g_s}}\right)^{2/3} {g_s}^{2/3} \csc ^2({\theta_1}) \left(2 N \cos ^2({\theta_1})+N \sin ^2({\theta_1}) \left(2 \cot
   ^2\left(\frac{{\theta_1}}{\sqrt[10]{N}}\right)+3\right)+6 \sin (2 {\theta_1}) \cot \left(\frac{{\theta_1}}{\sqrt[10]{N}}\right)\right)}{3 \sqrt[3]{3}
   N}\nonumber\\
& & = \frac{2 \left(\frac{1}{{g_s}}\right)^{2/3} {g_s}^{2/3} \sqrt[5]{N}}{3 \sqrt[3]{3} {\theta_1}^2} + {\cal O}(N^0),
\end{eqnarray}
implying the following most dominant term in the small-$\theta_{1,2}$ limit:
\begin{equation}
\label{psipsi}
G^{IIA}_{\psi\psi} = \frac{2 \sqrt{{g_s}} N^{7/10}}{3 \sqrt[3]{3} {\theta_1}^2}.
\end{equation}

\begin{eqnarray}
\label{xy}
& & (10)\ G^{IIA}_{xy} = -\frac{8 \sqrt{2} \left(\frac{1}{{g_s}}\right)^{2/3} {g_s}^{2/3} \cos ^2({\theta_1}) (\cos (2 {\theta_1})-5) \sin
   \left(\frac{{\theta_1}}{\sqrt[10]{N}}\right) \cos ^3\left(\frac{{\theta_1}}{\sqrt[10]{N}}\right)}{3 \sqrt[6]{3} \left(\cos
   \left(\left(2-\frac{2}{\sqrt[10]{N}}\right) {\theta_1}\right)+\cos \left(2 \left(\frac{1}{\sqrt[10]{N}}+1\right) {\theta_1}\right)-2\right)^2}\nonumber\\
   & & = \frac{\left(\frac{1}{{g_s}}\right)^{2/3} {g_s}^{2/3} {\theta_1} (\cos (2 {\theta_1})-5) \cot ^2({\theta_1}) \csc ^2({\theta_1})}{3 \sqrt{2}
   \sqrt[6]{3} \sqrt[10]{N}} + {\cal O}\left(\frac{1}{N^{\frac{3}{10}}}\right),
\end{eqnarray}
implying the following most dominant term in the small-$\theta_{1,2}$ limit:
\begin{equation}
\label{xy-ii}
G^{IIA}_{\phi_1\phi_2} = \frac{2 \sqrt{2} \sqrt{{g_s}} N^{3/10}}{3 \sqrt[6]{3} {\theta_1}}.
\end{equation}

\begin{eqnarray}
\label{yz}
& & (11)\ G^{IIA}_{yz} = -\frac{\sqrt{2} \csc ({\theta_1}) \left(\frac{3 \cos ({\theta_1})}{N}+\sin ({\theta_1}) \cot
   \left(\frac{{\theta_1}}{\sqrt[10]{N}}\right)\right)}{\sqrt[6]{3}} = -\frac{\sqrt{2} \sqrt[10]{N}}{\sqrt[6]{3} {\theta_1}} + {\cal O}\left(\frac{1}{N^{\frac{1}{10}}}\right),
\end{eqnarray}
implying the following most dominant term in the small-$\theta_{1,2}$ limit:
\begin{equation}
\label{phi2psi}
G^{IIA}_{\phi_2\psi} = -\frac{\sqrt{2} \sqrt{{g_s} N}}{\sqrt[6]{3}}.
\end{equation}

\begin{eqnarray}
\label{xz}
& &\hskip -0.8in (12)\ G^{IIA}_{xz} = \frac{1}{9 \sqrt[3]{3} N^2 (\cos (2 {\theta_1})-5)}\Biggl\{-4 N^2 \cos ^2({\theta_1})
\nonumber\\
& & \hskip -0.8in\times \left(-4 \cot ^2({\theta_1}) \cot ^2\left(\frac{{\theta_1}}{\sqrt[10]{N}}\right)-6 \cot
   ^2\left(\frac{{\theta_1}}{\sqrt[10]{N}}\right)+3 \sqrt{6} \cot ({\theta_1}) \left(2 \cot
   ^2\left(\frac{{\theta_1}}{\sqrt[10]{N}}\right)-3\right)\right)\nonumber\\
   & & \hskip -0.8in +162 \sin ^2({\theta_1}) \left(\sqrt{6} N \cot
   \left(\frac{{\theta_1}}{\sqrt[10]{N}}\right)+3\right)+54 N \sin ({\theta_1}) \cos ({\theta_1}) \left(\sqrt{6} N-4 \cot
   \left(\frac{{\theta_1}}{\sqrt[10]{N}}\right)\right)\Biggr\}\nonumber\\
   & & \hskip -0.8in = -\frac{4 \sqrt[5]{N} \cot ^2({\theta_1}) \left(9 \sqrt{2} \sin (2 {\theta_1})+\sqrt{3} \cos (2 {\theta_1})-5 \sqrt{3}\right)}{9 3^{5/6} {\theta_1}^2
   (\cos (2 {\theta_1})-5)},
\end{eqnarray}
implying the following most dominant term in the small-$\theta_{1,2}$ limit:
\begin{equation}
\label{phi1psi}
G^{IIA}_{\phi_1\psi} = -\frac{4 \sqrt{{g_s}} N^{7/10}}{9 \sqrt[3]{3} {\theta_1}^3}.
\end{equation}

The following are the three eigenvalues of (\ref{d=3}):
\begin{itemize}
\item
\begin{eqnarray}
\label{ev1-1}
& & \frac{1}{6} N^{7/10} \Biggl(\frac{6 \sqrt[3]{2} {g_{13}}^2 {g_s}}{\sqrt[3]{9 {g_{13}}^2 {g_{33}} {g_s}^{3/2} {\theta_1}^{10}+\sqrt{{g_s}^3
   {\theta_1}^{18} \left(-108 {g_{13}}^6+81 {g_{13}}^4 {g_{33}}^2 {\theta_1}^2+36 {g_{13}}^2 {g_{33}}^4 {\theta_1}^4+4 {g_{33}}^6
   {\theta_1}^6\right)}}}\nonumber\\
   & & +\frac{2^{2/3} \sqrt[3]{9 {g_{13}}^2 {g_{33}} {g_s}^{3/2} {\theta_1}^{10}+\sqrt{{g_s}^3 {\theta_1}^{18} \left(-108
   {g_{13}}^6+81 {g_{13}}^4 {g_{33}}^2 {\theta_1}^2+36 {g_{13}}^2 {g_{33}}^4 {\theta_1}^4+4 {g_{33}}^6
   {\theta_1}^6\right)}}}{{\theta_1}^6}\nonumber\\
   & & +\frac{2 {g_{33}} \sqrt{{g_s}}}{{\theta_1}^2}\Biggr) + {\cal O}\left(\frac{1}{N^{\frac{11}{10}}}\right),
\end{eqnarray}
whose small-$\theta_1$ expansion yields:
\begin{eqnarray}
\label{ev1-2}
&  & -\frac{{g_{33}}^2 {g_s} N^{7/10} \left(\left(-{g_{13}}^6 {g_s}^3\right)^{2/3}-{g_{13}}^4 {g_s}^2\right)}{24 \sqrt{3} {\theta_1}
   \left(-{g_{13}}^6 {g_s}^3\right)^{5/6}}+\frac{{g_{33}}^3 {g_s}^{3/2} N^{7/10} \left(\sqrt[3]{-{g_{13}}^6 {g_s}^3}-{g_{13}}^2
   {g_s}\right)}{54 \left(-{g_{13}}^6 {g_s}^3\right)^{2/3}}\nonumber\\
   & & +\frac{{g_{33}}^4 {g_s} N^{7/10} {\theta_1} \left(157 \left(-{g_{13}}^6
   {g_s}^3\right)^{2/3}+227 {g_{13}}^4 {g_s}^2\right)}{3456 \sqrt{3} {g_{13}}^2 \left(-{g_{13}}^6 {g_s}^3\right)^{5/6}}
   \nonumber\\
   & & +\frac{{g_{33}}
   N^{7/10} \left(-\left(-{g_{13}}^6 {g_s}^3\right)^{2/3}+2 {g_{13}}^4 {g_s}^2+{g_{13}}^2 {g_s} \sqrt[3]{-{g_{13}}^6 {g_s}^3}\right)}{6
   {g_{13}}^4 {g_s}^{3/2} {\theta_1}^2}\nonumber\\
   & & +\frac{{g_{13}}^2 {g_s} N^{7/10} \left(\left(-{g_{13}}^6 {g_s}^3\right)^{2/3}-{g_{13}}^4
   {g_s}^2\right)}{\sqrt{3} {\theta_1}^3 \left(-{g_{13}}^6 {g_s}^3\right)^{5/6}} + {\cal O}(N^{\frac{7}{10}}\theta_1^2).
\end{eqnarray}
Assuming $(-)^{\frac{1}{3}}=e^{\frac{i \pi}{3}}$, etc. , the leading-order term of (\ref{ev1-2}) obtains:
\begin{equation}
\label{ev1-3}
\frac{{g_{13}} \sqrt{{g_s}} N^{7/10}}{{\theta_1}^3}+\frac{0.5 {g_{33}} \sqrt{{g_s}} N^{7/10}}{{\theta_1}^2}.
\end{equation}
\item
\begin{eqnarray}
\label{ev2-1}
& & \frac{1}{12 {\theta_1}^6}\Biggl\{N^{7/10} \Biggl(i 2^{2/3} \left(\sqrt{3}+i\right) \sqrt[3]{9 {g_{13}}^2 {g_{33}} {g_s}^{3/2} {\theta_1}^{10}+\sqrt{81 {g_{13}}^4 {g_{33}}^2
   {g_s}^3 {\theta_1}^{20}-108 {g_{13}}^6 {g_s}^3 {\theta_1}^{18}}}\nonumber\\
   & & -\frac{6 i \sqrt[3]{2} \left(\sqrt{3}-i\right) {g_{13}}^2 {g_s}
   {\theta_1}^6}{\sqrt[3]{9 {g_{13}}^2 {g_{33}} {g_s}^{3/2} {\theta_1}^{10}+\sqrt{81 {g_{13}}^4 {g_{33}}^2 {g_s}^3 {\theta_1}^{20}-108
   {g_{13}}^6 {g_s}^3 {\theta_1}^{18}}}}+4 {g_{33}} \sqrt{{g_s}} {\theta_1}^4\Biggr)\Biggr\}\nonumber\\
& &    + {\cal O}\left(\frac{1}{N^{\frac{11}{10}}}\right)
   \end{eqnarray}
   whose small-$\theta_1$ expansion yields:
\begin{eqnarray}
\label{ev2-2}
& &\hskip -1in \frac{{g_{33}}^3 {g_s}^{3/2} N^{7/10} \left(i \left(\sqrt{3}+i\right) \sqrt[3]{-{g_{13}}^6 {g_s}^3}+{g_{13}}^2 \left({g_s}+i \sqrt{3}
   {g_s}\right)\right)}{108 \left(-{g_{13}}^6 {g_s}^3\right)^{2/3}} +\frac{{g_{33}}^2 N^{7/10} \left(\left(\sqrt{3}-3 i\right) \sqrt[3]{-{g_{13}}^6
   {g_s}^3}+\left(\sqrt{3}+3 i\right) {g_{13}}^2 {g_s}\right)}{144 {g_{13}}^2 {\theta_1} \sqrt[6]{-{g_{13}}^6 {g_s}^3}}
   \nonumber\\
   & &\hskip -1in -\frac{N^{7/10}
   \left(\left(\sqrt{3}-3 i\right) \sqrt[3]{-{g_{13}}^6 {g_s}^3}+\left(\sqrt{3}+3 i\right) {g_{13}}^2 {g_s}\right)}{6 {\theta_1}^3
   \sqrt[6]{-{g_{13}}^6 {g_s}^3}} +\frac{35 {g_{33}}^4 N^{7/10} {\theta_1} \left(\left(\sqrt{3}-3 i\right) \sqrt[3]{-{g_{13}}^6
   {g_s}^3}+\left(\sqrt{3}+3 i\right) {g_{13}}^2 {g_s}\right)}{20736 {g_{13}}^4 \sqrt[6]{-{g_{13}}^6 {g_s}^3}}\nonumber\\
   & &\hskip -1in +\frac{{g_{33}} \sqrt{{g_s}}
   N^{7/10} \left(\frac{\left(1+i \sqrt{3}\right) {g_{13}}^4 {g_s}^2}{\left(-{g_{13}}^6 {g_s}^3\right)^{2/3}}+\frac{i \left(\sqrt{3}+i\right)
   {g_{13}}^2 {g_s}}{\sqrt[3]{-{g_{13}}^6 {g_s}^3}}+4\right)}{12 {\theta_1}^2} + {\cal O}\left(N^{\frac{7}{10}}\theta_1^2\right).
\end{eqnarray}
Assuming $(-)^{\frac{1}{3}}=e^{\frac{i \pi}{3}}$, etc. , the leading-order term of (\ref{ev2-2}) obtains:
\begin{equation}
\label{ev2-3}
\frac{0.5 {g_{33}} \sqrt{{g_s}} N^{7/10}}{{\theta_1}^2}-\frac{{g_{13}} \sqrt{{g_s}} N^{7/10}}{{\theta_1}^3},
\end{equation}
assuming that $\frac{10 {g_{13}}}{{g_{33}}}<{\theta_1}\ll  1$ with $0<\frac{g_{13}}{g_{33}}\ll  1$ guaranteeing a positive eigenvalue.
\item
\begin{eqnarray}
\label{ev3-1}
& & \frac{1}{36 {\theta_1}^6}\Biggl\{N^{7/10} \Biggl(-3 2^{2/3} \left(1+i \sqrt{3}\right) \sqrt[3]{9 {g_{13}}^2 {g_{33}} {g_s}^{3/2} {\theta_1}^{10}+3 \sqrt{3} \sqrt{-4 {g_{13}}^6
   {g_s}^3 {\theta_1}^{18}-{g_{13}}^4 {g_{33}}^2 {g_s}^3 {\theta_1}^{20}}}\nonumber\\
   & & +\frac{2 i \sqrt[3]{2} 3^{2/3} \left(\sqrt{3}+i\right) {g_s}
   {\theta_1}^6 \left(3 {g_{13}}^2+{g_{33}}^2 {\theta_1}^2\right)}{\sqrt[3]{3 {g_{13}}^2 {g_{33}} {g_s}^{3/2} {\theta_1}^{10}+\sqrt{3}
   \sqrt{-4 {g_{13}}^6 {g_s}^3 {\theta_1}^{18}-{g_{13}}^4 {g_{33}}^2 {g_s}^3 {\theta_1}^{20}}}}+12 {g_{33}} \sqrt{{g_s}}
   {\theta_1}^4\Biggr)\Biggr\} + {\cal O}\left(\frac{1}{N^{\frac{11}{10}}}\right),\nonumber\\
   & &
\end{eqnarray}
whose small-$\theta_1$ expansion yields:
\begin{eqnarray}
\label{ev3-2}
& & \frac{N^{7/10} \left(-\frac{3 i \left(\sqrt{3}+i\right) {g_{13}}^4 {g_{33}} {g_s}^{5/2}}{\left(-{g_{13}}^6 {g_s}^3\right)^{2/3}}+\frac{3 \left(1+i
   \sqrt{3}\right) {g_{33}} \left(-{g_{13}}^6 {g_s}^3\right)^{2/3}}{{g_{13}}^4 {g_s}^{3/2}}+12 {g_{33}} \sqrt{{g_s}}\right)}{36
   {\theta_1}^2}\nonumber\\
   & & +\frac{1}{36} N^{7/10} \left(\frac{2 i \left(\sqrt{3}+i\right) {g_{13}}^2 {g_{33}}^3 {g_s}^{5/2}}{3 \left(-{g_{13}}^6
   {g_s}^3\right)^{2/3}}-\frac{2 \left(1+i \sqrt{3}\right) {g_{33}}^3 \left(-{g_{13}}^6 {g_s}^3\right)^{2/3}}{3 {g_{13}}^6
   {g_s}^{3/2}}\right)\nonumber\\
   & & +\frac{N^{7/10} \left(\frac{3 i \sqrt{3} \left(\sqrt{3}+i\right) {g_{33}}^2 {g_s}}{4 \sqrt[6]{-{g_{13}}^6 {g_s}^3}}-\frac{3
   \sqrt{3} \left(1+i \sqrt{3}\right) {g_{33}}^2 \sqrt[6]{-{g_{13}}^6 {g_s}^3}}{4 {g_{13}}^2}\right)}{36 {\theta_1}}\nonumber\\
   & & +\frac{N^{7/10} \left(\frac{6 i
   \sqrt{3} \left(\sqrt{3}+i\right) {g_{13}}^2 {g_s}}{\sqrt[6]{-{g_{13}}^6 {g_s}^3}}-6 \sqrt{3} \left(1+i \sqrt{3}\right) \sqrt[6]{-{g_{13}}^6
   {g_s}^3}\right)}{36 {\theta_1}^3} + {\cal O}\left(N^{\frac{7}{10}}\theta_1\right).
   \end{eqnarray}
Assuming $(-)^{\frac{1}{3}}=e^{\frac{i \pi}{3}}$, etc. , the leading-order term of (\ref{ev3-2}) obtains:
\begin{equation}
\label{ev3-3}
\frac{0.074 {g_{33}}^3 \sqrt{{g_s}} N^{7/10}}{{g_{13}}^2}.
\end{equation}
\end{itemize}
The following are the associated eigenvectors:
\begin{itemize}
\item
\begin{equation}
\label{evec1-i}
\left(
\begin{array}{c}
 \frac{0.25 {g_{33}}^2 {\theta_1}^2-{g_{13}}^2}{{g_{13}} ({g_{13}}+0.5 {g_{33}} {\theta_1})}+\frac{{g_{23}}^2 {\theta_1}^6}{{g_{13}}
   ({g_{13}}+0.5 {g_{33}} {\theta_1}) N^{2/5}}+{\cal O}\left(\left(\frac{1}{N}\right)^{11/10}\right) \\
 -\frac{{g_{23}} {\theta_1}^3}{({g_{13}}+0.5 {g_{33}} {\theta_1}) \sqrt[5]{N}}+{\cal O}\left(\left(\frac{1}{N}\right)^{6/5}\right) \\
 1
\end{array}
\right)
\end{equation}
whose small-$\theta_1$ expansion is given by:
\begin{equation}
\label{evec1-ii}
\left(
\begin{array}{c}
 -1+\frac{0.5 {g_{33}} {\theta_1}}{{g_{13}}}+\frac{{g_{23}}^2 {\theta_1}^6}{{g_{13}}^2 N^{2/5}}+O\left({\theta_1}^7\right) \\
 -\frac{{g_{23}} {\theta_1}^3}{{g_{13}} \sqrt[5]{N}}+\frac{0.5 {g_{23}} {g_{33}} {\theta_1}^4}{{g_{13}}^2 \sqrt[5]{N}}-\frac{0.25 \left({g_{23}}
   {g_{33}}^2\right) {\theta_1}^5}{{g_{13}}^3 \sqrt[5]{N}}+\frac{0.125 {g_{23}} {g_{33}}^3 {\theta_1}^6}{{g_{13}}^4
   \sqrt[5]{N}}+O\left({\theta_1}^7\right) \\
 1
\end{array}
\right).
\end{equation}
We will however use the following eigenvector normalized to unity:
\begin{equation}
\label{normevec1}
\left(
\begin{array}{c}
 \frac{0.002 {g_{33}}^4 {\theta_1}^4}{{g_{13}}^4}+\frac{0.07 {g_{33}}^3 {\theta_1}^3}{{g_{13}}^3}+\frac{0.07 {g_{33}}^2
   {\theta_1}^2}{{g_{13}}^2}+\frac{0.18 {g_{33}} {\theta_1}}{{g_{13}}}-\frac{1}{\sqrt{2}} \\
 \frac{\frac{0.18 {g_{23}} {g_{33}} {\theta_1}^4}{{g_{13}}^2}-\frac{{g_{23}} {\theta_1}^3}{\sqrt{2} {g_{13}}}}{\sqrt[5]{N}} \\
 -\frac{0.005 {g_{33}}^4 {\theta_1}^4}{{g_{13}}^4}-\frac{0.006 {g_{33}}^3 {\theta_1}^3}{{g_{13}}^3}+\frac{0.02 {g_{33}}^2
   {\theta_1}^2}{{g_{13}}^2}+\frac{0.18 {g_{33}} {\theta_1}}{{g_{13}}}+\frac{1}{\sqrt{2}}
\end{array}
\right).
\end{equation}

\item
\begin{equation}
\label{evec2-i}
\left(
\begin{array}{c}
 \frac{ {g_{13}}^2-0.25 {g_{33}}^2 {\theta_1}^2}{ {g_{13}}^2-0.5 {g_{13}} {g_{33}} {\theta_1}}+\frac{{g_{23}}^2 {\theta_1}^6}{{g_{13}}
   (0.5 {g_{33}} {\theta_1}- {g_{13}}) N^{2/5}}+O\left(\left(\frac{1}{N}\right)^{11/10}\right) \\
 \frac{ {g_{23}} {\theta_1}^3}{( {g_{13}}-0.5 {g_{33}} {\theta_1}) \sqrt[5]{N}}+O\left(\left(\frac{1}{N}\right)^{6/5}\right) \\
 1
\end{array}
\right)
\end{equation}
whose small-$\theta_1$ expansion is given by:
\begin{equation}
\label{evec2-ii}
\left(
\begin{array}{c}
 \left(1+\frac{0.5 {g_{33}} {\theta_1}}{{g_{13}}}+O\left({\theta_1}^7\right)\right)+\frac{-\frac{ {g_{23}}^2
   {\theta_1}^6}{{g_{13}}^2}+O\left({\theta_1}^7\right)}{N^{2/5}}+O\left(\left(\frac{1}{N}\right)^{11/10}\right) \\
 \frac{\frac{ {g_{23}} {\theta_1}^3}{{g_{13}}}+\frac{0.5 {g_{23}} {g_{33}} {\theta_1}^4}{{g_{13}}^2}+\frac{0.25 {g_{23}} {g_{33}}^2
   {\theta_1}^5}{{g_{13}}^3}+\frac{0.125 {g_{23}} {g_{33}}^3
   {\theta_1}^6}{{g_{13}}^4}+O\left({\theta_1}^7\right)}{\sqrt[5]{N}}+O\left(\left(\frac{1}{N}\right)^{6/5}\right) \\
 1
\end{array}
\right)
\end{equation}
We will however use the following eigenvector normalized to unity:
\begin{equation}
\label{normevec2}
\left(
\begin{array}{c}
 -\frac{0.002 {g_{33}}^4 {\theta_1}^4}{{g_{13}}^4}+\frac{0.02 {g_{33}}^3 {\theta_1}^3}{{g_{13}}^3}-\frac{0.07 {g_{33}}^2
   {\theta_1}^2}{{g_{13}}^2}+\frac{0.18 {g_{33}} {\theta_1}}{{g_{13}}}+\frac{1}{\sqrt{2}} \\
 \frac{\frac{0.18 {g_{23}} {g_{33}} {\theta_1}^4}{{g_{13}}^2}+\frac{{g_{23}} {\theta_1}^3}{\sqrt{2} {g_{13}}}}{\sqrt[5]{N}} \\
 -\frac{0.004 {g_{33}}^4 {\theta_1}^4}{{g_{13}}^4}+\frac{0.006 {g_{33}}^3 {\theta_1}^3}{{g_{13}}^3}+\frac{0.02 {g_{33}}^2
   {\theta_1}^2}{{g_{13}}^2}-\frac{0.18 {g_{33}} {\theta_1}}{{g_{13}}}+\frac{1}{\sqrt{2}}
\end{array}
\right).
\end{equation}

\item
The large-$N$ small-$\theta_1$ expansion of the third eigenvector is given by:
\begin{equation}
\label{evec3-i}
\left(
\begin{array}{c}
 -\frac{0.07 {g_{33}}^3 {\theta_1}^3}{{g_{13}}^3}+\frac{13.51 {g_{13}} {g_{23}}^2 {\theta_1}^3}{{g_{33}}^3 N^{2/5}}+\frac{{g_{33}}
   {\theta_1}}{{g_{13}}} \\
 -\frac{13.51 {g_{13}}^2 {g_{23}}}{{g_{33}}^3 \sqrt[5]{N}} \\
 1
\end{array}
\right)
\end{equation}
We will however use the following eigenvector normalized to unity:
\begin{equation}
\label{normevec3}
\left(
\begin{array}{c}
 -\frac{0.57 {g_{33}}^3 {\theta_1}^3}{{g_{13}}^3}+\frac{{g_{33}} {\theta_1}}{{g_{13}}}+\frac{\frac{156.79 {g_{13}} {g_{23}}^2
   {\theta_1}^3}{{g_{33}}^3}-\frac{91.26 {g_{13}}^3 {g_{23}}^2 {\theta_1}}{{g_{33}}^5}}{N^{2/5}} \\
 \frac{-\frac{6 {g_{23}} {g_{33}} {\theta_1}^4}{{g_{13}}^2}+\frac{6.76 {g_{23}} {\theta_1}^2}{{g_{33}}}-\frac{13.51 {g_{13}}^2
   {g_{23}}}{{g_{33}}^3}}{\sqrt[5]{N}} \\
 \frac{0.45 {g_{33}}^4 {\theta_1}^4}{{g_{13}}^4}-\frac{0.5 {g_{33}}^2 {\theta_1}^2}{{g_{13}}^2}+1
\end{array}
\right)
\end{equation}

\end{itemize}
Hence, the modal matrix whose columns are the afore-obtained eigenvectors, is given by:
\begin{eqnarray}
\label{M}
& & \hskip -0.7in {\cal M} = \nonumber\\
& & \hskip -0.7in \left(
\begin{array}{ccc}
 \frac{0.07 {g_{33}}^3 {\theta_1}^3}{{g_{13}}^3}+\frac{0.07 {g_{33}}^2 {\theta_1}^2}{{g_{13}}^2}+\frac{0.18 {g_{33}}
   {\theta_1}}{{g_{13}}}-\frac{1}{\sqrt{2}} & \frac{0.02 {g_{33}}^3 {\theta_1}^3}{{g_{13}}^3}-\frac{0.07 {g_{33}}^2
   {\theta_1}^2}{{g_{13}}^2}+\frac{0.18 {g_{33}} {\theta_1}}{{g_{13}}}+\frac{1}{\sqrt{2}} & \frac{{g_{33}} {\theta_1}}{{g_{13}}}-\frac{0.57
   {g_{33}}^3 {\theta_1}^3}{{g_{13}}^3} \\
   & & \\
 -\frac{{g_{23}} {\theta_1}^3}{\sqrt{2} {g_{13}} \sqrt[5]{N}} & \frac{{g_{23}} {\theta_1}^3}{\sqrt{2} {g_{13}} \sqrt[5]{N}} & \frac{\frac{6.76
   {g_{23}} {\theta_1}^2}{{g_{33}}}-\frac{13.51 {g_{13}}^2 {g_{23}}}{{g_{33}}^3}}{\sqrt[5]{N}} \\
   & & \\
 -\frac{0.006 {g_{33}}^3 {\theta_1}^3}{{g_{13}}^3}+\frac{0.02 {g_{33}}^2 {\theta_1}^2}{{g_{13}}^2}+\frac{0.18 {g_{33}}
   {\theta_1}}{{g_{13}}}+\frac{1}{\sqrt{2}} & \frac{0.006 {g_{33}}^3 {\theta_1}^3}{{g_{13}}^3}+\frac{0.02 {g_{33}}^2
   {\theta_1}^2}{{g_{13}}^2}-\frac{0.18 {g_{33}} {\theta_1}}{{g_{13}}}+\frac{1}{\sqrt{2}} & 1-\frac{0.5 {g_{33}}^2 {\theta_1}^2}{{g_{13}}^2}
\end{array}
\right).\nonumber\\
& &
\end{eqnarray}
Taking first the large-$N$ limit and then the small-$\theta_1$ limit, the inverse of modal matrix ${\cal M}$:
\begin{eqnarray}
\label{Inverse-M}
& & \hskip -1.1in {\cal M}^{-1} = \nonumber\\
& & \hskip -1.1in \left(
\begin{array}{ccc}
 -\frac{0.02 {g_{33}}^2 {\theta_1}^2}{{g_{13}}^2}+\frac{0.18 {g_{33}}
   {\theta_1}}{{g_{13}}}-\frac{1}{\sqrt{2}} & \sqrt[5]{N} \left(\frac{0.008
   {\theta_1}^2 {g_{33}}^5}{{g_{13}}^4 {g_{23}}}-\frac{0.04 {\theta_1} {g_{33}}^4}{{g_{13}}^3 {g_{23}}}+\frac{0.05 {g_{33}}^3}{{g_{13}}^2
   {g_{23}}}\right) &-\frac{0.07 {g_{33}}^2 {\theta_1}^2}{{g_{13}}^2}+\frac{0.18 {g_{33}}
   {\theta_1}}{{g_{13}}}+\frac{1}{\sqrt{2}} \\
 \frac{0.02 {g_{33}}^2 {\theta_1}^2}{{g_{13}}^2}+\frac{0.18 {g_{33}}
   {\theta_1}}{{g_{13}}}+\frac{1}{\sqrt{2}} & \sqrt[5]{N} \left(\frac{0.008
   {\theta_1}^2 {g_{33}}^5}{{g_{13}}^4 {g_{23}}}+\frac{0.04 {\theta_1} {g_{33}}^4}{{g_{13}}^3 {g_{23}}}+\frac{0.05 {g_{33}}^3}{{g_{13}}^2
   {g_{23}}}\right) &-\frac{0.07 {g_{33}}^2 {\theta_1}^2}{{g_{13}}^2}-\frac{0.18 {g_{33}}
   {\theta_1}}{{g_{13}}}+\frac{1}{\sqrt{2}} \\
 \frac{0.07 {g_{33}}^3 {\theta_1}^3}{{g_{13}}^3} & -\frac{0.04 {\theta_1}^2 {g_{33}}^5}{{g_{13}}^4 {g_{23}}}-\frac{0.07 {g_{33}}^3}{{g_{13}}^2
   {g_{23}}} & -\frac{0.012 {g_{33}}^4 {\theta_1}^4}{{g_{13}}^4}
\end{array}
\right).\nonumber\\
& &
\end{eqnarray}
The orthonormality of the sechsbeins in Sec. {\bf 5} is verified below.
\begin{itemize}
\item
We see that:
\begin{equation}
\label{phi1phi1-a}
G_{\phi_1\phi_1}\sim g_{11}\sqrt{g_sN}\theta_1^2\sim\alpha_\theta^{-1}\sqrt{g_sN}\epsilon^{5}\sim\sqrt{g_sN}N^{\frac{1}{5}}.
\end{equation}
Also,
\begin{eqnarray}
\label{phi1phi1-b}
& & \left(e^4_{\phi_1}\right)^2 + \left(e^5_{\phi_1}\right)^2 + \left(e^6_{\phi_1}\right)^2\nonumber\\
& & \sim {\cal O}(1)\sqrt{g_sN}N^{\frac{1}{5}}\biggl({\cal O}(1) + 0.07 \frac{g_{33}^6}{g_{13}^5}\biggr)\nonumber\\
& & \sim {\cal O}(1)\sqrt{g_sN}N^{\frac{1}{5}}\biggl({\cal O}(1) + 0.07 \alpha_\theta \biggr),
\end{eqnarray}
implying consistency.

\item
We see that:
\begin{equation}
\label{phi2phi2-a}
G^{IIA}_{\phi_2\phi_2}\sim0.
\end{equation}
Also
\begin{eqnarray}
\label{phi2phi2-b}
& & \left(e^4_{\phi_2}\right)^2 + \left(e^5_{\phi_2}\right)^2 + \left(e^6_{\phi_2}\right)^2\nonumber\\
& & \sim{\cal O}(1)\sqrt{g_s}N^{\frac{7}{10}}\left(N^{\frac{2}{5}}(0.04)^2\frac{g_{33}^8}{g_{23}^2g_{13}^2} + (0.07)^2\frac{g_{33}^6}{g_{13}^2g_{23}}\right)\nonumber\\
& & \sim{\cal O}(1)\sqrt{g_s}N^{\frac{7}{10}}\left(N^{\frac{2}{5}}\alpha_\theta^{10}\frac{1}{g_{23}^2} +
\alpha_\theta^6\frac{1}{g_{23}}\right)_{\alpha_\theta\sim N^{-\frac{1}{5}},\epsilon\stackrel{<}{\sim}1}\sim0,
\end{eqnarray}
hence consistent.
\item
We see that:
\begin{equation}
\label{psipsi-a}
G^{IIA}_{\psi\psi}\sim\frac{g_{33}\sqrt{g_s}N^{\frac{7}{10}}}{\theta_1^2}\sim\sqrt{g_s}N^{\frac{7}{10}}.
\end{equation}
Also,
\begin{eqnarray}
\label{psipsi-b}
& & \left(e^4_{\psi}\right)^2 + \left(e^5_{\psi}\right)^2 + \left(e^6_{\psi}\right)^2\nonumber\\
& & \sim {\cal O}(1)\sqrt{g_s}N^{\frac{7}{10}}\left({\cal O}(1) + (0.012)^2\times 0.074\frac{g_{33}^{11}}{g_{13}^3}\right)
\nonumber\\
& & \sim {\cal O}(1)\sqrt{g_s}N^{\frac{7}{10}}\left({\cal O}(1) + {\cal O}(1)\frac{(0.012)^2\times 0.074}{\alpha_\theta^8}\right)_{\alpha_\theta\sim N^{-\frac{1}{5}},N\sim10^2({\rm for}\ g_s\sim0.9}\nonumber\\
& & \sim {\cal O}(1)\sqrt{g_s}N^{\frac{7}{10}},
\end{eqnarray}
hence consistent.
\item
We see:
\begin{equation}
\label{phi1phi2-a}
G^{IIA}_{\phi_1\phi_2}\sim0.
\end{equation}
Also,
\begin{eqnarray}
\label{phi1phi2-b}
& & e^4_{\phi_1}e^4_{\phi_1} + e^5_{\phi_1}e^5_{\phi_2} + e^6_{\phi_1}e^6_{\phi_2}\nonumber\\
& & \sim {\cal O}(1)\sqrt{g_s}N^{\frac{7}{10}}\left(N^{\frac{1}{5}}\frac{g_{33}^4}{g_{13}^3g_{23}}
+ (0.07)^2\frac{g_{33}^6}{g_{13}^5g_{23}}\theta_1^3\right)\nonumber\\
& & \sim {\cal O}(1)\sqrt{g_s}N^{\frac{7}{10}}\left(\frac{10^{-2}}{\alpha_\theta g_{23}} + \frac{(0.07)^2\epsilon^3}{g_{23}}\right)_{N\sim 10^2\ {\rm for}\ \epsilon\sim0.9,\ \alpha_\theta\sim N^{-\frac{1}{5}}}\ll  1,
\end{eqnarray}
implying consistency.

 \item
 We see:
 \begin{equation}
 \label{phi1psi-a}
 G^{IIA}_{\phi_1\psi}\sim - \sqrt{g_s}N^{\frac{7}{10}}.
 \end{equation}
 Also,
 \begin{eqnarray}
 \label{phi1psi-b}
 & & e^4_{\phi_1}e^4_{\psi} + e^5_{\phi_1}e^5_{\psi} + e^6_{\phi_1}e^6_{\psi}\nonumber\\
 & & \sim {\cal O}(1)\sqrt{g_s}N^{\frac{7}{10}}\left( - \left({\cal O}(1)\right)^2 + \left({\cal O}(1)\right)^2 + \epsilon^7\times(0.07\times 0.012)\right)\sim - {\cal O}(1)\sqrt{g_s}N^{\frac{7}{10}},
 \end{eqnarray}
 implying consistency.

 \item
We see:
\begin{equation}
\label{phi2psi-a}
G^{IIA}_{\phi_2\psi}\sim g_{23}\sqrt{g_sN}\stackrel{g_{23}\sim N^{\frac{1}{5}}}{\longrightarrow}\sqrt{g_s}N^{\frac{7}{10}}.
\end{equation}
Also,
\begin{eqnarray}
\label{phi2psi-b}
& &  e^4_{\phi_2}e^4_{\psi} + e^5_{\phi_2}e^5_{\psi} + e^6_{\phi_2}e^6_{\psi}\nonumber\\
& & {\cal O}(1)\sqrt{g_s}N^{\frac{7}{10}}\left({\cal O}(1)\times 0.05 N^{\frac{1}{5}}\frac{g_{33}^3}{g_{13}^2g_{23}}
+ 0.07\times 0.012 \frac{g_{33}^7}{g_{13}^2g_{23}}\theta_1^4\right)_{g_{23}\sim N^{\frac{1}{5}}}\nonumber\\
& & \sim {\cal O}(1)\sqrt{g_s}N^{\frac{7}{10}},
\end{eqnarray}
which is consistent.
\end{itemize}

\end{document}